\documentclass{jfm}

\usepackage{amsmath}
\usepackage{subcaption}
\usepackage{mathtools}
\usepackage{color}
\usepackage{lineno}
\usepackage{booktabs}

\usepackage[plain]{algorithm}
\usepackage{algpascal}

    \usepackage{graphicx}

\usepackage[outdir=./]{epstopdf}
\graphicspath{{FIGURES_Compact/}}
\DeclareMathOperator{\Tr}{Tr}

    \makeatother
\begin{document}

\shorttitle{Multi-Scale Proper Orthogonal Decomposition of Complex Fluid Flows} 
\shortauthor{M. A. Mendez et al} 

\title{Multi-Scale Proper Orthogonal Decomposition of Complex Fluid Flows}

\author
 {
 M.A. Mendez\aff{1}
  \corresp{\email{mendez@vki.ac.be}},
M. Balabane\aff{2}
  \and 
  J.-M. Buchlin\aff{1}
  }

\affiliation
{
\aff{1}
von Karman Institute for Fluid Dynamics, \\Environmental and Applied Fluid Dynamics Department
\aff{2}
 Laboratoire Analyse, G\'{e}om\'{e}trie et Applications, Universit\'{e} Paris 13,  
}

\maketitle

\begin{abstract}
Data-driven decompositions are becoming essential tools in fluid dynamics, allowing for tracking the evolution of coherent patterns in large datasets, and for constructing low order models of complex phenomena. In this work, we analyze the main limits of two popular decompositions, namely the Proper Orthogonal Decomposition (POD) and the Dynamic Mode Decomposition (DMD), and we propose a novel decomposition which allows for enhanced feature detection capabilities. This novel decomposition is referred to as Multiscale Proper Orthogonal Decomposition (mPOD) and combines Multiresolution Analysis (MRA) with a standard POD.
Using MRA, the mPOD splits the correlation matrix into the contribution of different scales, retaining non-overlapping portions of the correlation spectra; using the standard POD, the mPOD extracts the optimal basis from each scale.
After introducing a matrix factorization framework for data-driven decompositions, the MRA is formulated via 1D and 2D filter banks for the dataset and the correlation matrix respectively.
The validation of the mPOD, and a comparison with the Discrete Fourier Transform (DFT), DMD and POD are provided in three test cases. These include a synthetic test case, a numerical simulation of a nonlinear advection-diffusion problem, and an experimental dataset obtained by the Time-Resolved Particle Image Velocimetry (TR-PIV) of an impinging gas jet. For each of these examples, the decompositions are compared in terms of convergence, feature detection capabilities, and time-frequency localization.
\end{abstract}

\section{Introduction and motivation}\label{sec:1}
The analysis of fluid flows hinges upon the identification of coherent flow features out of seemingly chaotic data. These structures could produce dynamic force loads \citep{Vibration_1,Vibration_2}, unstable patterns in the evolution of hydrodynamic instabilities \citep{Carru,Spectral_POD_2b,Insta_2,Insta_3}, transition to turbulence \citep{Trans_1,Trans_2}, enhancement of heat and mass transfer \citep{Heat_1,Heat_2}, noise generation \citep{Noise_1,Noise_2} and more. 

The development of data processing algorithms for detecting coherent features is continuously fostered by the availability of datasets with ever-growing resolutions, and by the ongoing big data revolution which permeates any area of applied science. Identifying the relevant features from high-dimensional datasets is the purpose of data-driven decompositions, which lay the foundation of model order reduction (MOR), data compression, filtering, and pattern recognition, and which are nowadays greatly enlarging the toolbox at the researcher's disposal. Recent reviews on data-driven decomposition with emphasis on fluid dynamics are proposed by \cite{Taira} and \cite{Rowley}, while an extensive overview on the impact of the big data revolution in turbulence research is presented by \cite{Whiter_BOOK}, while \cite{Thomas} give an overview of machine learning methods for flow control.

Data-driven decompositions represent the data as a linear combination of basis elements, referred to as \emph{modes}, having spatial and temporal structures with a certain energy contribution. These decompositions can be classified into two major classes: \emph{energy}-based and \emph{frequency}-based. This work presents a generalized formulation of both approaches, analyzes their respective limits, and proposes a novel hybrid decomposition referred to as Multiscale Proper Orthogonal Decomposition (mPOD).

In the following subsections, we present a literature review on the energy-based and frequency-based data-driven decompositions (\S\ref{1_1}) and outline the recent developments on hybrid methods (\S\ref{1_2}). An overview of the main contribution of this study and the manuscript organization is presented in \S\ref{1_3}.

\subsection{Energy optimality or spectral purity?}\label{1_1}

The fundamental energy based decomposition is the Proper Orthogonal Decomposition (POD). This decomposition originates from the Karhunen-Lo\'{e}ve (KL) theorem \citep{Loeve}, which formulates the optimal (in a least-square sense) representation of a stochastic process as a series of orthogonal functions. This decomposition has set the basis for data processing methods which are known, depending on the field, as Principal Component Analysis (PCA), Hotelling Transform, Empirical Orthogonal Function (EOF) or Proper Orthogonal Decomposition (POD). Literature reviews on these applications can be found in the work of \cite{Cordier}, and the monographs from \cite{Jackson} and \cite{Holmes_BOOK}.

The POD for data-driven analysis of turbulent flows was introduced to the fluid mechanic's community by \cite{Lumley2,Lumley1}, who focused on the spatial structures of the POD modes and their link to coherent structures in turbulent flows. The link between spatial and temporal structures was investigated by \cite{Aubry}, who presented a variant of the decomposition, referred to as bi-orthogonal decomposition (BOD, \citealt{Aubry1991}), which focused on the temporal structures of the modes. This link was made more explicit by \cite{Siro1,Siro2,Siro3}, who proposed a simple algorithm, known as `snapshot POD', to compute the entire decomposition.

It is nevertheless only thanks to the link between the POD and the well known Singular Value Decomposition (SVD) that the calculation of the POD modes of a discrete dataset could be formulated as a simple matrix factorization (see \citealt{Kunisch1999,Fahl, tut4}), in contrast to formulations based on statistics or dynamical system theory (such as \citealt{Holmes, Noack2003,Berkoz,Berkooz2}). For an introduction to the SVD and a historical account on its development, the reader is referred to \cite{Golub2013} and \cite{Stewart1993}.

This simple matrix factorization framework has made the POD an extremely popular tool for both experimental and numerical fluid mechanics. Typical applications include the identification of coherent structures from experimental data (e.g. \citealt{Andrea_2, Sciacchi_POF}), video analysis for adaptive masking or image pre-processing \citep{Mendez_Journal_2, Mendez_J_1}, flow sensing \citep{FLOW_SENS_1,FLOW_SENS_2}, flow control \citep{Brunton_CONTROL,Contr_2}, reduced order modeling \citep{ROM_CFD_4}, experimental validation of CFD data \citep{CFD_EXP_1}, {data-driven identification of non-linear systems \citep{SINDY, Loiseau} and more (see \citealt{Whiter_BOOK})}. Moreover, extended versions of the decomposition, in which the data-set to decompose is constructed by assembling different quantities (e.g. velocities and concentrations), have also found application in the correlation analysis of coupled phenomena \citep{POD_EXTENDED_0,POD_EXTENDED_1,POD_EXTENDED_2,POD_EXTENDED_3,Andrea}.

While the POD optimality guarantees the identification of the most energetic contributions within a few modes, the interpretation of these taken individually, and thus their usage for feature detection, can become difficult. As later illustrated with exemplary test cases, there exist situations in which different phenomena, possibly occurring at largely different frequencies, have very similar energy content. In such cases, an energy-based formulation cannot distinguish the various contributions, which consequently share the same POD structures. Frequency-based methods, conversely, assign to each mode a single frequency. Such methods are data-driven adaptations of the Discrete Fourier Transform (DFT) and infer the frequency of each mode from the data rather than defining it a priori (like in the DFT) from the time discretization. 

The fundamental frequency-based and data-driven method is the Dynamic Mode Decomposition (DMD). This decomposition originates from the Koopman theory \citep{Koopman}, which consists in studying a nonlinear system by mapping it onto a linear one of much larger size (generally infinite dimensional). This theory was then used by \cite{Mezic} to show how finite dimensional approximations of such Koopman (linear) dynamical system
could be used for MOR applications. The extraction of such finite approximation from the data was proposed by \cite{Rowley2} and \cite{Schmid}. An introduction to the Koopman framework is proposed by \cite{Brunton_Koopman}, while extensive reviews on its connection to the DMD are proposed by \cite{Mezic2}, \cite{Budisic} and \cite{Tu_DMD}.  After the first two formulations, briefly reviewed in this work, many extensions of the algorithm have been developed; examples are the optimized DMD \citep{Chen_DMD}, the sparsity-promoting DMD \citep{Opti}, the randomized DMD \citep{Rando}, or the multi-resolution DMD \citep{MultiDMD}. An overview of the DMD and its application is proposed by \cite{Book_DMD}.

Besides the data-driven selection of the frequency of each mode, the advantage of the DMD over the DFT is to allow these to grow or decay exponentially, being each frequency complex. This makes the DMD suited for data-driven stability analysis \citep{Insta_3} and feedback control design for linear systems \citep{Rowley}.

The constraint of fully harmonic modes, on the other hand, can pose significant challenges to frequency-based decompositions for nonlinear datasets. A purely harmonic decomposition cannot easily represent frequency modulations, frequency and phase jitter, or temporally localized events. A large number of modes required for the harmonic description of these phenomena in one domain (e.g., time) results in a substantial redundancy in the structures in the other domain (e.g., space). Moreover, since harmonics have infinite temporal support (they do not `start' nor `finish') their localization capabilities are particularly weak. Extending the frequency domain to the complex plane, as in the DMD, further amplifies these problems in the presence of nonlinear phenomena such as saturation of growth or decay rates.

The limits of harmonic bases have motivated the development of time-frequency analysis and Wavelet theory \citep{Wavelet1, Kaiser}, which has found many applications in the multi-scale modal analysis for fluid flows \citep{Berkooz_WAVE,Wavelet_FLU1,Wavelet_FLU2,Sousa,RINOSHIKA2005}. Wavelet decompositions are nevertheless not data-driven decomposition since one defines a priori the set of basis elements which have a given scale (frequency) and extension (duration). {As described in \S\ref{IV} and Appendix \ref{Annex1}, the proposed mPOD borrows several ideas from the Wavelet literature and combines them with the optimality constraints of the POD.}

\subsection{The need for hybrid decompositions}\label{1_2}

Focusing on data-driven methods, the limits of energy based (POD) or frequency based (DMD) approaches have been recently debated in the literature, and \cite{Focus} has recently discussed the quest for hybrid decompositions.
The first single-harmonic variant of the POD was proposed by \cite{Lumley1} and formulated as a natural extension of the POD for stationary and homogeneous flows (see \citealt{Glauser,George}). This approach consists in computing the POD structures as eigenvectors of the cross-spectral density matrix (see \citealt{Spectral_POD_1,Spectral_POD_2,Spectral_POD_2b}) and is referred by several authors (e.g \citealt{Picard2000,Taira,Towne}) to as Spectral Proper Orthogonal Decomposition (SPOD). \cite{Bourgeois} uses a pre-filtering of the data with a low-pass Morlet filter before computing the POD, to limit the frequency content of the resulting modes. \cite{Camilleri} proposes a combination of POD and DMD, referred to as Cronos-Koopman analysis, which consists in performing DMD on the temporal basis of the POD, to force purely harmonic temporal modes. \cite{Noack_RDMD} uses a recursive approach for forcing the orthogonality in the DMD modes while minimizing the loss of spectral purity.

It is therefore evident that both the energy maximization and the spectral purity are too restrictive constraints, and an ideal decomposition should match the two approaches, possibly allowing for switching from one to the other. A decisive step towards this direction was achieved by \cite{SPOD}, who proposed a novel decomposition blending the POD and the DFT. This decomposition was also named Spectral Proper Orthogonal Decomposition (SPOD), although it has nothing to do with the frequency-based POD that is also referred to as SPOD by \cite{Picard2000,Taira,Towne}.

Sieber's SPOD consists of using a low pass filter along all the diagonals of the correlation matrix, before proceeding the Sirovich's `snapshot POD'. The idea of such diagonal filtering arises from Szeg\"{o} theorem  \citep{Toep_1,Toep_2} which states that the eigenbasis of a Toeplitz Circulant matrix is a Fourier basis. Therefore, the more the correlation matrix resembles a Circulant Toeplitz matrix, the more a POD resembles a DFT. A correlation matrix, however, approaches a Toeplitz Circulant form only for a stationary process \citep{Stat_BOOK,Toep_2} and the diagonal filter used in the SPOD artificially forces such pattern, thus forcing the resulting eigenvectors to approach the DFT basis. Depending on the filter size, one moves from a DFT (very large filter kernel) to a POD (very short filter kernel). Between these two limits, a good compromise between energy optimality and spectral purity can be obtained.

The main limitation of this method, however, is that this filtering operation can significantly alter the correlation matrix of non-stationary processes, compromising its symmetry and thus the orthogonality of its eigenvectors. Such operation can be seen as a smoothing of the nonlinearities \citep{SPOD,Sieber2}, but the decision on the filter size is not supported by a mathematical link between the diagonal filtering and the spectral content of the resulting eigenvectors.

\subsection{Contribution of this work and manuscript organization}\label{1_3}

Section \ref{III} presents a general matrix factorization that is common to all the decompositions, and set the POD, DFT, and DMD in this generalized form. This framework is then used in \S\ref{IV} to derive the novel Multiscale Proper Orthogonal Decomposition (mPOD).

The mPOD proposed in this work combines ideas from the windowed SPOD in \cite{SPOD} and the SPOD in \cite{Towne}.
 Instead of \emph{filtering} the correlation matrix, as in \cite{SPOD}, the mPOD \emph{decomposes} it into the contributions of different scales using the multiresolution (MRA) architecture from Wavelet theory. Instead of computing various eigenbases from the spectra of different portions of the data, as in \cite{Towne}, the mPOD computes eigenbases on the correlation matrix of different scales. The derivation of the mPOD in \S \ref{IV} is carried out by analyzing the impact of the filtering process on the spectra of the correlation matrix and the POD modes. Finally, the mPOD is tested and compared to POD, DFT and DMD on an illustrative synthetic test case (\S\ref{VI}), a numerical test case (\S\ref{VII}) and an experimental test case (\S\ref{VIII}). Conclusions and perspectives for future applications are presented in \S \ref{X}.

\section{Data Decompositions as Matrix Factorizations}\label{III}

Let ${D}(\mathbf{{x}_i},t_k)={D}[\mathbf{i},k]\in \mathbb{R}^{n_s\times n_t}$ be a matrix collecting the set of time realizations (snapshot) of a real variable. In the time domain, the data is sampled with a uniform temporal discretization $\{t_k=(k-1)\Delta t\}_{k=1}^{n_t}$. In the spatial domain, the data is sampled on a Cartesian grid $\mathbf{{x}_i}\in\mathbb{R}^{n_x\times n_y}$, with $\mathbf{i}\in[1,\dots, n_s]$ a matrix linear index, $n_s=n_C\,n_x\,n_y$ the number of points, and $n_C$ the number of components of the data in case of vector quantities (e.g. $n_C=1$ for a pressure field, $n_C=3$ for a 3D velocity field). After reshaping each snapshot of the data into a column vector $d_k[\mathbf{i}]\in\mathbb{R}^{n_s\times 1}$, the data matrix reads

\begin{equation}
\label{Di}
{D}\left [ \mathbf{i}, k\right ]=
\begin{bmatrix}
d_1[ {1}] & \dots & d_{k}[ {1}]& \dots  & d_{n_t}[ 1] \\
    \vdots & \vdots & \vdots &  \vdots & \vdots\\
    d_1[ {n_s}] & \dots & d_{k}[ {n_s}]& \dots  & d_{n_t}[ n_s] 
\end{bmatrix}\,\, .
\end{equation}

The criterion followed to reshape each snapshot into a vector is irrelevant, provided that the same is used when reshaping back the results of the decomposition. The scope of any discrete decomposition is to break this matrix into the sum of rank-1 contributions, referred to as \emph{modes}, written in variable separated form. Each mode has a spatial structure $\phi_r [\mathbf{i}]$, a temporal structure $\psi_r [k]$ and an amplitude $\sigma_r$:

\begin{equation}
\label{F_DECO}
{D}\left [\mathbf{i},k\right ]= \sum_{r=1}^{rk(D)} \sigma_r \phi_r \left [ \mathbf{i}\right ] {\psi}_r [k] \,\,.
\end{equation}

A data matrix of rank $rk(D)=min(n_t,n_s)$ is said to have a full rank, while truncating the summation at $r_c<rk(D)$ leads to an approximation $\tilde{D}$ of rank $r_c$. In the MOR community, this approximation is referred to as \emph{reduced order model} of the data.

In order to let the scalars $\sigma_r$ be fully representative of the energy of each mode, both the spatial and the temporal structures must have unitary energy (norm). The notion of energy is defined by the choice of inner product in the space and in the time domain. Although different definitions have been proposed in the literature \citep{Cordier,Holmes_BOOK}, we herein focus on the classical $L^2$ inner product in its continuous and discrete forms. We shall moreover assume that any data realization (and any spatial or temporal structure) is the discrete representation (sampling) of a continuous and square integrable function. Therefore, let the vector $\phi_r[\mathbf{i}]\in\mathbb{C}^{n_s\times 1}$ collect $n_s$ samples of its continuous counterpart $\phi_r(\mathbf{x})$, over a spatial domain $x\in \Omega$ discretized with a Cartesian grid $\mathbf{x}_i$. An estimation of the average energy content of this function is provided by the Euclidean inner product 

\begin{equation}
\label{inner_space}
\bigl|\bigl|\phi_r(\mathbf{x})\bigr|\bigr|^2_2=\frac{1}{\Omega}\int_{\Omega} \phi_r(\mathbf{x})\,\overline{\phi}_r(\mathbf{x})\, d\Omega\approx \frac{1}{n_s} \sum^{n_s}_{\mathbf{i}=1}\phi_r[\mathbf{i}] \overline{\phi}_r[\mathbf{i}]=\frac{1}{n_s}\langle \phi_r,\phi_r \rangle =\frac{1}{n_s} \phi^\dag_r\,\phi_r  \,,
\end{equation}

{\parindent0pt where} the spatial structure is also reshaped as a column vector, the over-line denotes complex conjugation, $\langle a,b\rangle=a^\dag b$ is the Euclidean inner product of two vectors $a,b$, and $^\dag$ represents the Hermitian transpose. Similarly, assuming that the vector $\psi_r[k]\in\mathbb{C}^{n_t\times1}$ collects $n_t$ samples of the function $\psi_r(t)$, defined in a time domain $t\in[0,T]$, the averaged energy content in the time domain reads:

\begin{equation}
\label{inner_time}
\bigl|\bigl|\psi_r({t})\bigr|\bigr|^2_2=\frac{1}{T}\int_{T} \psi_r(t) \overline{\psi}_r(t)\, dT\approx \frac{1}{n_t} \sum^{n_t}_{k=1}\psi_r[k] \overline{\psi}_r[k]=\frac{1}{n_t}\langle \psi_r,\psi_r \rangle =\frac{1}{n_t} \psi^\dag_r\,\psi_r \,.
\end{equation}   

Equations \eqref{inner_space} and \eqref{inner_time} are numerical approximations of the continuous inner products using the right endpoint rule, and are valuable only if both the spatial and the temporal discretizations are Cartesian. More advanced quadrature methods or non-uniform discretization requires the definition of a weighted inner product \citep{Volkwein,Cordier,Spectral_POD_1}.  To simplify the following derivations, we here focus on Euclidean inner products and uniformly sampled data.

From \eqref{inner_space}-\eqref{inner_time}, it is clear that the normalization $||\phi_r||_2=||\psi_r||_2=1$, at a discrete level, implies that the amplitudes $\sigma_r$ in \eqref{F_DECO} must be normalized by $\sqrt{n_s n_t}$ to produce a grid-independent estimation of the energy of each mode. Before discussing further the link between the energy contributions $\sigma_r$ and the energy of the entire dataset, it is useful to arrange the spatial and the temporal structures into matrices $
{\Phi}=[\phi_1[\mathbf {i}],\dots, \phi_{rk(D)}[\mathbf{i}]]\in \mathbb{C}^{n_s\times rk(D)}$, $
{\Psi}=[\psi_1[k],\dots, \psi_{rk(D)}[k]]\in \mathbb{C}^{n_t\times rk(D)}$. Any decomposition of the form \eqref{F_DECO} becomes a matrix factorization of the form

\begin{equation}
\label{F_DECO_MATRIX}
{D}=  \sum_{r=1}^{rk(D)} \sigma_r \phi_r {\psi}_r^T =
{\Phi} \, 
{\Sigma} \,
{\Psi}^T \, \, ,
\end{equation}

{\parindent0pt where} $\Sigma=diag[\sigma_1,\sigma_2,\dots,\sigma_{rk(D)}]$ is the diagonal matrix containing the energy contribution of each mode, and the superscript $^T$ denotes matrix transposition. Besides allowing for a compact notation, \eqref{F_DECO_MATRIX} allows for visualizing the operation as a projection of the dataset onto a spatial basis $\Phi$ and a temporal basis $\Psi$; the first is a base for the columns of $
{D}$; the second is a base for its rows. 

It is worth noticing that since ${\Sigma}$ is diagonal and ${\Phi}$ and ${\Psi}$ have normalized columns, the factorization in \eqref{F_DECO_MATRIX} is entirely defined by either ${\Phi}$ or ${\Psi}$. In this work, we focus on projections in the time (row) domain of the dataset, i.e., \eqref{F_DECO_MATRIX} is defined by the temporal structures $\Psi$.  
From \eqref{F_DECO_MATRIX}, a matrix inversion and a column-wise normalization allow for computing the spatial structures and the amplitudes:

\begin{equation}
\label{PHI_CALC}
\Phi \,\Sigma={D} \bigl ({\Psi}^T\bigr)^{-1}\,=C\rightarrow \Phi = C\,\Sigma^{-1}\,,
\end{equation}

{\parindent0pt where} $\sigma_r=||C_r||_2$ and $C_r$ is $r^{th}$ column of the matrix $C={D} \bigl ({\Psi}^T\bigr)^{-1}$.

Particularly convenient is the case of an orthonormal temporal basis, for which $\Psi^\dag\,\Psi=\Psi^T\, \overline{\Psi}=I$, with $I$ the identity matrix. In this case, the inversion becomes $\bigl ({\Psi}^T\bigr)^{-1}=\overline{\Psi}$ and the energy contributions, which become $\sigma_r=||D\,\psi_r||_2=||\sigma_r\,\phi_r||_2$, can be computed independently for each mode. 

{Among the infinite possible bases $\Psi$, we focus on the three most popular ones: the POD, the DFT and the DMD.} Their matrix form is discussed in the following subsections, using the subscripts $\mathcal{P}$, $\mathcal{F}$ and $\mathcal{D}$ are used to distinguish the decompositions. 

{Before proceeding, it is worth commenting on two important aspects of any decompositions: 1) the treatment of the time (column-wise) average, 2) the link between the contributions $\sigma_r$ and the total energy content in $D$.} Concerning the first, one can see from \eqref{F_DECO} and \eqref{F_DECO_MATRIX} that removing the time average is equivalent to taking the normalized constant vector $\psi_{\mu}=\underline{1}/\sqrt{n_t}\in\mathbb{R}^{n_t\times 1}$ as one of the temporal structures, such that

\begin{equation}
\label{MEAN_REM}
{D}\left [\mathbf{i},k\right ]=d_\mu[\mathbf {i}]\,\psi_{\mu}[k]+\breve{D}\left [\mathbf{i},k\right ]= \sigma_\mu\,\phi_{\mu}[\mathbf{i}]\psi_{\mu}[k]+\sum_{r=1}^{rk(D)-1} \breve \sigma_r \breve \phi_r \left [ \mathbf{i}\right ] \breve {\psi}_r [k] \,\,
\end{equation}


{\parindent0pt where $d_\mu[\mathbf {i}]=D\,\psi_{\mu}$} is proportional to the average column of the dataset, and the energy content $\sigma_\mu=||d_\mu||$ is proportional to the averaged correlation level, to be normalized by $\sqrt{n_s\,n_t}$ to provide a grid independent estimation. The $\breve{D}\left [\mathbf{i},k\right]$ is the zero-mean shifted dataset and the last summation is its modal decomposition (distinguished by an accent $\breve{\bullet}$). The zero-mean shifted dataset has $rk(\breve{D})=n_t-1$ and all its temporal structures --if an orthogonal decomposition is considered-- must have zero mean. This is naturally the case for the DFT, but not for the POD or the DMD.

Concerning the second point, the energy content of the dataset can be defined via different matrix norms. Using the most classical Frobenius norm yields 

\begin{equation}
\label{FROB}
||D||^2_F={\Tr (\underbrace{D^\dag D}_{K})}={\Tr (\underbrace{D D^\dag}_C)}={\sum^{rk(D)}_{r=1}\,\lambda_r}\,,
\end{equation}

{\parindent0pt {where}} {$\Tr$ indicates the trace of the matrices $K=D^\dag D\in \mathbb{R}^{n_t\times n_t}$ and $C=D D^\dag \in \mathbb{R}^{n_s\times n_s}$.} {These are the finite dimensional estimators of the two-point temporal and spatial correlation tensors \citep{Holmes_BOOK, Cordier} respectively, using the inner products in \eqref{inner_space}-\eqref{inner_time}. These symmetric positive definite matrices share the same non zero eigenvalues $\lambda_r$}. Introducing \eqref{F_DECO_MATRIX} in \eqref{FROB} gives two ways of measuring the total energy from the contribution of each mode:

\begin{equation}
\label{FROB2}
||D||^2_F={\Tr (\underbrace{\Psi\,\Sigma\,\Phi^\dag\,\Phi\,\Sigma\,\Psi^\dag}_K)}={\Tr (\underbrace{\Phi\,\Sigma\,\Psi^{\dag}\,\Psi\,\Sigma\,\Phi^\dag}_C)}={\sum^{rk(D)}_{r=1}\,\lambda_r},
\end{equation}

 It is therefore evident that it is not possible to compute the total energy of the data by solely using the energy contributions of the modes, unless \emph{both} the spatial and the temporal structures are orthonormal ($\Phi^\dag\, \Phi=\Psi^\dag\, \Psi=I$). This case is described in \ref{POD_INTRO}.

\subsection{Matrix form of the Proper Orthogonal Decomposition (POD)}\label{POD_INTRO}

The POD temporal structures $\Psi_\mathcal{P}$ are the eigenvectors of the temporal correlation matrix $K=D^\dag D$. If $D$ is real, as in all the applications considered in this work, these are orthonormal and real vectors:

\begin{equation}
\label{Key_eq}
K={\Psi}_\mathcal{P} \Lambda_\mathcal{P} \Psi_\mathcal{P}^T=\sum^{rk(D)}_{r=1}\,\lambda_{\mathcal{P}_r}\,\psi_{\mathcal{P}_r}\,\psi^T_{\mathcal{P}_r}\,.
\end{equation}

Introducing this definition in \eqref{FROB2} is particularly revealing: a comparison of \eqref{Key_eq} to the first factorization in \eqref{FROB2} yields $\Lambda_{\mathcal{P}}=\Sigma_{P}\Phi_{\mathcal{P}}^T\Phi_{\mathcal{P}}\Sigma_{\mathcal{P}}$. Since this eigenvalue matrix is diagonal, one recovers $\Phi_{\mathcal{P}}^T \Phi_{\mathcal{P}}=I$ and $\Lambda_{\mathcal{P}}=\Sigma_{\mathcal{P}}^2$. Therefore, also the spatial structures are real and orthonormal and thus eigenvectors of the spatial correlation matrix $C=D\,D^\dag={\Phi}_\mathcal{P} \Lambda_\mathcal{P} \Phi_\mathcal{P}^T$. Therefore, the energy contribution of each mode becomes $\sigma_{\mathcal{P}_r}=\sqrt{\lambda_r}$, and the energy of the data-set can be recovered from the $\Sigma_{\mathcal{P}}$.

{For the choice of temporal basis in \eqref{Key_eq}, \eqref{F_DECO_MATRIX} yields the well-known Singular Value Decomposition (SVD), here denoted as $D=\Phi_{\mathcal{P}}\,\Sigma_{\mathcal{P}}\Psi^T_{\mathcal{P}}$. The optimality of the decomposition is thus guaranteed by the Eckart-Young-Mirsky theorem, which states that the rank $r_c<rk(D)$ approximation of a matrix $D$ obtained from a truncated SVD minimizes the norm of the error 

\begin{equation}
\label{EMY}
Err(r_c)=\biggl|\biggl|{D}-\sum^{r_c}_{r=1}\sigma_{\mathcal{P}r}\phi_{\mathcal{P}r}\psi_{\mathcal{P}r}^T\biggr|\biggr|_2=\sigma_{\mathcal{P}r_c+1}\,.
\end{equation}

Therefore, each coefficients $\sigma_{\mathcal{P}_{r}}$ represents \emph{also} the $L^2$ error produced by a rank $r-1$ approximation, and the error decay $Err(r_c)\rightarrow 0$ as $r_c\rightarrow rk(D)$ is the strongest possible.

The POD from \cite{Lumley2,Lumley1} computes the spatial structures ${\Phi_\mathcal{P}}$ from the eigendecomposition of $C$; the BOD from \cite{Aubry} computes the temporal structures (referred to as BOD modes) $\Psi_\mathcal{P}$ from the eigendecomposition of $K$. The POD from \cite{Siro1,Siro2,Siro3} computes both $\Phi_\mathcal{P}$ and $\Psi_\mathcal{P}$ by first solving \eqref{Key_eq} (less demanding if $n_t\ll n_s$) and then using \eqref{PHI_CALC}(which reads $\Phi_{\mathcal{P}}=D\,\Psi_{\mathcal{P}}\,\Sigma_{\mathcal{P}}^{-1}$).

Concerning the impact of the mean removal before computing the POD from \eqref{MEAN_REM}, the POD is equipped with a vector of constant in its temporal basis only if such a vector is an eigenvector of $K$. This implies:

\begin{equation}
\label{MEAN_SUB}
K \underline{1}= \sigma_{\mu}^2 \,\underline{1} \quad  \Longleftrightarrow  \quad \sum_{j=1}^{n_t} K[i,j]= \sigma_{\mu}^2 \quad \forall j\in[1,n_t]\,
\end{equation}

{\parindent 0pt that is the sum over any row of $K$ is equal to the corresponding eigenvalue $\sigma_{\mu}^2$.} This occurs for a stationary process, for which the temporal correlation matrix tends towards a circulant form. For a non-stationary process, the mean removal imposes an additional constraint that alters the connection with the SVD. In such condition, one can still recover the result on optimality by considering the decomposition process as a linear affine transformation (see \citealt{Miranda2008} for a derivation).

Finally, it is worth highlighting that the energy optimality is linked to the decomposition uniqueness: in case of repeated singular values, which occurs when different coherent phenomena have a similar energy content (as shown in \S\ref{VI}), there exist infinite possible choices of singular vectors $\Phi_{\mathcal{P}}$ and $\Psi_{\mathcal{P}}$. 

\subsection{Matrix form of the Discrete Fourier Transform (DFT)}

The DFT temporal structures $\Psi_{\mathcal{F}}$ are defined a priori, regardless of the dataset at hand, as columns of the well known Fourier Matrix

\begin{equation}
\label{MATRICION_F}
\Psi_\mathcal{F}=\frac{1}{\sqrt{n_t}}\begin{bmatrix}
   1& 1 & 1 & \dots & 1 \\
    1& w & w^2 & \dots & w^{n_t-1} \\
    \vdots & \vdots &  \ddots  & \vdots \\
      1& w^{n_t-1} & w^{2 (n_t-1)} & \dots & w^{(n_t-1)^2} \\
\end{bmatrix}
\end{equation}

{\parindent 0pt where $w=\exp \bigl (2 \pi \mathrm {j} /{n_t}\bigr )$,} with $\mathrm {j}=\sqrt{-1}$. Each temporal structure is thus a complex exponential $\psi_r[k]=\exp \bigl (2 \pi \mathrm {j}f_r\,(k-1)\Delta t\bigr )=\exp \bigl (2 \pi \mathrm {j}(r-1)(k-1)/n_t\bigr)$ with real frequency discretization $f_r=(r-1)\Delta f$, with $\Delta f=1/(n_t\Delta t)=f_s/n_t$ the frequency resolution and $f_s$ the sampling frequency. 

{Besides being orthonormal ($\Psi^{-1}_{\mathcal{F}}={\Psi}^{\dag}_{\mathcal{F}}$), this Vandermonde matrix is also symmetric ($\Psi_\mathcal{F}^T=\Psi_\mathcal{F}$). The resulting spatial structures $\Phi_\mathcal{F}=D\overline{\Psi}_\mathcal{F}\,\Sigma_\mathcal{F}^{-1}$ from \eqref{PHI_CALC} are complex and generally not orthogonal.} Finally, since the DFT satisfies \eqref{MEAN_REM} by construction, the zero-mean shifting has no effects on the decomposition.

\subsection{Matrix form of the Dynamic Mode Decomposition (DMD)}\label{DMD}

{The DMD temporal structures $\Psi_{\mathcal{D}}$ are computed under the assumption that a linear dynamical system can well approximate the dataset. It is thus assumed that a propagator matrix $ P\in \mathbb{R}^{n_s\times n_s}$ maps each column of the data matrix $ D$ onto the following one via simple matrix multiplication}

\begin{equation}
\label{Prop}
d_{k}= \, {P} d_{k-1}=\, {P}^{k-1} d_{1}\, \,.
\end{equation}

As each time step $k$ involves the $k-1$ power of $P$ acting on the initial data $d_1$, the evolution of such linear system depends on the eigendecomposition $P=S\Lambda S^{-1}$ of the propagator, since ${P}^{k-1}={S} \, {\Lambda}^{k-1} {S}^{-1}$:

\begin{equation}
\label{Eigen}
d_{k}[\mathbf{i}]= {S} \, {\Lambda}^{k-1} \Bigl({S^{-1}} \, d_1[\mathbf{i}]\Bigr) ={S} \, {\Lambda}^{k-1}{a_0} =\sum_{r=1}^{n_s} a_0[r]  s_{r}[\mathbf{i}] \lambda_r^{k-1} \,\,.
\end{equation}  

The columns of the eigenvector matrix $S=[s_1,s_2,\dots, s_{n_s}]$ represent the spatial basis of such evolution. Observe that the possible growth/decay of each mode makes the notion of mode `amplitude' particularly cumbersome in the DMD. Instead, the vector $a_0={S}^{-1} \, d_1$ is the projection of the initial data $d_1$ onto the eigenvectors of the propagator. To arrange \eqref{Eigen} into the factorization form \eqref{F_DECO_MATRIX}, one should first arrange the vector ${a_0}$ into a diagonal matrix ${A}_0=diag(a_0)$ and then build the temporal basis from the powers of the eigenvalues. The resulting factorization can be written as

\begin{equation}
\label{Eigen_MATRIX}
{D}=S\, {A}_0 \,Z^T   \quad with \quad {Z}^T=\begin{bmatrix}
    1& \lambda_1 & \lambda^2_1 & \dots & \lambda^{n_t}_1 \\
    1& \lambda_2 & \lambda^2_2 & \dots & \lambda^{n_t}_2 \\
    \vdots & \vdots &  \ddots  & \vdots \\
    1& \lambda_{n_s} & \lambda^{2}_{n_s}& \dots & \lambda^{n_t}_{n_s} \\
\end{bmatrix}\,\,.
\end{equation}  

The temporal basis of the DMD is thus obtained by normalizing each column of $Z$ independently, that is $\psi_{\mathcal{D}r}=Z_r/||Z_r||$, and the spatial structures can be computed using the projection in \eqref{PHI_CALC} with no need for computing the eigenvectors $S$. In this final step, it is important to observe that the DMD temporal basis is generally not orthogonal and thus the matrix inversion is generally unavoidable.

%
%

To compute the DMD, one needs to calculate the eigenvalues of the best (in a $L^2$ sense) propagator describing the dataset. This propagator can be easily defined by arranging the data matrix into two shifted portions, ${D}_1=[d_1,d_2,\dots d_{n_t-1}]$ and ${D}_2=[d_2,d_3,\dots d_{n_t}]$, containing $n_t-1$ realization:

\begin{equation}
\label{Prop_M}
{D}_2\approx [{P} d_1, {P} d_2, \dots {P} d_{n_t-1}]= {P} {D}_1 \Longleftrightarrow P={D}_2\,{D}^{+}_1\,,
\end{equation}

{\parindent0pt where ${D}^{+}_1$ denotes the pseudo-inverse of the first portion of the dataset. Except for the very special case in which $n_s=n_t$ and the matrix ${D}_1$ is invertible, the definition of this propagator is not well posed in the classical sense of Hadamard. Moreover, even resorting to the classical Moore-Penrose inverse ${D}^{+}_1=\Psi_{\mathcal{P}}\,\Sigma_{\mathcal{P}}^{-1}\,\Phi^T_{\mathcal{P}}$}, the size of this matrix is usually prohibitively large ($P\in\mathbb{R}^{n_s \times n_s}$) and thus the DMD decomposition based on the eigenvalues of $P$ (referred to as \emph{exact} DMD in \citealt{Tu_DMD}) is rarely of practical interest. Many variants of the DMD have been developed in the last few years as reviewed in \S.\ref{1_2}. In this work, we only consider the two original DMD algorithms to compare with the proposed mPOD.

The first DMD algorithm, herein denoted as cDMD, reproduces the action of the propagator $P$ with a much smaller matrix $\mathcal{C}\in \mathbb{R}^{(n_t-1)\times(n_t-1)}$ acting on the right side \citep{Rowley2}. This matrix is the Companion matrix, defined so that

\begin{equation}
\label{Compa}
\mathcal{C}=\begin{bmatrix}
   0& 0 & \dots & 0&  c_1 \\
   1& 0 &  &  0&  c_2 \\
   0& 1 &  &  0& \vdots \\
   \vdots&  & \ddots &  & \vdots \\
   0& 0 & \dots & 1 &  c_{n_t-1} \\
\end{bmatrix}\Longleftrightarrow {D}_2={P} {D}_1\approx  {D}_1 \, \mathcal{C}
\end{equation}
\vspace{1mm}

The ones in the sub-diagonals shift the columns of $D_1$ to obtain those of $D_2$; the last column contains the set of coefficients $\mathbf{c}=[c_1,c_2\dots,c_{n_t-1}]^T$ that approximates the last temporal realization $d_{n_t}$ as a linear combination of the previous ones. As an eigenvalue $\lambda$ of $\mathcal{C}$, with eigenvector $v$ such that $\mathcal{C} v = \lambda v$, is also an eigenvalue of ${P}$ with eigenvector $D\,v$, this formulation of the DMD relies on the eigendecomposition of $\mathcal{C}$ and thus the calculation of coefficients $\mathbf{c}$. This is a least square minimization problem, extremely sensitive to noise and the choice of the final dataset. Using a QR factorization $D_1=Q\,R$, the solution can be written as $\mathbf{c}=R^{-1}\,Q^T\,d_{n_t}$. More robust variants of this method are the optimized DMD by \cite{Chen_DMD} or the sparsity promoting DMD by \cite{Opti} in which the $L_2$ minimization problem is extended to the entire dataset and not just on the last realization.

The second DMD algorithm, herein denoted as sDMD, avoids the calculation of $\mathcal{C}$ and projects the problem onto a space of much lower dimension \citep{Schmid}, in which hopefully only a few eigenvectors are relevant. This is the space spanned by the first $r_c$ dominant POD modes of the dataset. 

This propagator is therefore defined as $\mathcal{S}=\tilde{\Phi}_{\mathcal{P}}^T \, {P} \,\tilde{\Phi}_{\mathcal{P}} \in \mathbb{R}^{r_c\times r_c}$, where the tilde denotes an approximation $\tilde{\Phi}_{\mathcal{P}}=[\phi_{\mathcal{P}1},\dots\phi_{\mathcal{P}r_c}]$. Writing the full size propagator using the Moore-Penrose inverse, this projected propagator reads

\begin{equation}
\label{PROPA_Schmidt}
\mathcal{S}=\tilde{\Phi}_{\mathcal{P}}^T \, {P} \, \tilde{\Phi}_{\mathcal{P}}=\tilde{\Phi}_{\mathcal{P}}^T \, {D}_2 \,\tilde{\Psi}_{\mathcal{P}} \tilde{\Sigma}^{-1}_{\mathcal{P}}\,.
\end{equation}

 Observe that \eqref{PROPA_Schmidt} is a similarity transform, and thus $\mathcal{S}$ and $P$ share the same eigenvalues, only if  $\tilde{\Phi}_{\mathcal{P}}$ is a square $n_s\times n_s$ matrix (that is the full POD basis plus its $n_s-n_t$ orthogonal complements are taken into account so that $\tilde{\Phi}_{\mathcal{P}}\,\tilde{\Phi}_{\mathcal{P}}^T=I $). That is never the case, and one heuristically hopes that at least the first $r_c$ eigenvalues of $P$ are available in $\mathcal{S}$, which implies that $\tilde{\Phi}_{\mathcal{P}}\,\tilde{\Phi}_{\mathcal{P}}^T\approx I$. 

Although numerically more robust, the ill-posedness of this POD-based projection is easily illustrated with the test cases in \S\ref{VI} and \S\ref{VII}. Specifically, major problems occur when the POD basis captures the energy content of the dataset within a few modes, each of which possibly having several essential frequencies. In this condition, the projected propagator in \eqref{PROPA_Schmidt} acts on a space which is too small to allow for a sufficient number of frequencies, and the DMD decomposition diverges exponentially fast.

Finally, concerning the mean-shifted form in \eqref{MEAN_REM}, one should notice that the DMD can provide the constant vector $\psi_{\mu}=\underline{1}/\sqrt{n_t}$ as basis element only if $\lambda=1$ is an eigenvalue of the propagator ($\mathcal{C}$ for the cDMD, $\mathcal{S}$ for the sDMD). While there is no simple relation to know when this is the case for $\mathcal{S}$, one can see from the characteristic polynomial of $\mathcal{C}$ that this condition sets $\langle \underline{1},\mathbf{c} \rangle=0$, that is the sum of all the coefficients should be null.
 This condition is not generally imposed in the calculation of the DMD, so the corresponding basis might not necessarily have a mode to represent the temporal average of data. On the other hand, removing the mean from the dataset results in $\mathbf{c}=-\underline{1}$ regardless of the dataset, since the last realization must cancel the summation. In this case, as shown by \cite{Chen_DMD}, the solution of the characteristic polynomials are the $n_t-1$ roots of unity, and the DMD reduces to the DFT.


\section{The Multiscale Proper Orthogonal Decomposition (mPOD)}\label{IV}

{The temporal basis $\Psi_{\mathcal{M}}$ for the Multiscale Proper Orthogonal Decomposition (mPOD) proposed in this work is computed by setting frequency constraints to the classical POD.} 

{The algorithm consists of three major steps. First, the temporal correlation matrix $K=D^{\dag} D$ is split into the contributions of different scales via multi-resolution analysis (MRA). This can be performed via 2D Wavelet Transform or more generally via an appropriate filter bank that preserves the symmetry of the correlation matrix.} {Second, each of these contributions is diagonalized as in the classical POD, to obtain the orthonormal basis corresponding to each scale. Third, the bases of each scale are merged into the final temporal basis $\Psi_{\mathcal{M}}$, and the decomposition is completed via \eqref{PHI_CALC}.}

The theoretical foundation of the algorithm is described in \S\ref{SUB1}, \S\ref{SUB2} and \S\ref{SUB3n}. In particular, \S\ref{SUB1} investigates the link between the frequency content in $D$, in $K$ and in its eigenvectors $\Psi_{\mathcal{P}}$. Section \ref{SUB2} introduces the generalized MRA of $K$ via filter banks, discussing its impact on the eigenvectors of different scales and the necessary conditions to keep these mutually orthogonal. \{Section \ref{SUB3n} discusses the link between POD, mPOD and DFT}, while \S\ref{SUB3} presents the proposed mPOD algorithm.

\subsection{Mode Spectra and Correlation Spectra}\label{SUB1}

{The multiresolution analysis (MRA) is introduced in \S\ref{SUB2} in the frequency domain, that is along the $2D$ Fourier transform of the correlation matrix. It is thus interesting to first analyze the standard POD in the frequency domain.} 
{At the scope, we denote as $\widehat{A}$ the Discrete Fourier Transform (DFT) of a matrix $A$, whether this is done along its rows, columns, or both. Using \eqref{MATRICION_F}, it is easy to see that the DFT} of a column vector is obtained via left multiplication by $\overline{\Psi}_\mathcal{F}$, while the DFT of a row vector is a right multiplication by $\overline{\Psi}_\mathcal{F}$.

{The three Fourier pairs to be linked in this section are} related to the time evolution of the data ($D$, row-wise), the temporal structures of the POD modes ($\Psi_{\mathcal{P}}$, column-wise) {and the temporal correlation matrix ($K$, over both columns and rows)}. These are:

\begin{subequations}
\begin{equation}
\label{Pair_2}
 \widehat{D} =D\,\overline{\Psi}_{\mathcal{F}}  \Longleftrightarrow  D=\widehat{D} \, {\Psi}_{\mathcal{F}}\,
\end{equation}
\begin{equation}
\label{Psi_hat}
 \widehat{{\Psi}}_{\mathcal{P}} =\overline{\Psi}_{\mathcal{F}} \,{\Psi}_{\mathcal{P}} \Longleftrightarrow  \Psi_{\mathcal{P}}= {\Psi}_{\mathcal{F}}\,\widehat{{\Psi}}_{\mathcal{P}} 
\end{equation}
\begin{equation}
\label{K_hat} \widehat{K} =\overline{\Psi}_{\mathcal{F}}\, K\,\overline{\Psi}_{\mathcal{F}}  \Longleftrightarrow K= {\Psi}_{\mathcal{F}}\, \widehat{K}\,{\Psi}_{\mathcal{F}}\,.
\end{equation}
\end{subequations}

A first link between the DFT of the dataset \eqref{Pair_2} and that of the POD modes \eqref{Psi_hat} is identified by the cross-spectral density matrix, defined and linked to the temporal correlation matrix as follows:

\begin{equation}
\label{K_F_i}
K_{\mathcal{F}}=\widehat{D}^\dag\,\widehat{D}= \Psi_{\mathcal{F}} \,  \bigl [D^\dag\, D\bigr ] \,\overline{\Psi}_{\mathcal{F}}=\Psi_{\mathcal{F}} \,  K \,\overline{\Psi}_{\mathcal{F}} \Longleftrightarrow  K=\overline{\Psi}_{\mathcal{F}} \,  K_{\mathcal{F}} \,{\Psi}_{\mathcal{F}}\,.
\end{equation}

This matrix is obtained from the temporal correlation matrix $K$ via a similarity transform (since $\Psi_{\mathcal{F}}\,\overline{\Psi}_{\mathcal{F}}=I$), and therefore share the same eigenvalues.

To link its eigenvectors to the POD temporal structures, it is important to observe that all the transformed quantities in \ref{Pair_2}-\ref{K_hat} (that is $D$, $\Psi_{\mathcal{P}}$ and $K$) are real, and thus their Fourier transforms are Hermitian symmetric. For a continuous signal $x(t)$, having Fourier transform $\mathcal{F}\{x(t)\}=X(\omega)$, the Hermitian symmetry sets $X(\omega)=\overline{X(-\omega)}$. For a discrete signal, this property can be introduced via a matrix permutation $P_{\pi}$, which flips the spectra along the zero frequency (first column of $\Psi_{\mathcal{F}}$ in \eqref{MATRICION_F}).

This permutation matrix can be obtained by applying the DFT operator twice:

\begin{equation}
\label{Permu}
P_\pi=\Psi_\mathcal{F}\Psi_\mathcal{F}=\overline{\Psi}_\mathcal{F}\overline{\Psi}_\mathcal{F}=\begin{bmatrix}
   1& 0 & \dots & 0&  0 \\
   0& 0 &  &  0&  1 \\
   0& 0 &  &  1& \vdots \\
   \vdots& \vdots  & \scalebox{-1}[1]{$\ddots$} &  & \vdots \\
   0& 1 & 0 & \dots &  0 \\
\end{bmatrix}\,\,.
\end{equation}

Therefore, for any discrete real vector $a$ with Discrete Fourier Transform $\hat{a}=\Psi_{\mathcal{F}}\,a$, the Hermitian symmetry can be written as $\hat{a}=\overline{P_\pi \hat{a}}$. The fact that the permutation matrix can be obtained by transforming the discrete vector twice reflects the symmetry (or duality) property of the Fourier transform.

Using the Hermitian symmetry on the first two Fourier pairs in (3.1a)-(3.1b) gives:

\begin{subequations}
\begin{equation}
\label{symmetry}
\widehat{\Psi}_{\mathcal{P}}=\overline{P_\pi\,\widehat{\Psi}_{\mathcal{P}}} \Longleftrightarrow  \overline{\widehat{\Psi}_{\mathcal{P}}}=P_\pi\,\widehat{\Psi}_{\mathcal{P}}\, \Longleftrightarrow  \Psi_{\mathcal{F}}\,\Psi_{\mathcal{P}}=P_\pi\,\widehat{\Psi}_{\mathcal{P}}
\end{equation}
\begin{equation}
\label{symmetry2}
\widehat{D}=\overline{\widehat{D}\,P_\pi} \Longleftrightarrow  \overline{\widehat D}=\widehat{D}\,P_\pi\, \Longleftrightarrow  D\,\Psi_{\mathcal{F}}\,=\widehat{D}\,P_\pi\,\,
\end{equation}
\end{subequations}

{Introducing the eigenvalue decomposition of $K$ in \eqref{K_F_i} and using \eqref{symmetry} yields}

\begin{equation}
\label{K_F}
K_{\mathcal{F}}=\Psi_{\mathcal{F}} \, \bigl [{{\Psi}}_{\mathcal{P}}  \Sigma_\mathcal{P}^2 {{\Psi}}_{\mathcal{P}}^\dag\bigr ] \,\overline{\Psi}_{\mathcal{F}}=\bigl(\Psi_{\mathcal{F}} \, {{\Psi}}_{\mathcal{P}}\bigr)  \Sigma_\mathcal{P}^2 \Bigl(\Psi_{\mathcal{F}} \, {{\Psi}}_{\mathcal{P}}\bigr) ^\dag=\bigl(P_\pi \, \widehat{{\Psi}}_{\mathcal{P}}\bigr)  \Sigma_\mathcal{P}^2 \bigl(P_\pi \, {\widehat{\Psi}}_{\mathcal{P}}\bigr) ^\dag\,.
\end{equation}

This shows that the eigenvectors of $K_{\mathcal{F}}$ are the permuted spectra of the POD modes. While the similarity argument is essential in any spectral formulation of the POD \citep{Glauser,Spectral_POD_1,Spectral_POD_2,Spectral_POD_2b, Sieber2,Towne}, \eqref{K_F} shows that care must be taken with the conjugation (or the flipping of) the eigenvectors before the inverse DFT, if one seeks to compute POD modes from the cross-spectral density matrix in \eqref{K_F_i}.

{
The DFT of the correlation matrix $K$ in \eqref{K_hat}, on the other hand, is not obtained via similarity transform and} is not self-adjoint ($\widehat{K}\neq\widehat{K}^\dag$). Hence, since it does not share the same eigenvalues of $K$ and $K_{\mathcal{F}}$, its diagonalization is of no interest for the purpose of this work. However, because of its relevance on the filtering process, it is important to highlight -- by introducing \eqref{K_F_i} into \eqref{K_hat} -- its link to $K_{\mathcal{F}}$:

\begin{equation}
\label{K_HAT_flip}
\widehat{K}=\overline{\Psi}_{\mathcal{F}} \, \bigl [\overline{\Psi}_{\mathcal{F}} \,  K_{\mathcal{F}} \,{\Psi}_{\mathcal{F}}\bigr ] \,\overline{\Psi}_{\mathcal{F}}=\overline{\Psi}_{\mathcal{F}} \, \overline{\Psi}_{\mathcal{F}} \,  K_{\mathcal{F}}= P_\pi\, K_{\mathcal{F}}\,.
\end{equation}


\subsection{Decomposing Data and Decomposing Correlations}\label{SUB2}

It is now of interest to understand how temporal filtering of the dataset influences the frequency content of the POD modes. In particular, we are interested not in just one filter, but an array of filters, constructed to isolate portions of the frequency spectra.

\begin{figure}
\centering
\includegraphics[width=6.2cm]{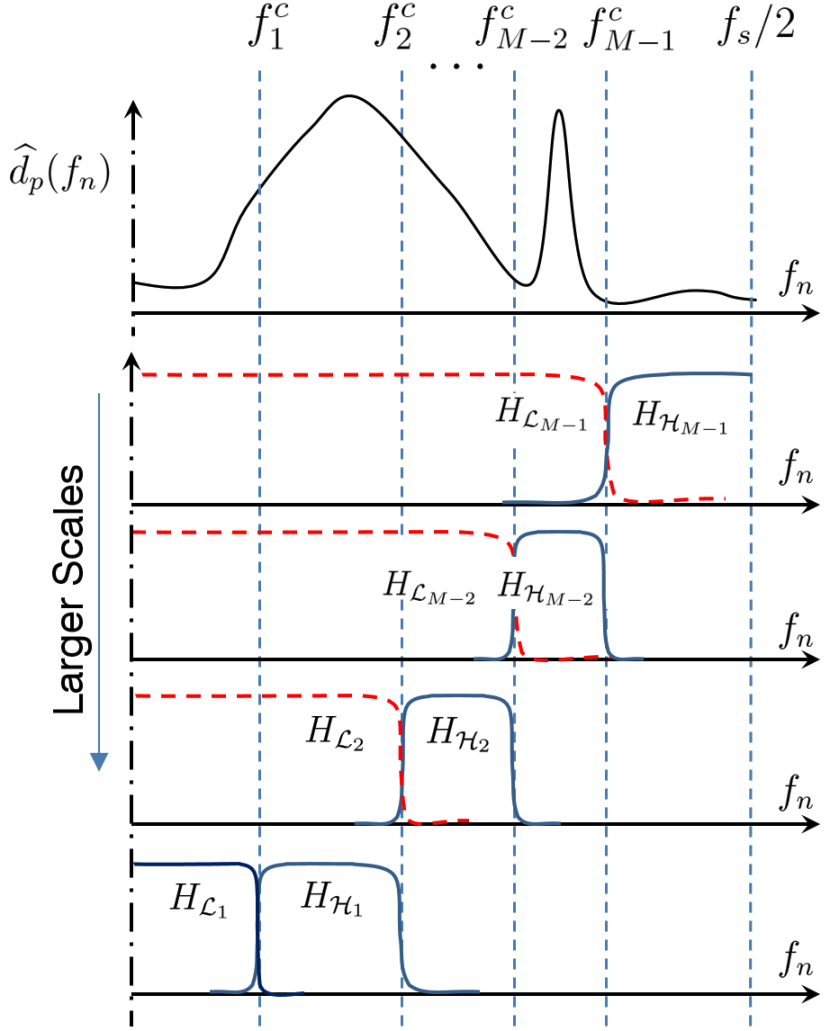}
\caption{Multi-Resolution Analysis (MRA) of a signal $d_p[k]$ along its spectrum $\widehat{d}_p[n]$. {Using a filter bank, the MRA splits the spectra into a set of $M$ scales. This set consists of a large scale (transfer function ${H}_{\mathcal{L}_1}$), a fine scale (transfer function ${H}_{\mathcal{H}_{M-1}}$) and $M-2$ intermediate scales (transfer functions from ${H}_{\mathcal{H}_1}$ to ${H}_{\mathcal{H}_{M-2}}$)}. Only the positive portion of the spectra is shown, the negative one being mirrored along the ordinates.}
\label{Uno}
\end{figure}

To illustrate the idea, let us consider the Fourier transform $\widehat{d}_p=d_p\,\overline{\Psi}_{\mathcal{F}}$ of the temporal evolution {(row-wise)} of the data {for a given point spatial location $\mathbf{i}_p$}, that is $d_p[k]=D[\mathbf{i}_p,k]$. We seek to split the spectra $\widehat{d}_p$ into $M$ scales, each retaining a portion with negligible overlapping as pictorially illustrated in Figure \ref{Uno}. The frequency bandwidths {of these scales} $\Delta f_m=f^c_{m+1}-f^c_{m}$, {with $m\in[1,\dots,M-1]$}, are defined by a frequency splitting vector $F_V=[f^c_1,f^c_2,\dots f^c_{M-1}]$. This vector is bounded by the minimum ($f_0^{c}=0$) and the maximum ({$f^c_M=f_s/2$}) possible frequencies. 

This splitting requires the definition of a filter bank \citep{DWT_Bank_2,Strang_WAVELET,DWT_Bank_1}, consisting of a low-pass filter with cut off $f^c_1$, a high-pass filter with cut off {$f^c_{M-1}$}, and $M-2$ band-pass filters between these. Following the lossless formulation from multiresolution analysis (MRA), these filters are constructed from the set of $m$ lowpass filters with transfer function $\mathcal{H}_{\mathcal{L}_m}$. Of these low pass filters, only the first one ($\mathcal{H}_{\mathcal{L}_1}$) is retained, while the others (shown in figure \ref{Uno} with dashed line) are only used to build the bandpass filters as complementary differences
$H_{\mathcal{H}m}=H_{\mathcal{L}_{m+1}}-H_{\mathcal{L}_{m}}$, the finest one being $H_{\mathcal{H}_{M}}=1-\mathcal{H}_{\mathcal{L}_{M}}$. The resulting set of transfer functions (with modulus sketched in Figure \ref{Uno} with a continuous line) is by construction such that ${H}_{\mathcal{L}_1}+ {H}_{\mathcal{H}_1}+\dots+ {H}_{\mathcal{H}_{M-1}}=\underline{1}$ and thus the entire spectra of the dataset is retained. The design of the low-pass filters in the filter bank used in this work is presented in Appendix \ref{Annex1}; in what follows, we focus on the algebra of the filtering process and the theoretical foundation of the proposed mPOD.

Considering the filtering in the frequency domain, this operation is performed via simple multiplication (entry by entry), which in matrix notation corresponds to the Hadamard product $\odot$. Moreover, since each of these filters is applied in the entire spatial domain (along each row of $D$), one should copy the transfer functions $H_{\mathcal{L}_m}$ or $H_{\mathcal{H}_m}$ (row vectors $\in \mathbb{C}^{1\times n_t}$) to obtain $n_s\times n_t$ matrices having all the rows equal to $H_{\mathcal{L}_m}$ or $H_{\mathcal{H}_m}$. Let $H'_m$ be the general transfer function (whether low-pass or band-pass), we denote with an apex the bi-dimensional extension of the 1D transfer function. The contribution of a scale $m$ from figure \ref{Uno} in each of the spatial points (the filtered portion of the dataset) is 

\begin{equation}
\label{DM}
D_m=\overbrace{\Bigr[\underbrace{\bigl(D\, \overline{\Psi}_\mathcal{F}\bigr)}_{\widehat{D}}  \odot{H}'_m\Bigr]}^{\widehat{D}_m}{\Psi}_\mathcal{F}\,.
\end{equation}

 The first multiplication is the DFT along the time domain ($\widehat{D}$), the result of the Hadamard multiplication in the square bracket is the filtered spectra of the data ($\widehat{D}_m$, the frequency contribution of the given scale) and the last right multiplication is the inverse DFT from \eqref{Pair_2}. Using \eqref{DM}, the correlation matrix $K_m=D_m^\dag\,D_m$ from each scale contribution reads

\begin{equation}
\label{Km}
K_m=\overline{\Psi}_\mathcal{F}\Bigl[\Bigl(\widehat{D} \odot {H}'_m\Bigr)^\dag\Bigl(\widehat{D} \odot{H}'_m\Bigr)\Bigr]{\Psi}_\mathcal{F}=\overline{\Psi}_\mathcal{F}\Bigl[\Bigl(\widehat{D}^\dag  \widehat{D} ) \odot\Bigl(({H}'_m)^\dag \odot{H}'_m\Bigr)\Bigr]{\Psi}_\mathcal{F}\,,
\end{equation}

{\parindent0pt  where} the associative and commutative properties of the Hadamard product are used to move the transfer functions. Introducing the cross-spectral density matrix $K_{\mathcal{F}}$ in \eqref{K_F_i} and writing the 2D transfer function as $\underline{\mathcal{H}}_m=({H}'_m)^\dag \odot{H}'_m$ yields

\begin{equation}
\label{Km_2}
K_m=\overline{\Psi}_\mathcal{F}\underbrace{\Bigl[K_{\mathcal{F}}\, \odot\underline{\mathcal{H}}_m\Bigr]}_{K_{\mathcal{F}m}}{\Psi}_\mathcal{F}={\Psi}_\mathcal{F}\underbrace{\Bigl[\widehat{K}\, \odot\underline{\mathcal{H}}_m\Bigr]}_{\widehat{K}_m}{\Psi}_\mathcal{F}\,,
\end{equation}

{\parindent0pt  having introduced} \eqref{K_HAT_flip} in the last step. 

This important result lays the foundation of the mPOD. First, the expression on the left of \eqref{Km_2} is a similarity transform: the filtered cross-spectral density $K_{\mathcal{F}m}=K_{\mathcal{F}}\, \odot\underline{\mathcal{H}}_m$ shares the same eigenvalues of the correlation of filtered data $K_m=D_m^\dag D_m$. Second, introducing the eigenvalue decomposition of $K_m$ as in \eqref{K_F}, shows that --up to a permutation-- the eigenvectors of $K_{\mathcal{F}}$ are the DFT of the POD modes of $D_m$:

\begin{equation}
\label{K_Fm}
K_{\mathcal{F}m}=\Psi_{\mathcal{F}} \, K_m \,\overline{\Psi}_{\mathcal{F}}=\Psi_{\mathcal{F}} \, \bigl [{{\Psi}}_{\mathcal{P}m}  \Sigma_{\mathcal{P}m}^2 {{\Psi}}_{{\mathcal{P}m}}^\dag\bigr ] \,\overline{\Psi}_{\mathcal{F}}=\bigl(P_\pi \, \widehat{{\Psi}}_{\mathcal{P}m}\bigr)  \Sigma_{\mathcal{P}m}^2 \bigl(P_\pi \, {\widehat{\Psi}}_{\mathcal{P}m}\bigr) ^\dag\,,
\end{equation}

{\parindent0pt  having introduced} \eqref{symmetry} in the last step. 
 The impact of the filter on the POD modes is revealed by the diagonal entries of $K_{\mathcal{F}m}$ and $K_{\mathcal{F}}$, which are summations of nonnegative real numbers. From \eqref{K_Fm}, using \eqref{symmetry}, these are:

\begin{equation}
\label{K_Fm_diag}
K_{\mathcal{F}m}[i,j]=K_{\mathcal{F}}[i,j]\, \odot\underline{\mathcal{H}}_m[i,j]=\sum_{r=1}^{n_t} \sigma^2_{\mathcal{P}m\,r}\overline{\widehat{\psi}}_{\mathcal{P}m\,r}[i]\,\overline{\widehat{\psi}}_{\mathcal{P}m\,r}^\dag[j]\,.
\end{equation}

If the filter at a given scale $m$ removes one of the entries along the diagonal of $K_{\mathcal{F}}$ ($\underline{\mathcal{H}}_m[i,i]=0$), the associated frequency $\widehat{\psi}_{\mathcal{P}m\,r}[i]$ cannot be present in any of the $r$ POD modes of the filtered data: the POD operating at this scale is constrained within the frequencies allowed by the transfer function.

Moreover, while the expression on the right side of \eqref{Km_2} is not a similarity transform, it shows that the correlation matrix associated to the filtered data $K_m$ can be computed by a simple 2D filtering of the original correlation matrix, saving considerable computational time with respect to $n_s$ filters along the rows of $D$. Equation \eqref{Km_2} shows that the MRA of the dataset can be carried out via 2D MRA of the temporal correlation matrix.

\begin{figure}
\vspace{2mm}
\begin{minipage}[c][6.3cm][t]{.49\textwidth}
   \centering
Approximation $\mathcal{H}_{\mathcal{L}m}=(H^{'}_{\mathcal{L}_m})^\dag\,\odot H^{'}_{\mathcal{L}_m}$\\
   \includegraphics[width=6.6cm]{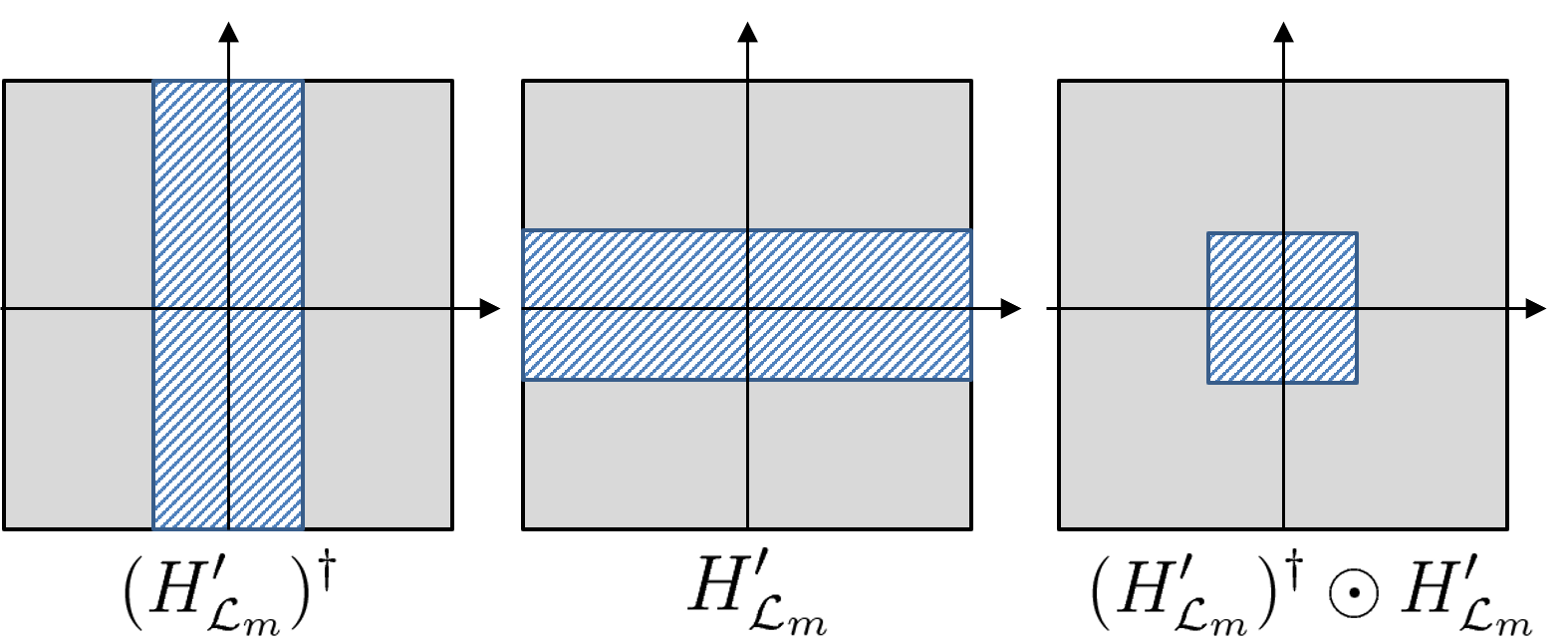}\\
\vspace{1mm}
Diagonal Detail $\mathcal{H}_{\mathcal{L}m}=(H^{'}_{\mathcal{H}_m})^\dag\,\odot H^{'}_{\mathcal{H}_m}$\\
  \includegraphics[width=6.6cm]{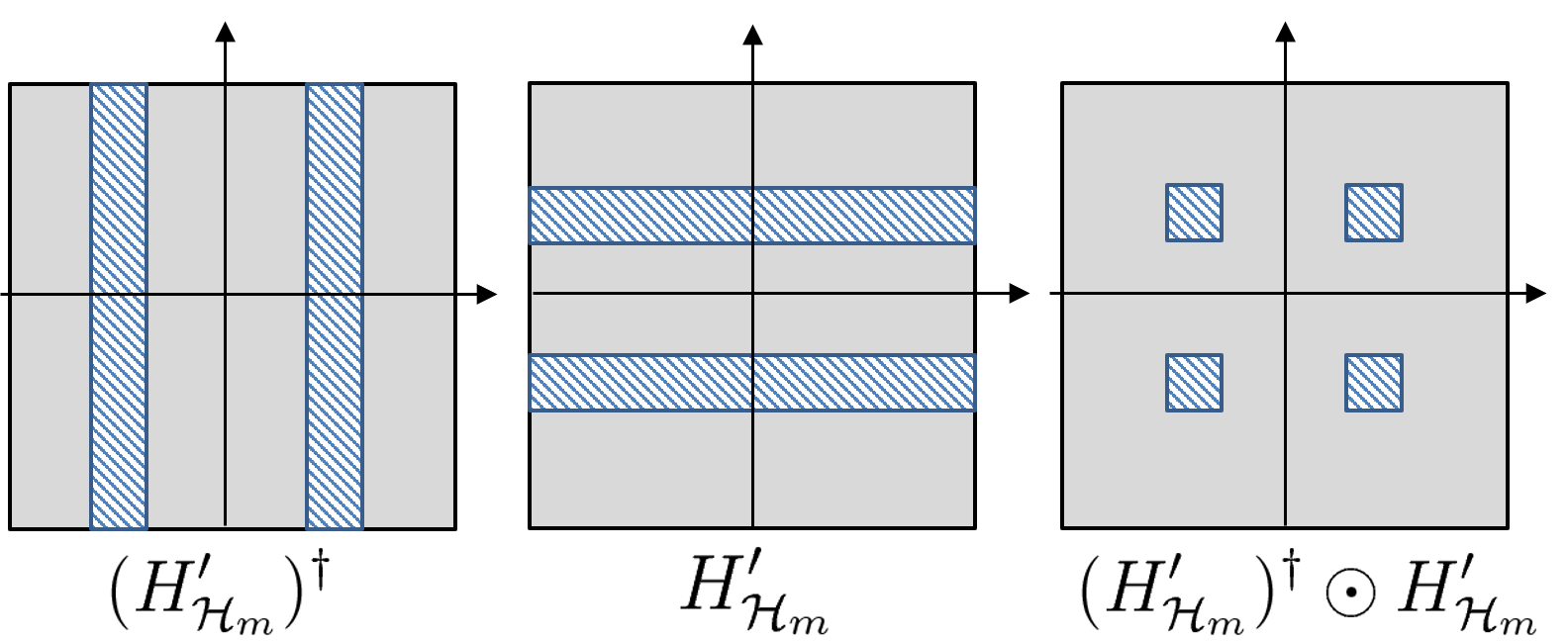}\\
\end{minipage}
\vline{}
\begin{minipage}[c][6.3cm][t]{.49\textwidth}
   \centering
Horizontal Detail $\mathcal{H}_{\mathcal{L}m}=(H^{'}_{\mathcal{L}_m})^\dag\,\odot H^{'}_{\mathcal{H}_m}$\\
   \includegraphics[width=6.6cm]{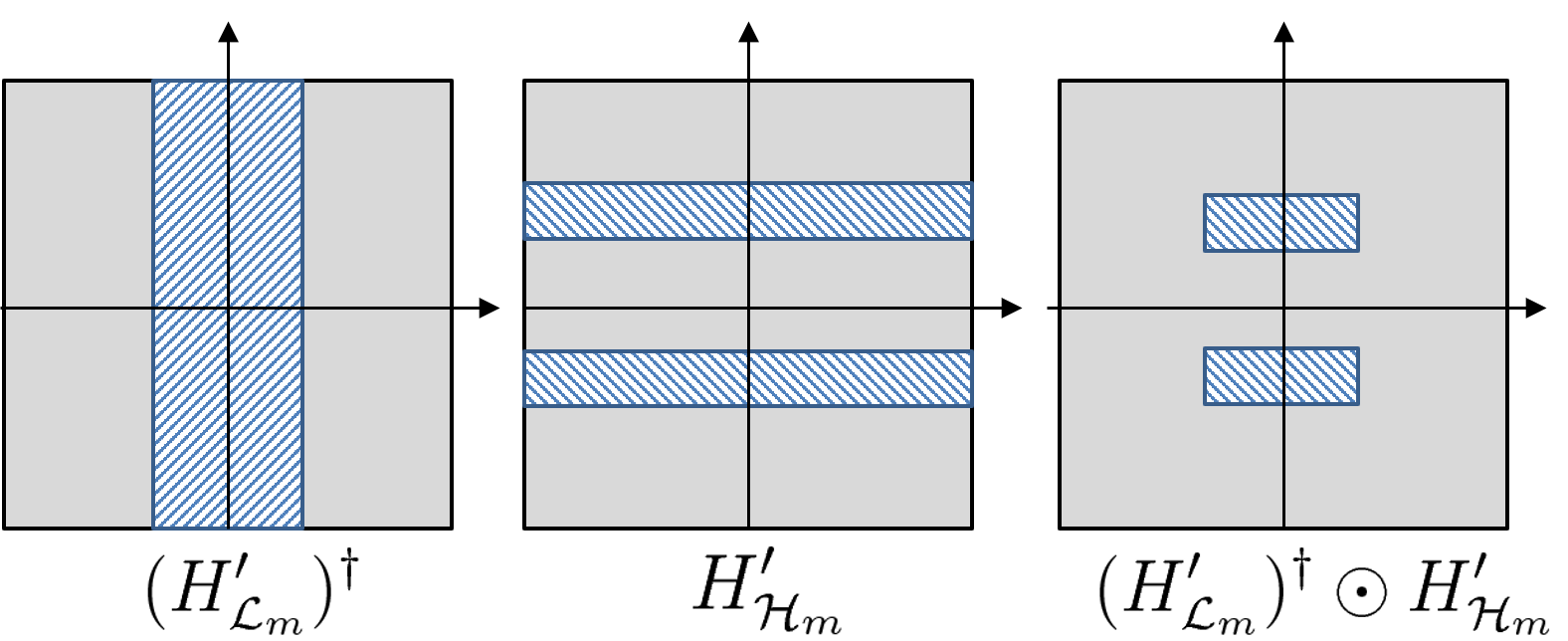}\\
\vspace{1mm}
  Vertical Detail $\mathcal{H}_{\mathcal{L}m}=(H^{'}_{\mathcal{H}_m})^\dag\,\odot H^{'}_{\mathcal{L}_m}$\\
  \includegraphics[width=6.6cm]{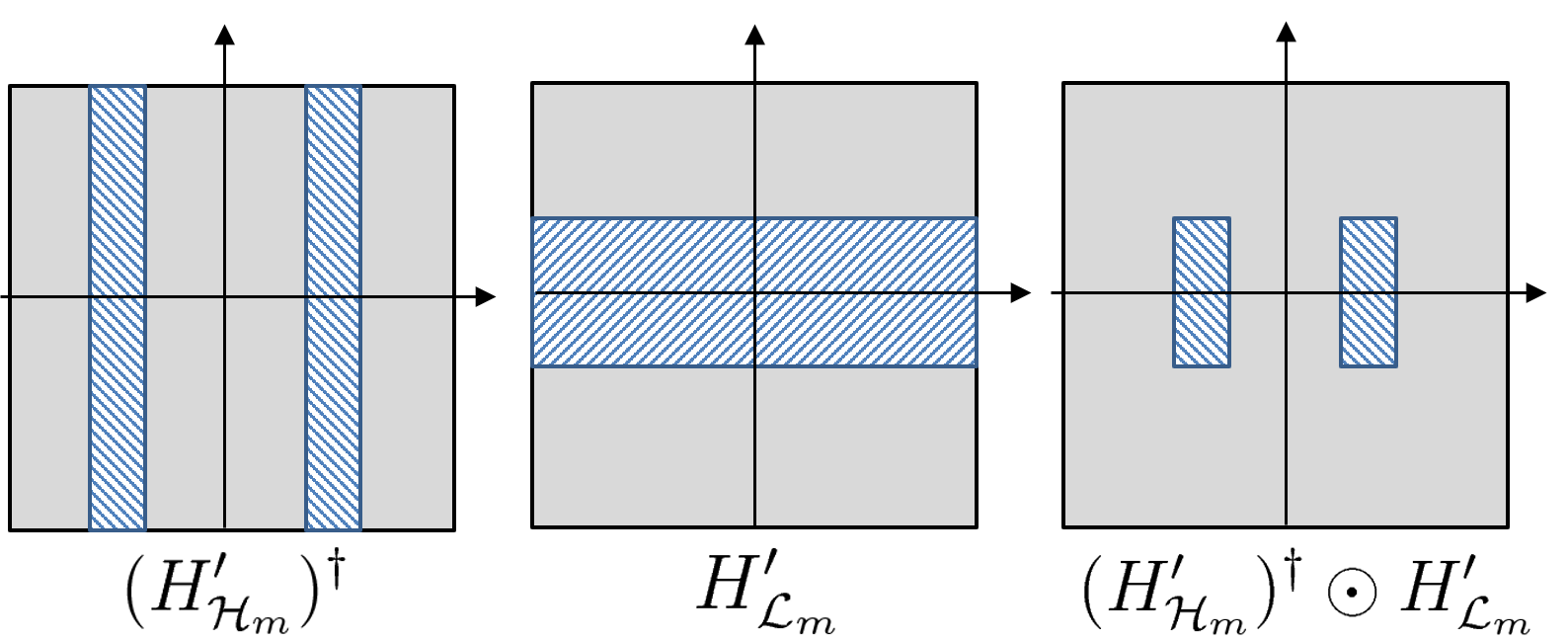}\\
\end{minipage}
\caption{Modulus of the four possible 2D transfer functions $\underline{\mathcal{H}}_m$ that can be obtained by combining the low-pass ($H_{\mathcal{L}_m}$) and the high-pass ($H_{\mathcal{H}_m}$) filters associated to a given $m$ scale of $K$, following the multi-scale architecture in figure \ref{Uno}. The areas where the transfer function is $\sim 1$ and $\sim 0$ are indicated with blue dashed lines and full gray respectively.}
\label{Due}
\end{figure}

The four possible structures for the 2D transfer functions $\underline{\mathcal{H}}_m$ that arise from a combination of the 1D kernels in the time domain are illustrated graphically in Figure \ref{Due}. For a given scale, composed of a low frequency and a high frequency $D_m=D_{\mathcal{L}_m}+D_{\mathcal{H}_m}$, these correspond to the contributions of four temporal correlation matrices:

\begin{equation}
\label{ContriK}
\begin{split}
K_m=D^\dag_m\,D_m&=D^\dag_{\mathcal{L}_m}D_{\mathcal{L}_m}+
D^\dag_{\mathcal{H}_m}D_{\mathcal{H}_m}+D^\dag_{\mathcal{L}_m}D_{\mathcal{H}_m}
+D^\dag_{\mathcal{H}_m}D_{\mathcal{L}_m}\\
&={\Psi}_\mathcal{F}\Bigl[\widehat{K}\odot\underline{\mathcal{H}}_{\mathcal{L}m}+\widehat{K}\odot\underline{\mathcal{H}}_{\mathcal{H}m}+\widehat{K}\odot\underline{\mathcal{H}}_{\mathcal{L}\mathcal{H}m}+\widehat{K}\odot\underline{\mathcal{H}}_{\mathcal{H}\mathcal{L}m}\Bigr]{\Psi}_\mathcal{F}
\end{split}
\end{equation}

The `pure' terms $D^\dag_{\mathcal{L}_m}D_{\mathcal{L}_m}$ and $D^\dag_{\mathcal{H}_m}D_{\mathcal{H}_m}$ can be obtained by first filtering the data \eqref{DM} and then computing the related correlations; the mixed terms can only be revealed by a MRA of the full correlation matrix $K_m$. 

It is worth noticing that the filter bank architecture described in this section is a generalized version of the MRA via Discrete (Dyadic) Wavelet Transform (DWT) proposed in previous works \citep{Mendez_Journal_2,Mendez_ICNAM}. In the 2D wavelet terminology, the four terms in \eqref{ContriK}, with spectral band-pass region illustrated in figure \ref{Due}, correspond to the \emph{approximation} and the \emph{diagonal}, \emph{horizontal} and \emph{vertical} \emph{details} of each scale.

Particular emphasis should be given to the `pure' terms $D^\dag_{\mathcal{L}_m}D_{\mathcal{L}_m}$ and $D^\dag_{\mathcal{H}_m}D_{\mathcal{H}_m}$. These contributions have no frequency overlapping and their eigenspaces are orthogonal complements by construction: their set of eigenvectors can be used to assemble a complete and orthogonal basis as illustrated in \S\ref{SUB3}. The mixed terms $D^\dag_{\mathcal{L}_m}D_{\mathcal{H}_m}$ and $D^\dag_{\mathcal{H}_m}D_{\mathcal{L}_m}$, on the other hand, generates eigenspaces that can potentially overlap with those of the `pure' contributions and must be disregarded in the MRA of the correlation matrix.

\subsection{The mPOD as bridge between POD and DFT}\label{SUB3n}

Neglecting the mixed terms prevents the full reconstruction of the correlation matrix, but not the full reconstruction of the dataset since the final temporal basis is complete. The consequences of this operation can be analyzed while studying how the choice of the frequency splitting vector $F_V$ allows the mPOD to move from the energy optimality of the POD to the spectral purity of the DFT.

\begin{figure}
\centering
\includegraphics[width=3.27cm]{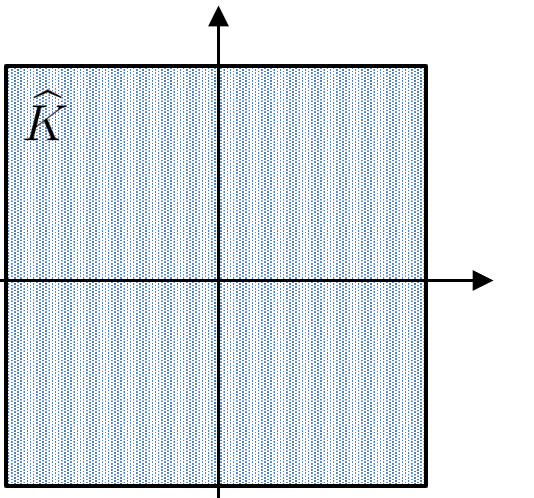}
\includegraphics[width=3.27cm]{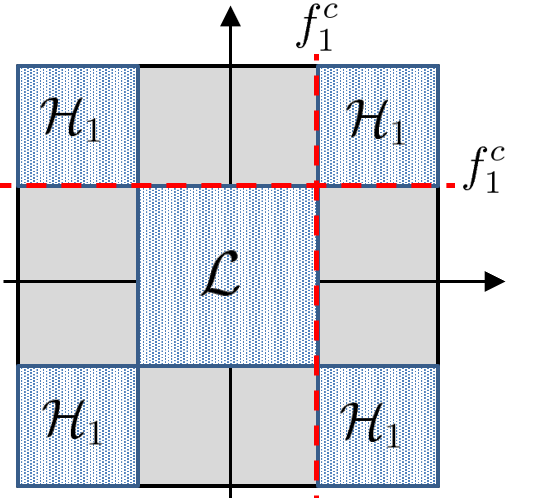}
\includegraphics[width=3.27cm]{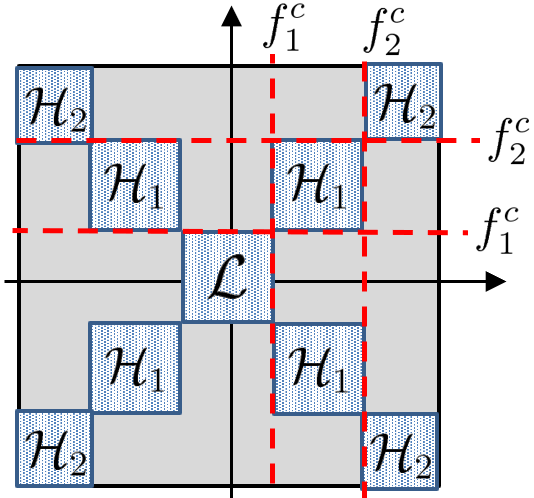}\hspace{1mm}
\includegraphics[width=3.27cm]{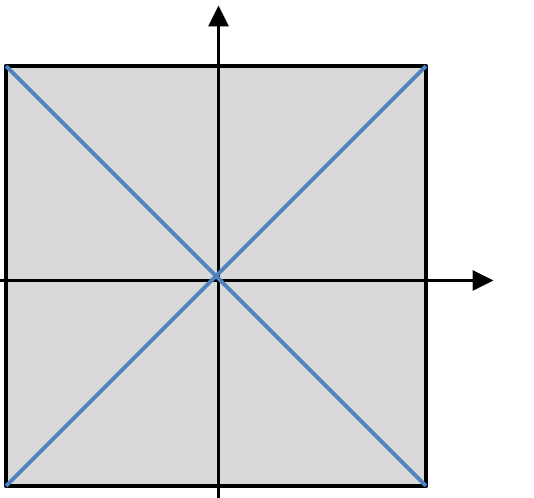}
\\\hspace{0.1cm}a) \hspace{2.9cm} b) \hspace{2.9cm} c) \hspace{2.9cm} d)\hspace{0.6cm}
\caption{Partition of the correlation matrix $\widehat{K}$ for different choices of the splitting vector $F_V$ in the MRA step of the mPOD. The same color notation of Fig.\ref{Due} is used for the band-pass/band-stop regions. Case a) consists of a single scale, thus the mPOD becomes a POD. Case b) and c) consists of two and three scales respectively. Case d) shows the limit of $n_t/2$ scales, in which the mPOD becomes a DFT.}
\label{Added}
\end{figure}

Figure \ref{Added} shows the band-pass region from the 2D spectra of the correlation matrix $\widehat{K}$ for four choices of $F_V$. The case (a) corresponds to a single scale mPOD  (empty $F_V$), that is the standard POD. All the eigenvectors $\psi_{\mathcal{P}_r}$ are allowed to span the entire set of discrete frequencies available. This configuration corresponds to perfect reconstruction of the correlation matrix and maximum time localization capabilities: if an impulse $\delta(t_k)$ is produced at a certain time, the POD is free to choose an impulse (which has a frequency spectra spanning the \emph{entire} frequency axis) as one of its temporal structures. On the other hand, all the POD modes share the full set of frequencies, potentially leading to spectral mixing, and phenomena spanning the entire frequency and energy spectra (such as random noise) are spread over all the modes.

The case (b) corresponds to a two-scale mPOD, that is with $F_V=f_1^c$. Each scale has its own correlation matrix ($K_{\mathcal{L}}$ and $K_{\mathcal{H}_1}$) with no common frequencies. Phenomena occurring at these two scales are spectrally separated and necessarily assigned to different modes. Moreover, the amount of random noise in each mode is distributed proportionally to the spectral bandwidth of the corresponding scale and becomes less important. Part of the information on the correlation matrix is lost, but the eigenspaces of the two scales, together, span the entire $\mathbb{R}^{n_t}$ space and thus preserves the decomposition convergence. On the other hand, the temporal localization capabilities are reduced: as no eigenvector is allowed to span the full frequency range, no temporal structure can capture impulses nor other phenomena requiring the entire frequency range (e.g. shocks).

As the number of scales is increased, their spectral bandwidth reduces and the decomposition becomes spectrally more localized, at the cost of an increased loss of information in the correlation and a reduction of time localization capabilities. The case (c) has three scales, with $F_V=[f^c_1,f^c_2]$, while the case (d) corresponds to the limit $F_V\rightarrow \{r\,f_s/(\Delta t\, n_t)\}^{n_t-1}_{r=1}$, with $r\in[1,\dots,n_t-1]$. In this limiting case, every scale is constrained to a single frequency and the mPOD reduces to the DFT, inheriting its finest spectral resolution and its limits (windowing problems, Gibbs phenomena, no time localization, and poor convergence, \citealt{WINDOWING,Wavelet0, Fourier_BOOK}).

\subsection{The mPOD algorithm}\label{SUB3}

The steps of the proposed mPOD algorithm are listed in the Algorithm \ref{alg:the_alg}. As for any data-driven decomposition, the first step consists in assembling the data matrix as in \eqref{Di}. The second step consists in computing the temporal correlation matrix $K$ and its Fourier transform $\widehat{K}$ from \eqref{K_hat}. From the analysis of the frequency content in the correlation spectra $\widehat{K}$, the third step consists in defining the frequency splitting vector $F_V$ and in constructing the set of associated transfer functions. 

This choice is the fundamental step which will set the spectral constraints to the mPOD modes. The frequency splitting could be user-defined if prior knowledge on the investigated data is available, or automatically computed by centering the band-pass windows of each scale on the dominant peaks in the diagonal of $\widehat{K}$ and splitting the band-pass regions accordingly. In this step, only the `pure' (approximation and diagonal detail) terms of the 2D transfer functions are constructed, that is $\mathcal{H}_{\mathcal{L}_1}=(H'_{\mathcal{L}_1})^\dag \odot H'_{\mathcal{L}_1}
$ and $\mathcal{H}_{\mathcal{H}_m}=(H'_{\mathcal{H}_m})^\dag \odot H'_{\mathcal{H}_m}$. Disregarding the mixed terms (horizontal and vertical details) in each scale, the correlation matrix is approximated as:

\begin{equation}
\label{ALGO}
K\approx {\Psi}_\mathcal{F}\Bigl[\widehat{K}\, \odot\mathcal{H}_{\mathcal{L}_1} \Bigr]{\Psi}_\mathcal{F}+\sum^M_{m=1}{\Psi}_\mathcal{F}\Bigl[\widehat{K}\, \odot\mathcal{H}_{\mathcal{H}_m} \Bigr]{\Psi}_\mathcal{F}\\
\approx K_{\mathcal{L}_1}+\sum^{M-1}_{m=1}K_{\mathcal{H}_m}\,.
\end{equation}

Each of these contributions is a symmetric, real and positive definite matrix, equipped with its orthonormal eigenspace, computed in the fourth step

\begin{equation}
\label{EIGEN_SCALE}
K\approx \Psi_{\mathcal{L}_1}\,\Sigma^2_{\mathcal{L}_1}\Psi^T_{\mathcal{L}_1}+\sum^{M-1}_{m=1}\Psi_{\mathcal{H}m}\,\Sigma^2_{\mathcal{H}_m}\Psi^T_{\mathcal{H}_m}\,.
\end{equation}

If the frequency-overlapping is identically null, the eigenspaces of these pure terms are orthogonal complements, and therefore there are at most $n_t$ non-zero eigenvalues among all the scales. This is a major difference between the mPOD and other multi-scale methods such as Continuous Wavelet Transform (CWT), in which the temporal basis is constructed by shifting and dilating a `mother' function, or the multi-resolution DMD (mrDMD) proposed by \cite{MultiDMD}, in which the temporal basis is constructed by performing DMD on different portions of the datasets. These decompositions potentially produce high redundancy and poor convergence since the basis is larger than $n_t$. Moreover, each of the basis elements in the mPOD exists over the entire time domain (although they could be null in an arbitrarily large portion of it), while CWT or mrDMD produce different bases for different portions of the time domain, leading to decompositions more complicated than those analyzed in this work.

\begin{algorithm}[t!]
{
Input: Set of $n_t$ snapshots of a dataset $D(\mathbf{x}_i,t_k)$ on a Cartesian grid $\mathbf{x}_i\in\mathbb{R}^{n_x\times n_y}$}
\begin{algorithmic}[1]
\hrule 
\vspace{1mm}
\State \textit{Reshape snapshots into column vectors $d_k[\mathbf{i}]\in \mathbb{R}^{n_s\times 1}$ and assemble $D[i,k]$ in \eqref{Di}}
\State \textit{Compute time correlation matrix $K=D^\dag D$ and $\widehat{K}=\overline{\Psi}_{\mathcal{F}}\,K\,\overline{\Psi}_{\mathcal{F}}$ in \eqref{K_hat}}
\State \textit{Prepare the filter bank $\mathcal{H}_{\mathcal{L}_1},\dots \mathcal{H}_{\mathcal{H}_M}$, and split $K$ into M contributions via \eqref{ALGO} }
\State \textit{Diagonalize each `pure' term $K_m=\Psi_m \, \Sigma_m^2 \, \Psi_m^T$ as in \eqref{EIGEN_SCALE}}
\State \textit{Sort the contribution of all the scales into $\Psi^0_\mathcal{M}$ as in \eqref{mPOD_0}}
\State \textit{Enforce orthogonality via QR factorization, $\Psi^0_{\mathcal{M}}=\Psi_{\mathcal{M}} R\rightarrow \Psi_{\mathcal{M}}=\Psi^0_{\mathcal{M}}\,R^{-1}$}
\State \textit{Compute the spatial basis $\Phi_{\mathcal{M}}=D\,\Psi_{\mathcal{M}}\,\Sigma^{-1}_{\mathcal{M}}$ from \eqref{PHI_CALC} and sort the results in descending order of energy contribution.}
\end{algorithmic}
\vspace{1mm}
\hrule 
\vspace{1mm}
{Output: Spatial $\Phi_{\mathcal{M}}$ and temporal $\Psi_{\mathcal{M}}$ structures with corresponding amplitudes $\Sigma_{\mathcal{M}}$ of the $n_t$ mPOD modes.}
\caption{{Multi-scale Proper Orthogonal Decomposition of a dataset $D\in{\mathbb{R}^{n_s\times n_t}}$.}}
\label{alg:the_alg}
\end{algorithm}

Because of the limited size of the filter impulse response (see Annex \ref{Annex1}) of each filter, the number of non-zero eigenvalues among the scales is in practice usually slightly larger than $n_t$. Therefore, only the first $n_t$ dominant eigenvectors are selected. This is done in the fifth step, by first sorting the full set of eigenvalues $[diag(\Sigma_{\mathcal{L}_1}), diag(\Sigma_{\mathcal{H}_1}), \dots,diag(\Sigma_{\mathcal{H}_{M-1}})]$ in decreasing order and storing the required permutation matrix $P_\Sigma$. This matrix is then used to re-arrange the eigenvector matrices $\Psi_{\mathcal{L}},\dots,\Psi_{\mathcal{H}m}$ into one single matrix (temporal basis) regardless of their scale of origin

\begin{equation}
\label{mPOD_0}
{\Psi^0_\mathcal{M}=\biggl [ {\Psi}_{\mathcal{L}_1},\,\Psi_{\mathcal{H}_1}, \,\Psi_{\mathcal{H}_2}\dots\Psi_{\mathcal{H}_{M-1}}\biggr]\,P_{\Sigma}\,\,.}
\end{equation}

As {discussed} in \S\ref{SUB2}, the temporal matrix \eqref{mPOD_0} is orthonormal if a perfect spectral separation is achieved. This is a major difference between the mPOD and the SPOD by \cite{SPOD}, in which the filtering procedure can eventually result in a non-orthogonal temporal basis, or the mrDMD by \cite{MultiDMD}, in which the loss of orthogonality can be produced by growing/decaying modes.

Due to the finite size of the transition band of the filters, this is generally not the case in practice, and a minor loss of orthogonality usually occurs. The sixth step treats this imperfection, polishing this temporal structure via a reduced QR factorization $\Psi^0_\mathcal{M}=\Psi_\mathcal{M}\,R$, so that $\Psi_{\mathcal{M}}=\Psi^0_{\mathcal{M}}\,R^{-1}\in\mathcal{R}^{n_t\times n_t}$. This compensates for the losses of orthogonality and the upper triangular matrix $R$ offers an indication of the quality of the filtering process, being close to the identity when the spectral overlapping is negligible. Finally, in the last step, the decomposition is completed via the projection and the normalization in \eqref{PHI_CALC}, as for any orthogonal decomposition.

\begin{table}
  \begin{center}
\def~{\hphantom{0}}
{
  \begin{tabular}{cccccccc}
        & POD  & DFT  & sDMD  & cDMD  & mPOD (M=3) & mPOD (M=10) & mPOD (M=50)\\
Test Case 1 (\S \ref{VI})   & 1.7  & 5.1  & -     & 15.1  & 2.3     & 2.9       & 6.3  \\
Test Case 2 (\S \ref{VII})  & 10.1 & 14.5 & 53.45 & 63.3  & 24.1  & 54.1    & 235        \\
Test Case 3 (\S \ref{VIII}) & 15.8 & 48.8 & -     & 113.0 & 51.6    & 135    & 595     \\ 
\end{tabular}}
  \caption{ {CPU time (in seconds) for the test cases analyzed in this work and the decompositions tested. The analysis is carried out using Matlab R2018b on a Intel(TM) i7-3770 @3.40 GHz processor with a 32 GB RAM.}}
  \label{tab1}
  \end{center}
\end{table}

To conclude the presentation of the mPOD algorithm, Table \ref{tab1} collects the CPU time for all the decompositions used in this work and the test cases in \S 4-\S5-\S6. The POD is naturally the fastest requiring only an eigenvalue decomposition and a projection, with no need for amplitude normalization and sorting. The DFT is slightly more expensive, mostly due to the normalization and the sorting steps, while the DMD requires the additional cost of a minimization problem or an SVD depending on the chosen algorithm. The computational cost of the mPOD is shown for $M=3,10,50$ and drastically increases with the number of scales considered. 

While the cost of the MRA is usually negligible when using an FFT-based formulation in the frequency domain, the major additional cost is in the diagonalizations for each scale. 
For $M=3$, the CPU time for the proposed mPOD algorithm is well below the CPU time of the DMD and become comparable to it for $M\sim 10$. For a much larger number of scales (see table for $M=50$), the mPOD requires a more efficient implementation of the diagonalization of the various scales. An improved algorithm which performs this step in the frequency domain is under development and will be presented in future work.

\section{Example I: Illustrative Synthetic Test }\label{VI}

\subsection{Dataset Description}

\begin{figure}
\centering
\begin{subfigure}[!b]{.32\textwidth}
  \centering
  \includegraphics[width=2.9cm]{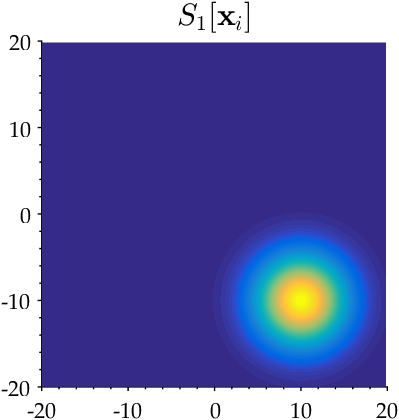}\\
\vspace{3mm}
\includegraphics[width=4.2cm]{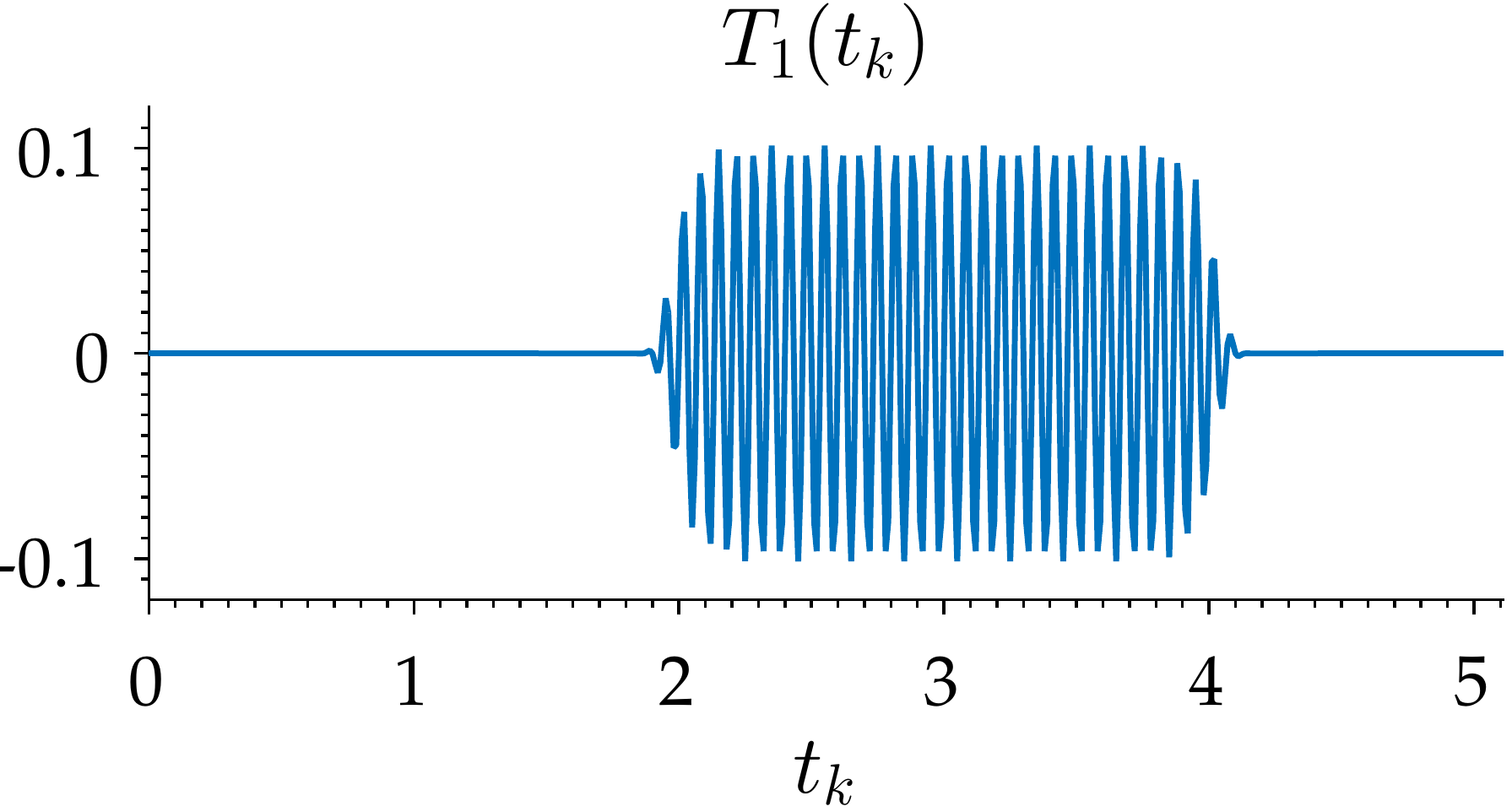}
\end{subfigure}
\begin{subfigure}[!b]{.32\textwidth}
  \centering
  \includegraphics[width=2.9cm]{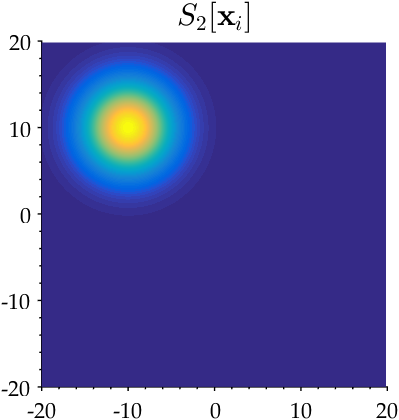}\\
\vspace{3mm}
  \includegraphics[width=4.2cm]{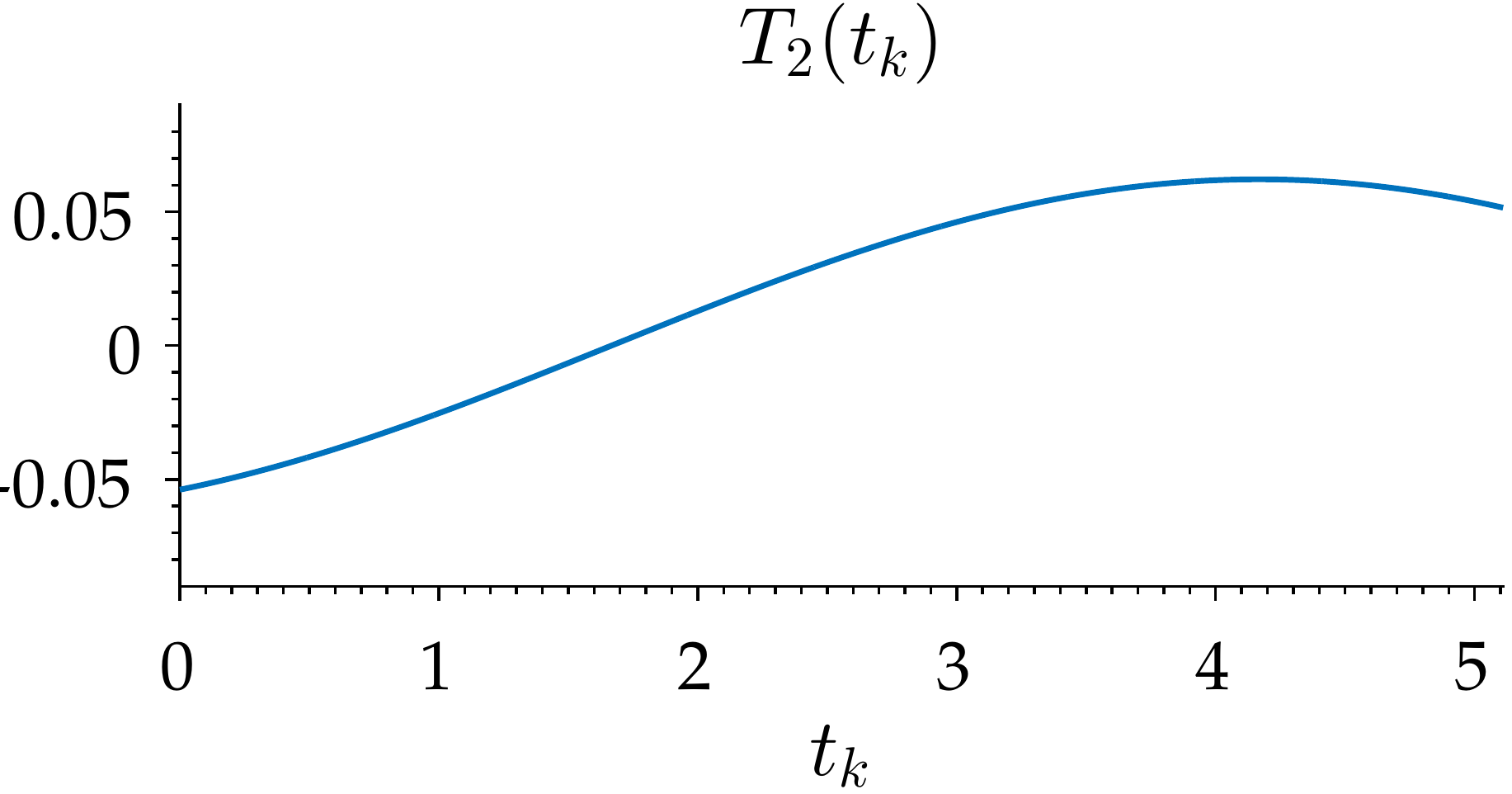}
\end{subfigure}
\begin{subfigure}[!b]{.32\textwidth}
  \centering
  \includegraphics[width=2.9cm]{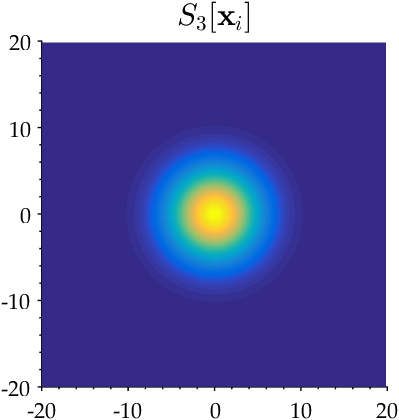}\\
\vspace{3mm}
  \includegraphics[width=4.2cm]{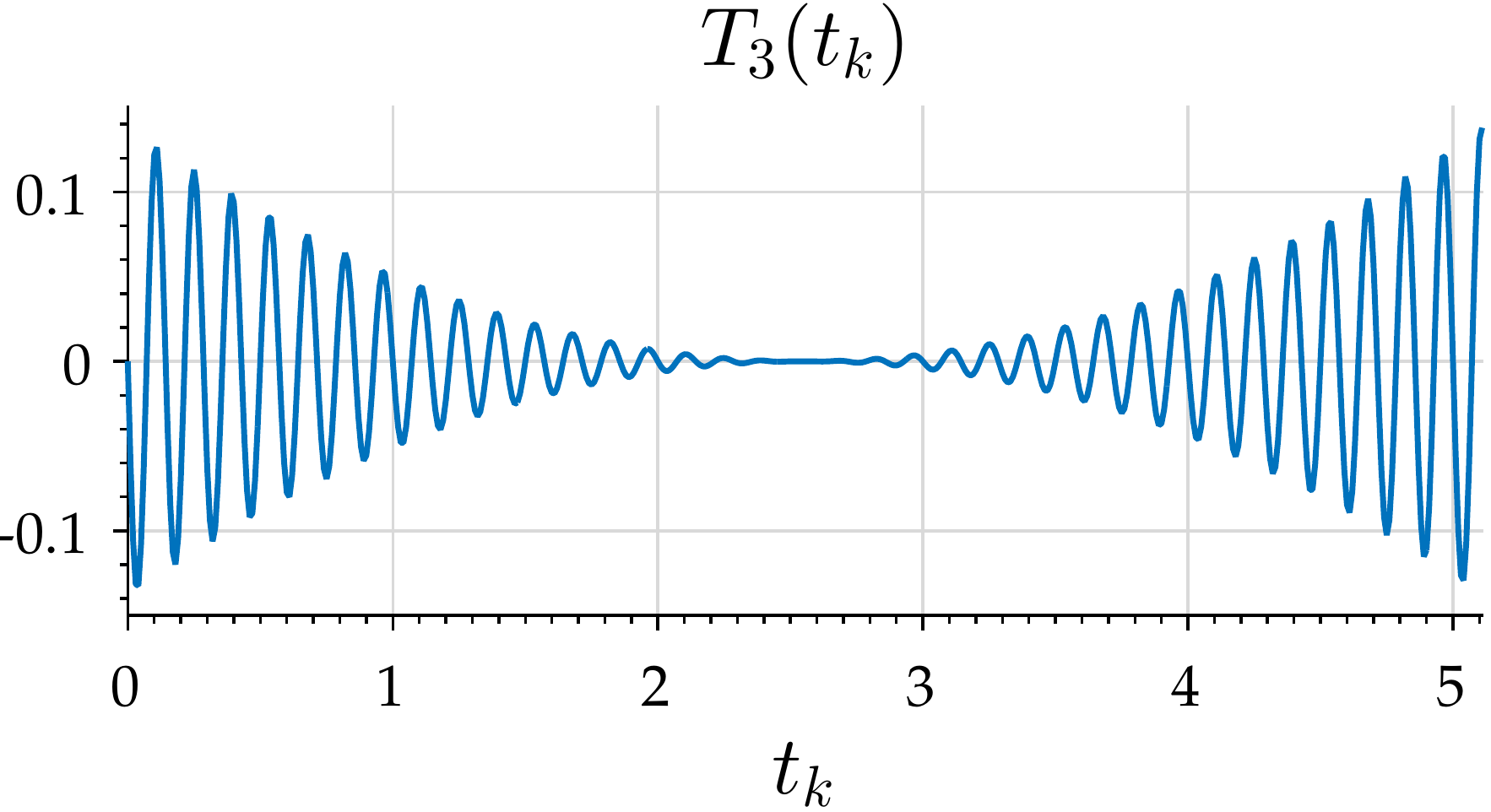}
\end{subfigure}
\caption{Spatial (top row) and corresponding temporal (bottom row) structures of the modes considered in the simplified test case. The amplitude of each contribution is computed so as to give the energy content. {An animation of this test case is available as supplemental Movie 1.}}
\label{Synthetic_Modes}
\end{figure}

As a first example, we propose a simple synthetic test case which is already in the decomposed form of \eqref{F_DECO_MATRIX} and yet capable of severely challenging standard decompositions. The spatial domain consists of a square Cartesian grid of $n_s=256\times256$ points over a domain $\mathbf{x}_i\in[-20,20]\times[-20,20]$, while the time domain spans $t_k\in[0, 5.11]$ with a total of $n_t=256$ points with sampling frequency $f_s=100$. The dataset is composed of the sum of three modes, having spatial and temporal structures as illustrated in Figure \ref{Synthetic_Modes}. These modes consist of identical Gaussians with standard deviations of $\sigma=5$, located in $\mathbf{x}_{1}=[10, -10]$, $\mathbf{x}_{2}=[-10, 10]$ and $\mathbf{x}_{3}=[0, 0]$, pulsing as

\begin{subequations}
\label{TIME}
\begin{equation}
T_1(t_k)= A_1 \,\sin \Bigr(2 \pi f_1 t_k\Bigr)\exp \Biggl [0.5\,(t_k-3)^{20}\Biggr]
\end{equation}
\begin{equation}
T_2(t_k)=  A_2 \,\sin \Bigl(2 \pi f_2 t_k-{\pi}/{3}\Bigr)
\end{equation}    
\begin{equation}
T_3(t_k)=A_3 \,\sin \Bigl(2 \pi f_3 t_k\Bigr)  (t_k-2.55)^2\,.
\end{equation}
\end{subequations}

The first mode has a smoothed square box modulation, the second has a harmonic with period longer than the observation time and the third has a parabolic modulation. The amplitudes $A_k$ are chosen so as to equal each contribution, that is $A_r=1/||S_r\,T^T_r||_2$, having reshaped both spatial and temporal structures as column vectors. The three frequencies are $f_1=15$, $f_2=0.1$ and $f_3=7$. The resulting dataset ${D}[\mathbf{i},k]$ is a matrix of size $65536\times 256$ of rank $rk(D)=3$.

\subsection{Modal Analysis}

In this test case, we compare the performances of the DFT, POD, cDMD and mPOD. The sDMD algorithm, based on the reduced propagator $\mathcal{S}$ in \eqref{PROPA_Schmidt}, is severely challenged by the rank deficiency of this dataset. Since $rk(D)=3$, one has $\mathcal{S}\in \mathbb{R}^{3\times3}$, that is a set of eigenvalues incapable of reproducing the data: each of the dominant frequencies $f_1,f_2,f_3$ requires two complex conjugate modes and thus at least a $6\times6$ propagator. Low dimensional propagators $\mathcal{S}$ that are inadequately too small occur for datasets which are close to being rank deficient, in which few coherent patterns occur at different frequencies.

For all the decomposition, the $L^2$ convergence is calculated as

\begin{equation}
\label{L2}
\frac{Err[r_c]}{||D||
_2}=\frac{||D-\sum^{r_c}_{r=1}\,\sigma_{r}\,\phi_{r}\,\psi_{r}||_2}{||D||_2}\,,
\end{equation}

{\parindent0pt where all the modal contributions} are first sorted in descending order $\sigma_{a}>\sigma_{b}\,,\, \forall a>b$. For the POD, \eqref{L2} reduces to $Err[r_c]/||D||
_2=\sigma_{\mathcal{P}{r}_{c+1}}/\sigma_{\mathcal{P}_{1}}$ since the $L^2$ norm of a matrix is equivalent to its largest singular value (see eq. \ref{EMY}). This is not the case for other decompositions, for which the spatial basis is not orthogonal.

\begin{figure}
\centering
\includegraphics[width=7.5cm]{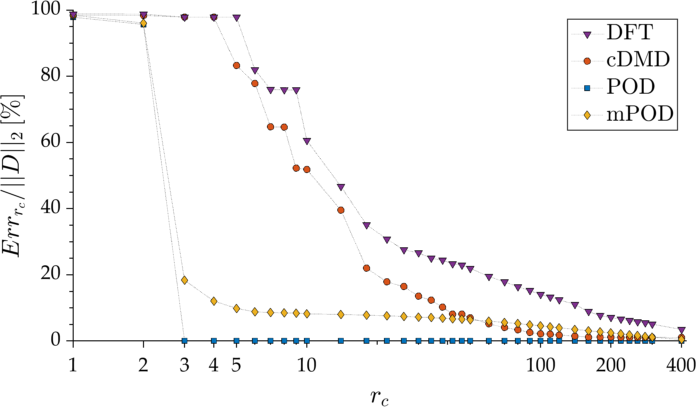}
\caption{Convergence of the POD, cDMD, mPOD and DFT for the illustrative synthetic test case in terms of $L^2$ relative error, defined as in \eqref{L2}. }
\label{Conv_1}
\end{figure}

\begin{figure}
\centering
\begin{subfigure}[c]{.45\textwidth}
  \centering
\includegraphics[width=5cm]{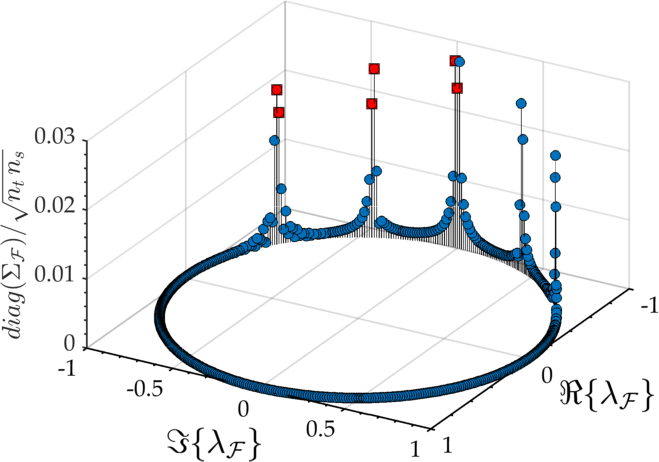}
 \end{subfigure}
 \begin{subfigure}[c]{.45\textwidth}
\includegraphics[width=5cm]{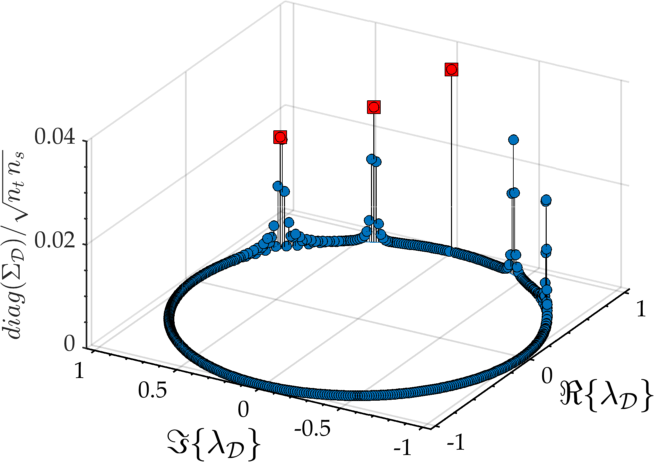}\ \end{subfigure}
\caption{Amplitude spectra of the DFT (left) and cDMD (right) spectra mapped in the complex plane. Both decompositions have temporal evolutions distributed along the unit circle. For the DFT this is true by definition; for the DMD this is the result from the eigenspectra of the Companion matrix assembled from the data.}
\label{COM_1}
\end{figure}

The convergence results of this test case are shown in figure \ref{Conv_1}. As expected, the POD produces no reconstruction error when $r_c\geq 3$, while the convergence of frequency based methods is considerably weaker. The finite duration of the first mode, the aperiodicity of the second, and the nonlinear modulation of the third are not easily derivable as a linear combination of infinite harmonics, and the convergence of the DFT requires up to $150$ modes to bring the relative error below $10\%$. The DMD performs better than the DFT, although yet not comparable to the POD. 
The mPOD of this synthetic test case is performed with $M=4$, $F_V=[0.4,10,20]$ and leads to a convergence much closer to that of the POD than the harmonic decompositions. 
To analyze the detection capabilities of these decompositions, several representative modes for each are now considered. First, the energy contributions of the DFT ($diag(\Sigma_{\mathcal{F}}$)) and the cDMD ($diag(\Sigma_{\mathcal{D}})$) are mapped in the complex plane in Figure \ref{COM_1}. 
By definition, all the DFT modes lie in the unit circle and are Hermitian symmetric; DMD eigenvalues can lay inside the unit circle (if $|\lambda_r|<1$) and exponentially decay, or outside (if $|\lambda_r|>1$) and exponentially grow.

\begin{figure*}
\begin{subfigure}{.29\textwidth}
  \centering
  \includegraphics[width=2.8cm]{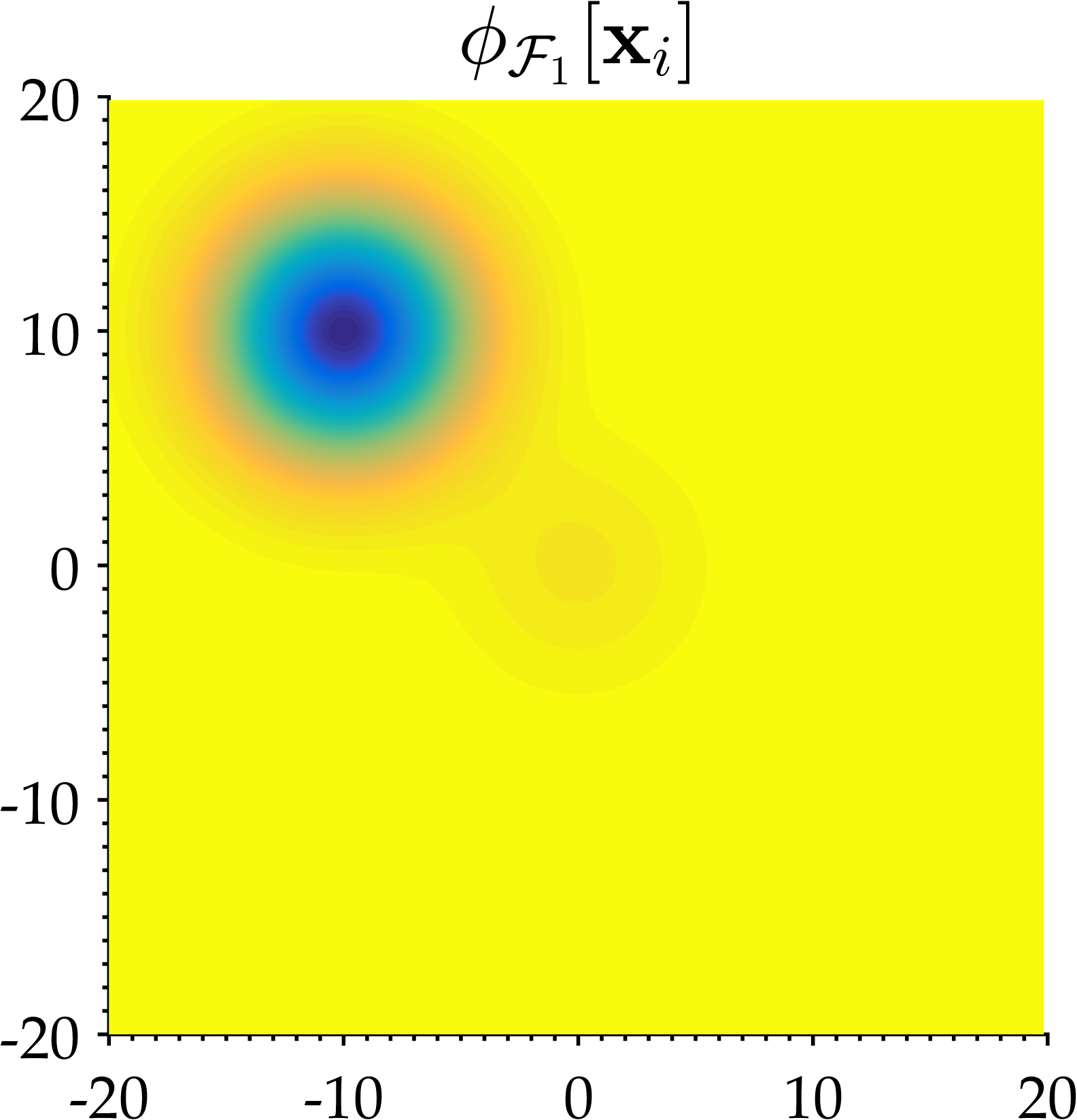}\\
\vspace{1mm}
\includegraphics[width=2.8cm]{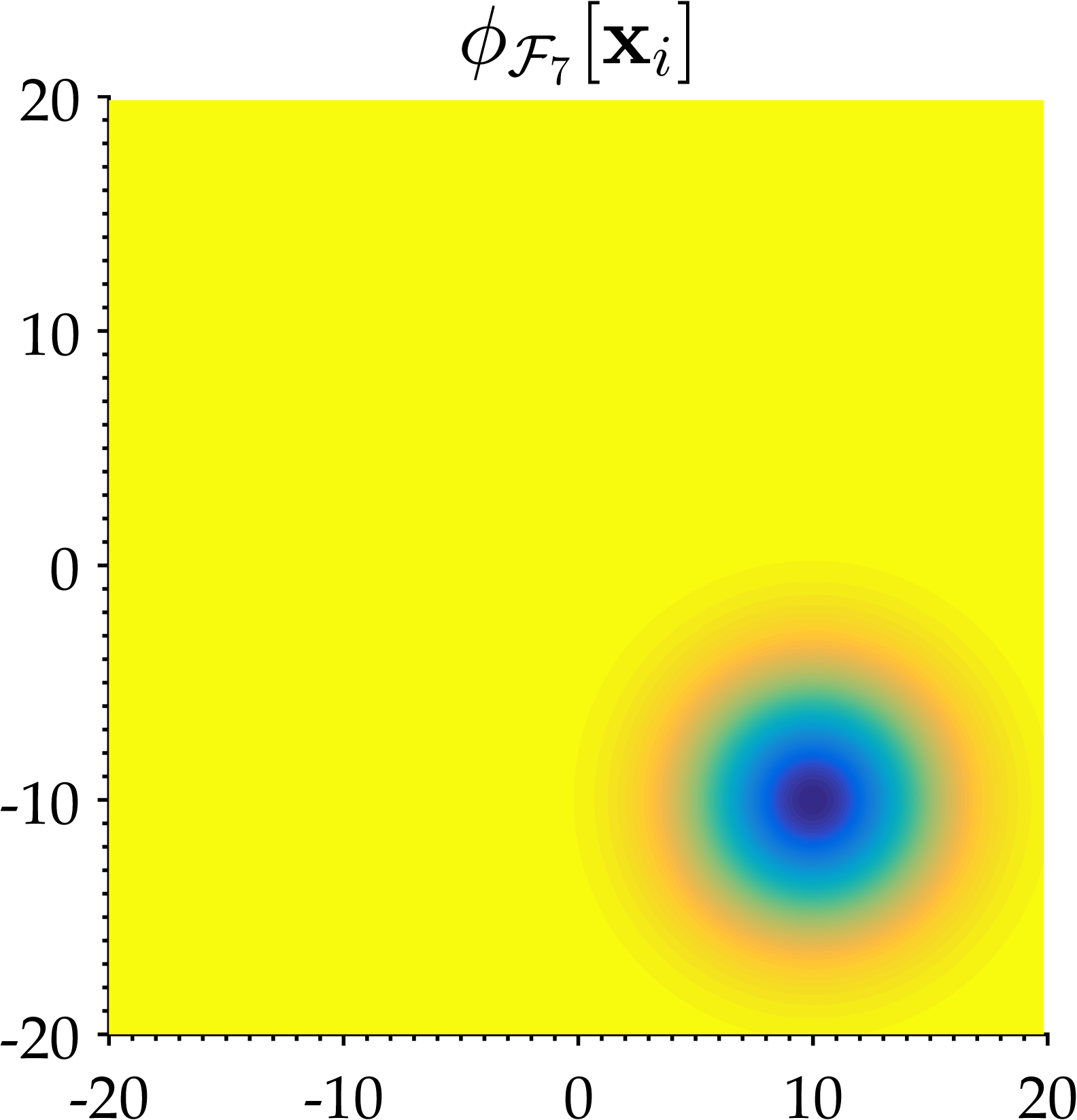}
\end{subfigure}
\begin{subfigure}{.29\textwidth}
  \centering
  \includegraphics[width=2.8cm]{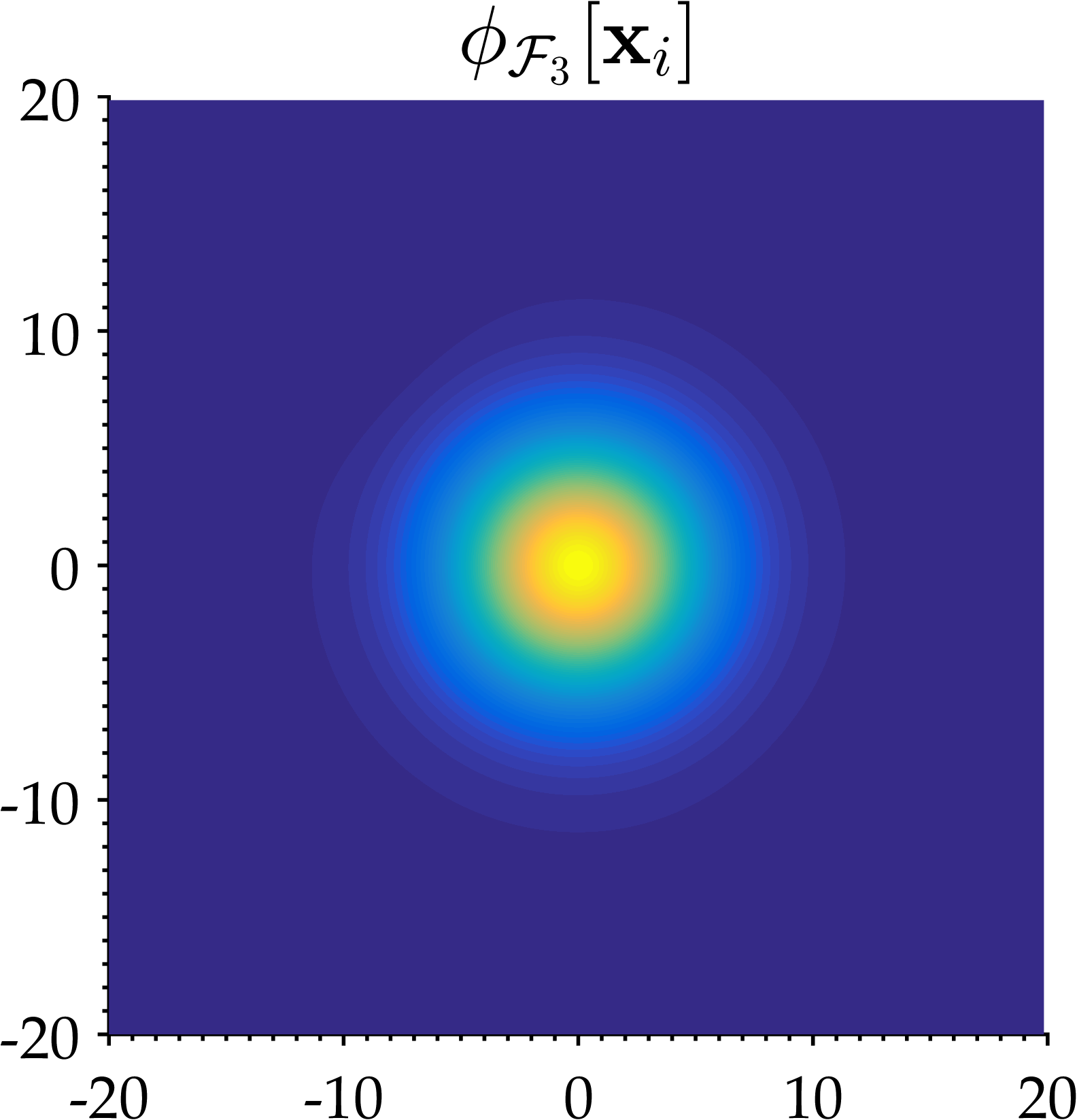}\\
\vspace{2mm}
\includegraphics[width=2.8cm]{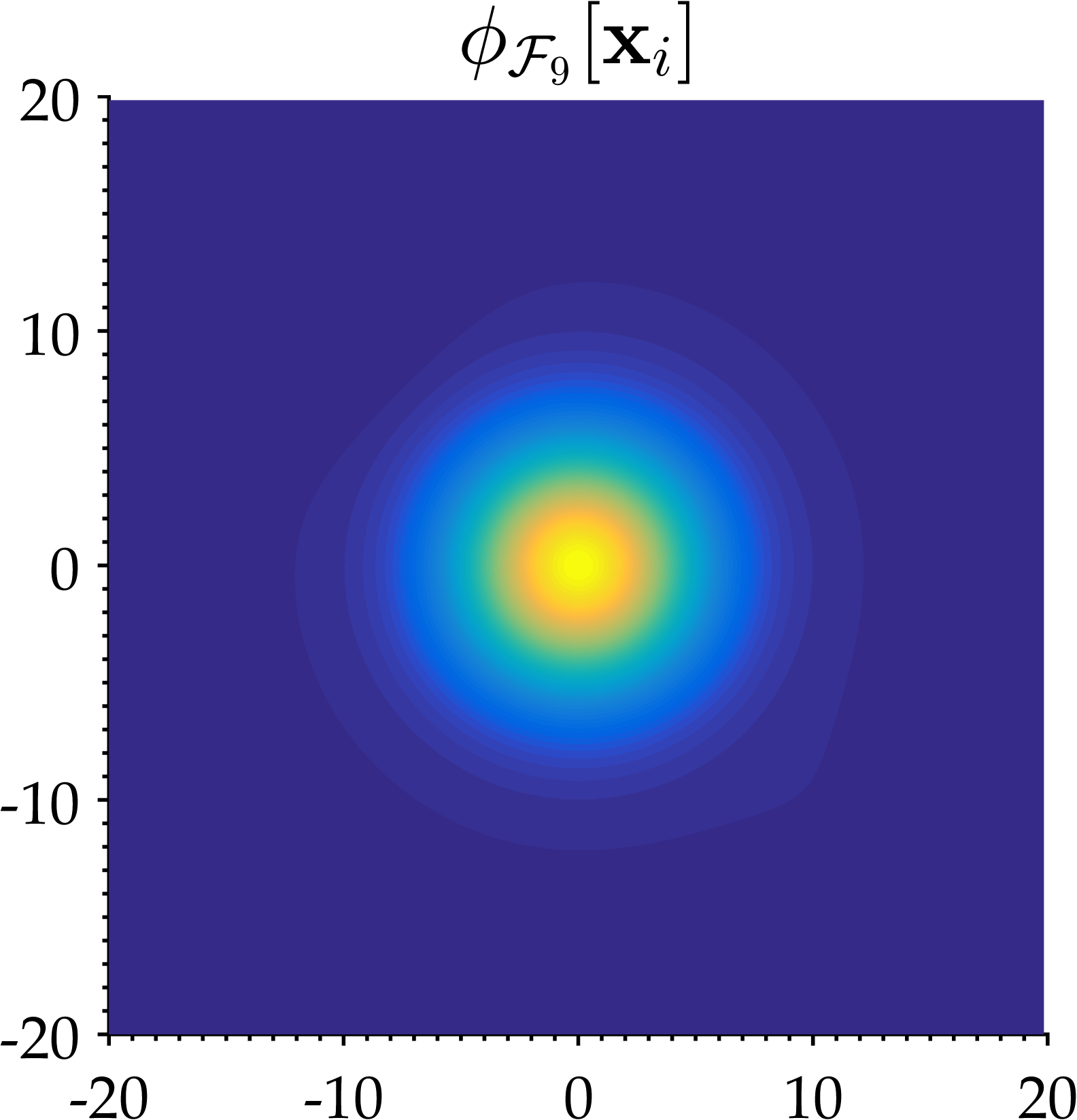}
\end{subfigure}
\begin{subfigure}{.29\textwidth}
  \centering
  \includegraphics[width=2.8cm]{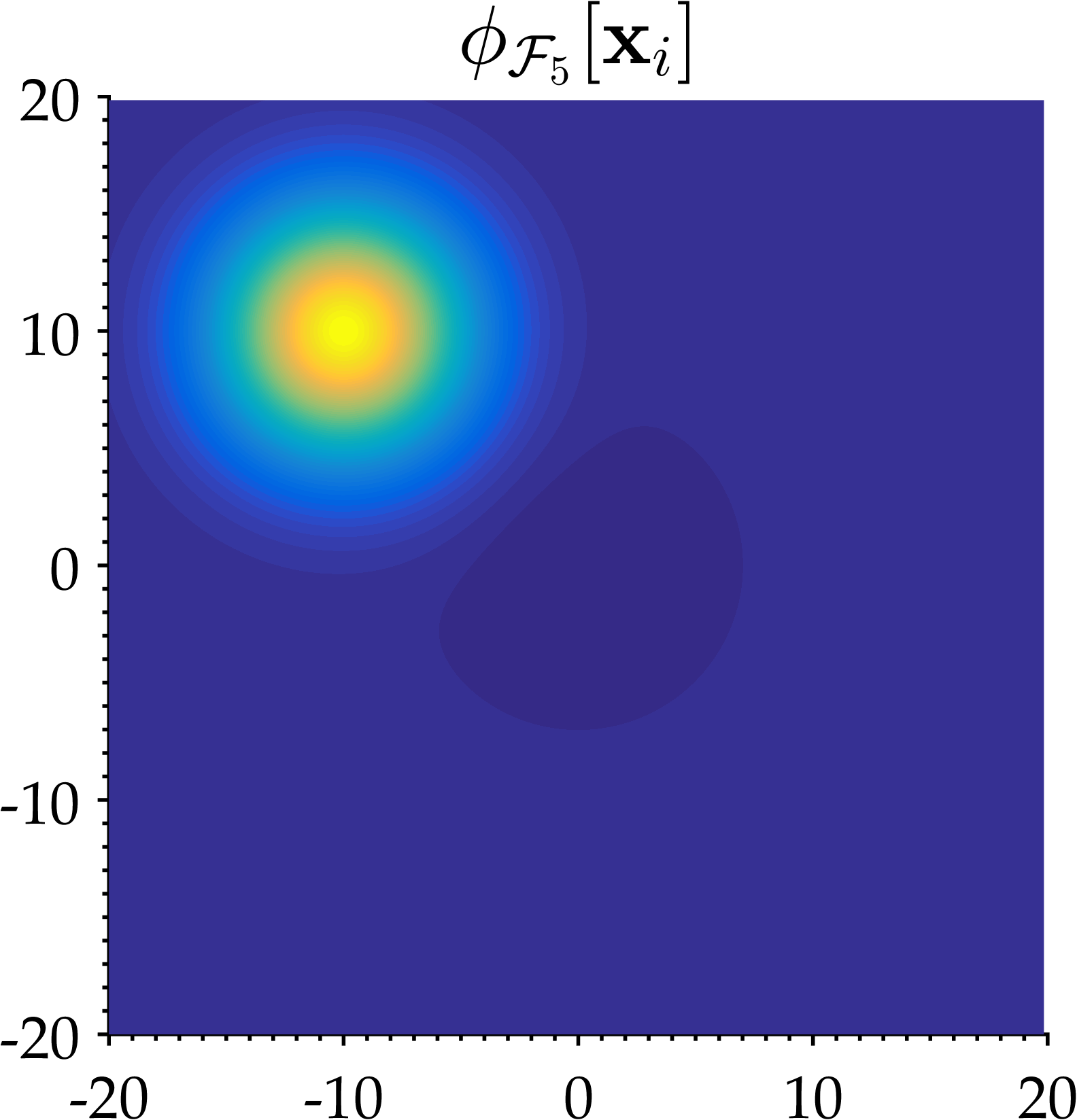}\\
\vspace{1mm}
\includegraphics[width=2.8cm]{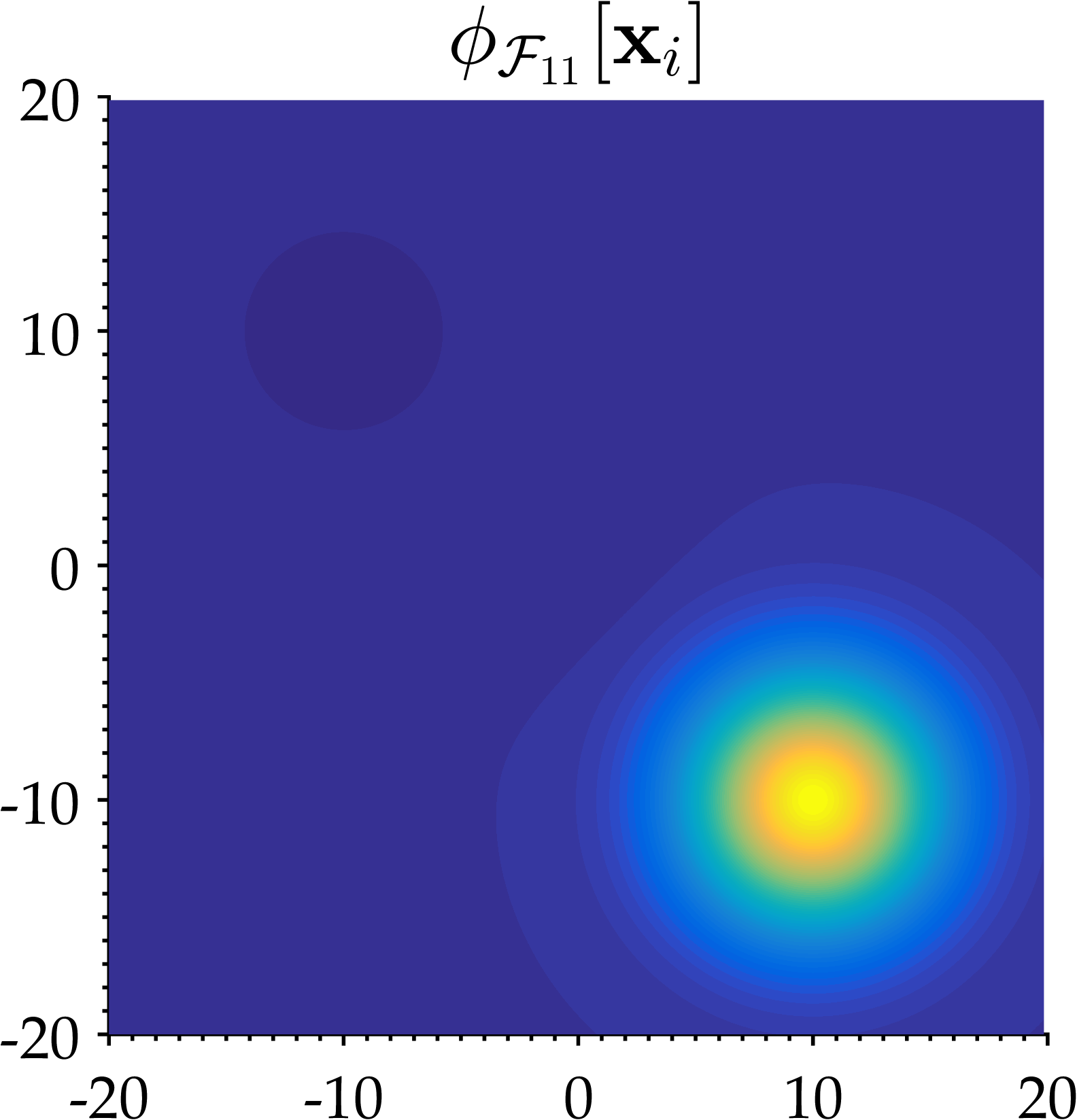}
\end{subfigure}
\begin{subfigure}{.08\textwidth}
\includegraphics[width=1cm]{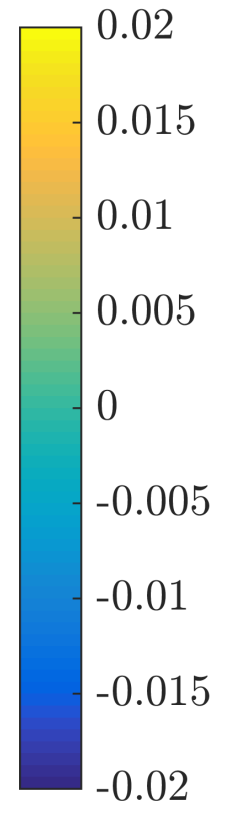}
\end{subfigure}\\
\caption{Spatial structures $\phi_\mathcal{F}[\mathbf{x}_i]$ of the first six DFT modes, with amplitude indicated in Figure \ref{COM_1} with a square marker.}
\label{DFT_RES}
\end{figure*}

Although the two spectra are similar, this additional degree of freedom gives to the DMD better convergence and reduced windowing problems, especially on the lower frequency. The spatial structures for the first $6$ odd DFT modes (the even one are phase-shifted copies) are shown in Figure \ref{DFT_RES}. The amplitude of these modes is indicated with a square marker in Figure \ref{COM_1} while their temporal structures, being simple harmonics, are not shown. The frequency resolution of the DFT is of the order of $f_s/n_t\approx 0.2$, with the leading modes pulsing at $f_n=[0.19,7.03,15.04]$. The spectral leakage is more evident at the lowest frequency, for which a full period of the oscillation is not captured. These modes have no time localization capabilities: none of these can locate in time the identified structures.

\begin{figure*}
\begin{subfigure}{.29\textwidth}
  \centering
  \includegraphics[width=2.8cm]{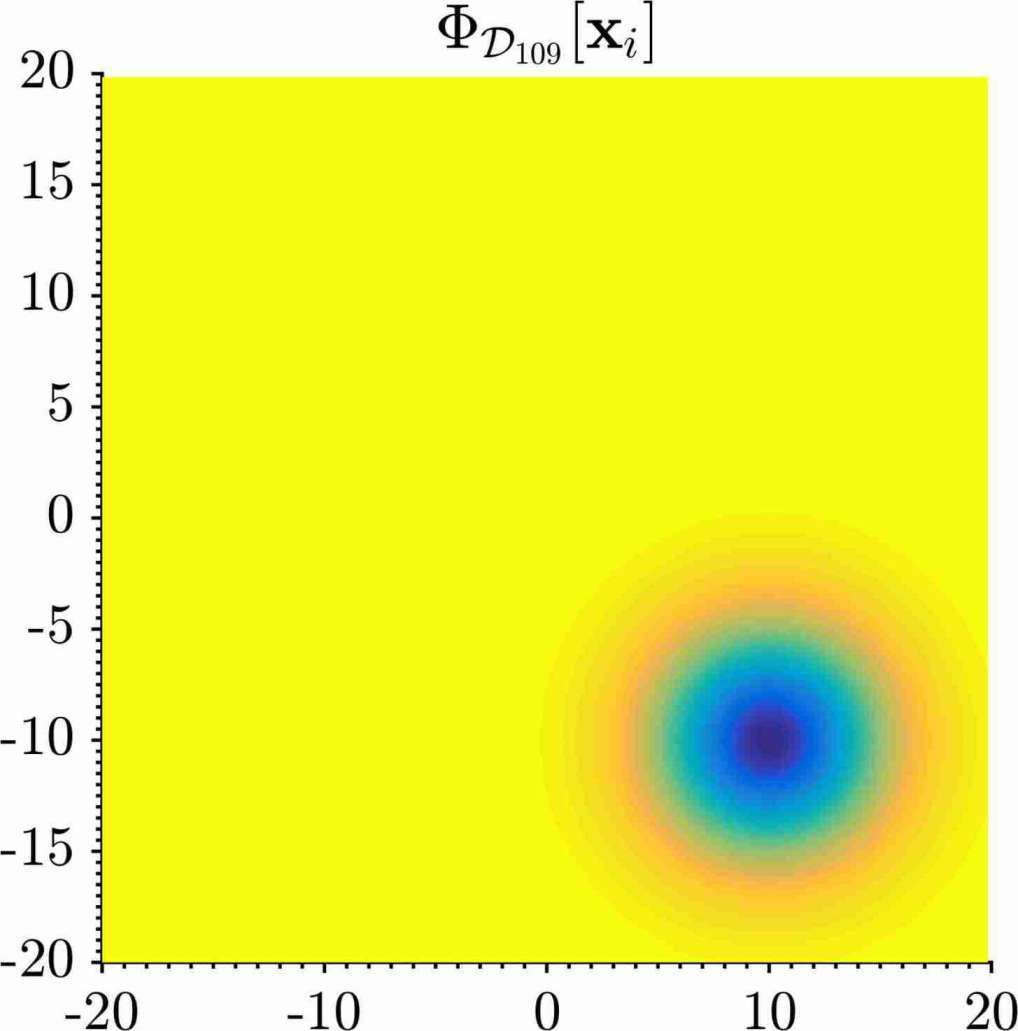}\\
  \label{Synthetic_1}
\end{subfigure}
\begin{subfigure}{.29\textwidth}
  \centering
  \includegraphics[width=2.8cm]{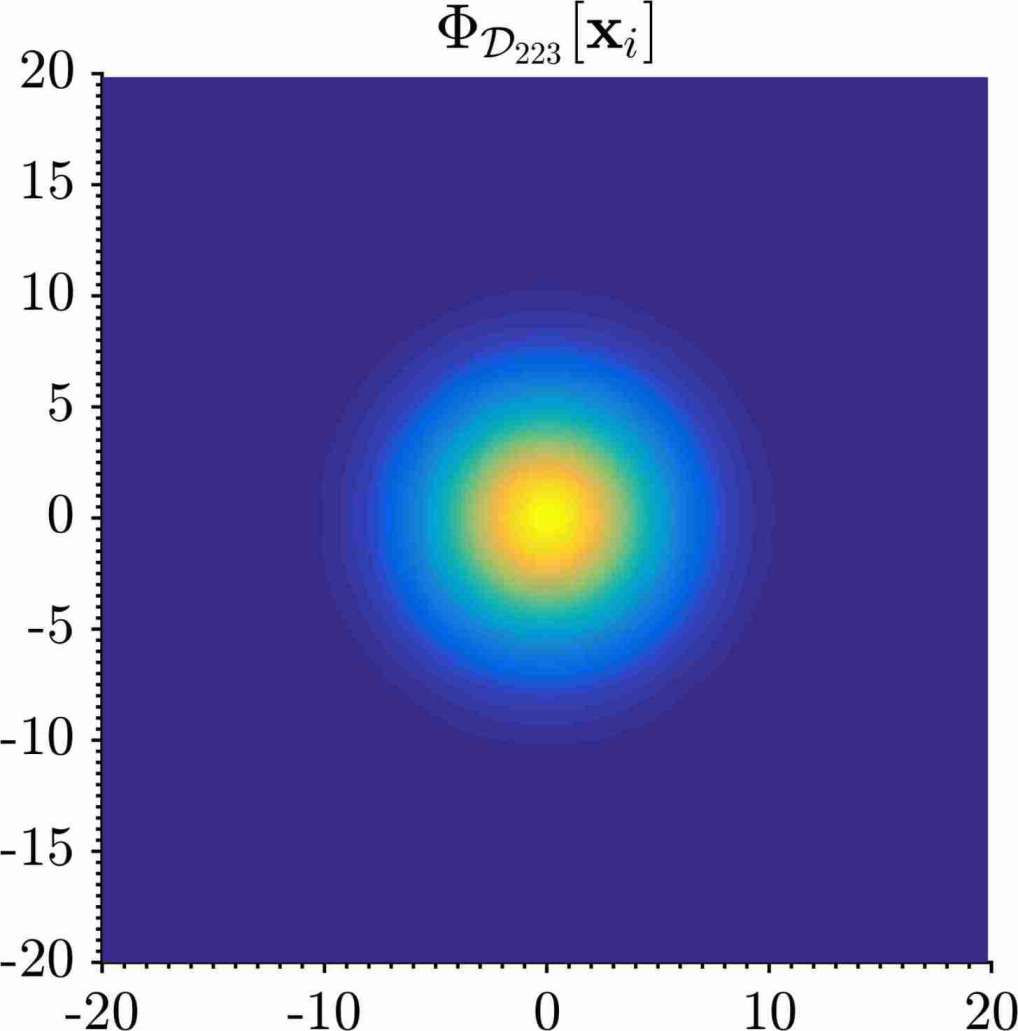}\\
  \label{Synthetic_3}
\end{subfigure}
\begin{subfigure}{.29\textwidth}
  \centering
  \includegraphics[width=2.8cm]{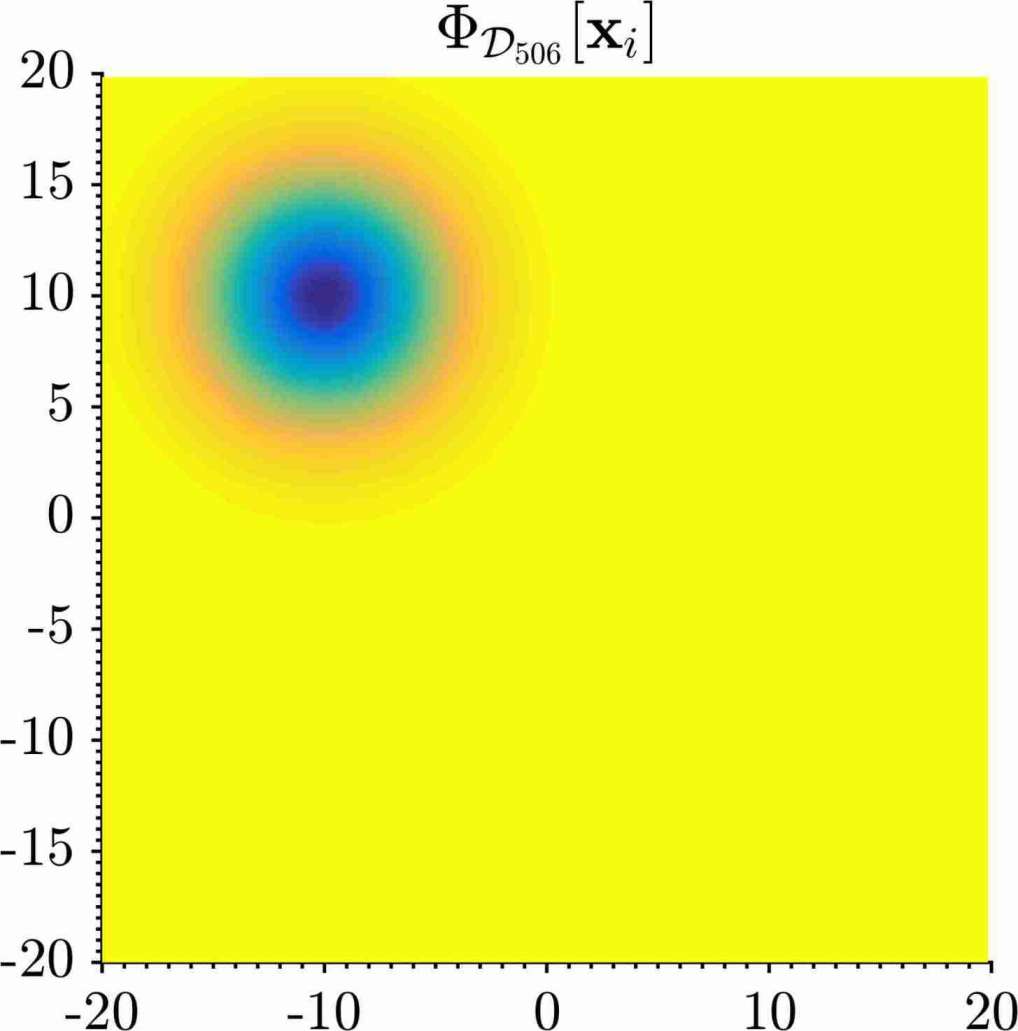}\\
\end{subfigure}
\begin{subfigure}{.08\textwidth}
\includegraphics[width=0.7cm]{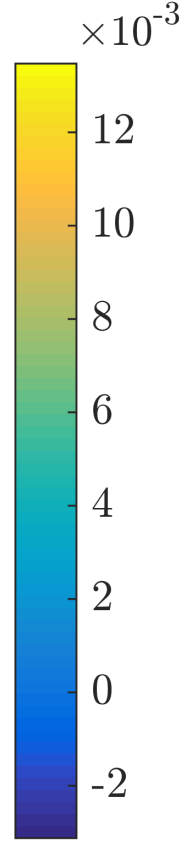}
\end{subfigure}
\caption{Spatial structures $\phi_\mathcal{D}[\mathbf{x}_i]$ of the first three cDMD modes, with amplitude indicated in Figure \ref{COM_1} with a square marker.}
\label{DMD_RES}
\end{figure*}

This localization limit, stemming from the Heisenberg uncertainty principle (see \cite{Wavelet0}) is the main responsible for the poor convergence, which in turns results in a high redundancy in the spatial modes: many modes (besides the conjugate pairs) have identical spatial structures (e.g., $\phi_{\mathcal{F}_3}\approx\phi_{\mathcal{F}_9}$, $\phi_{\mathcal{F}_5}\approx-\phi_{\mathcal{F}_{11}}$) associated, in the time domain, to harmonic corrections to represent finite duration events.

The DMD modes, having a complex frequency, give a valid alternative when describing events that vanish from the initial condition, but not for events that occur after the initial step. A natural way of overcoming such limit is to break the data into multiple windows and perform the DMD in each of these as in the mrDMD \citep{MultiDMD}, at the cost of significantly increasing the decomposition complexity as discussed in \S\ref{SUB3}. 

Even in its simplest formulation, the cDMD improves the convergence and has better frequency resolution than the DFT. The cDMD spectrum highlights the three dominant modes pulsing at $f_n=[0.101,7.00,15.09]$, and the corresponding spatial structures, shown in Figure \ref{DMD_RES}, correctly localize in space these pulsations. The time localization, of course, remains impossible as for any frequency-based method.

\begin{figure*}
 \centering
  \includegraphics[height=3.6cm]{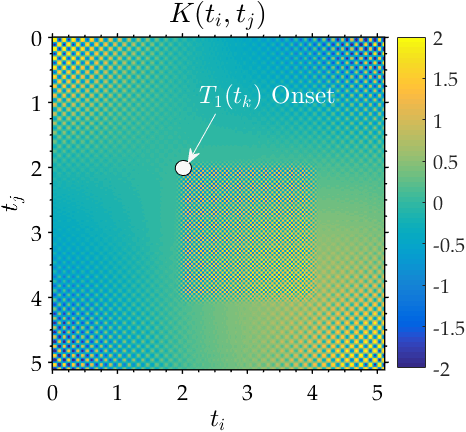}
\hspace{2mm}
  \includegraphics[height=3.6cm]{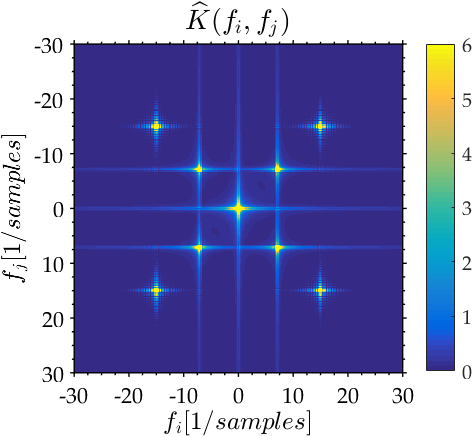}\\
\begin{subfigure}{.29\textwidth}
  \centering
  \includegraphics[width=2.95cm]{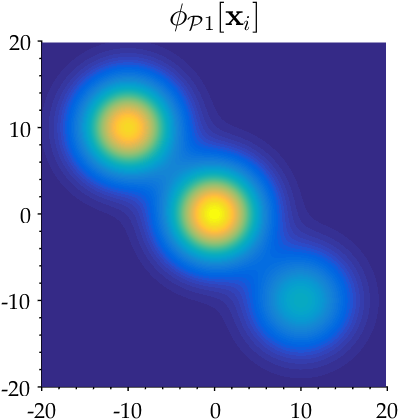}\\
\vspace{2mm}
\includegraphics[width=3.75cm]{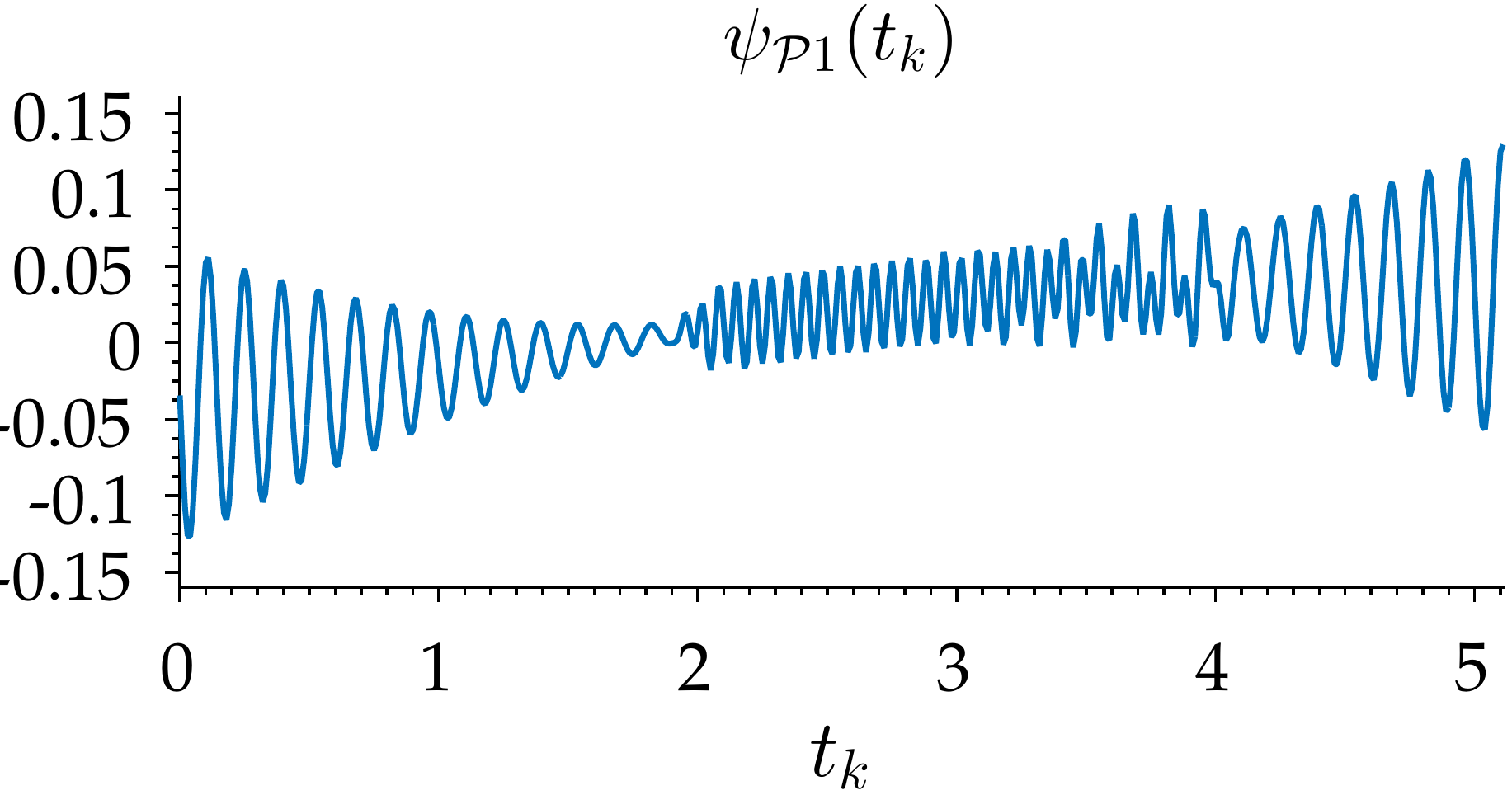}
\end{subfigure}
\begin{subfigure}{.29\textwidth}
  \centering
  \includegraphics[width=2.95cm]{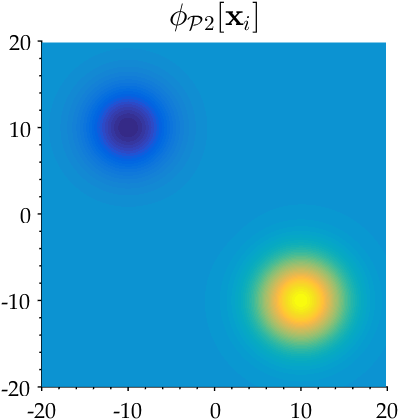}\\
\vspace{2mm}
\includegraphics[width=3.75cm]{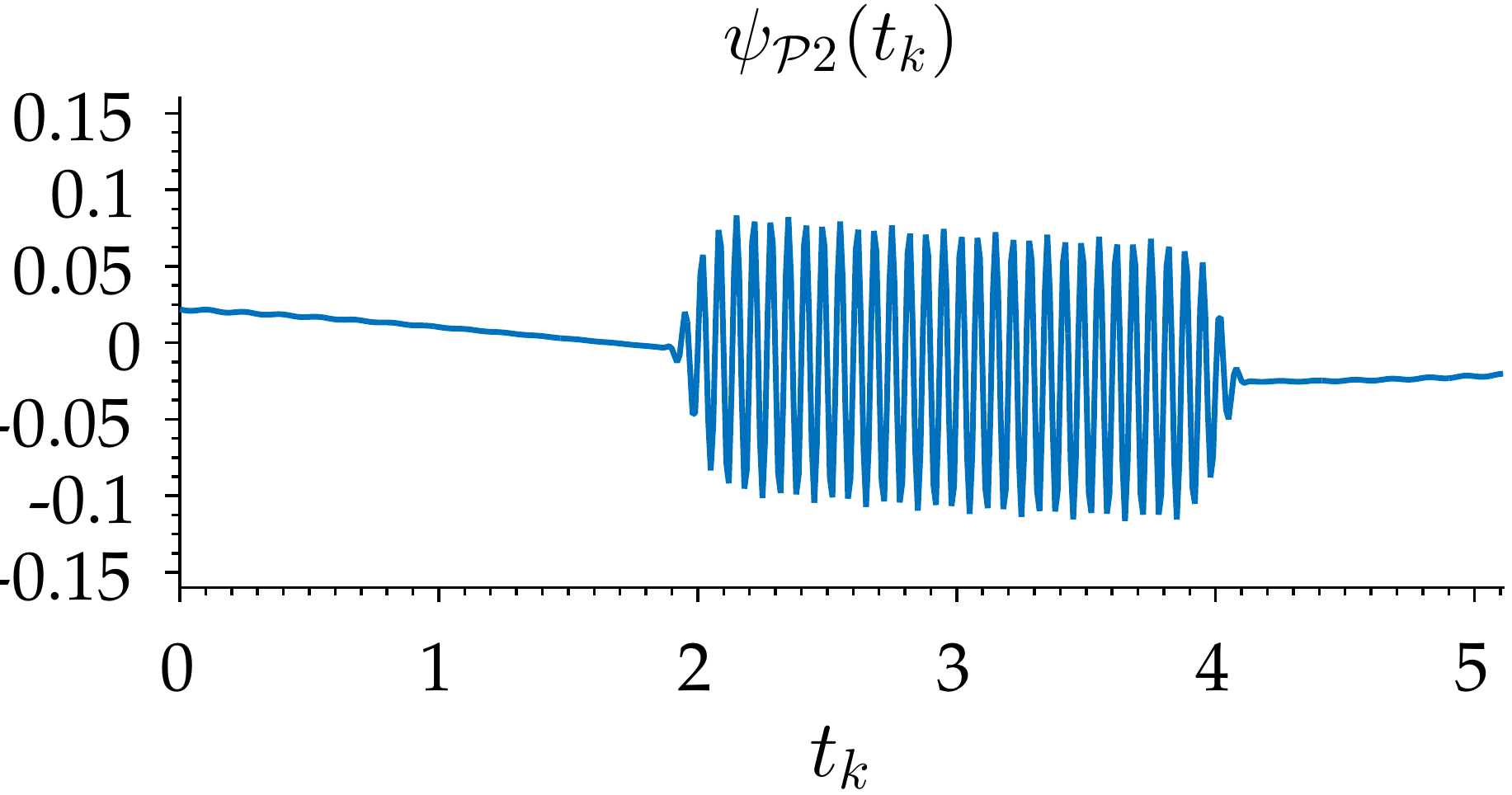}
\end{subfigure}
\begin{subfigure}{.29\textwidth}
  \centering
  \includegraphics[width=2.95cm]{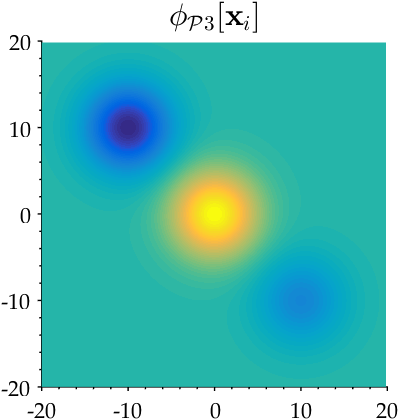}\\
\vspace{2mm}
\includegraphics[width=3.75cm]{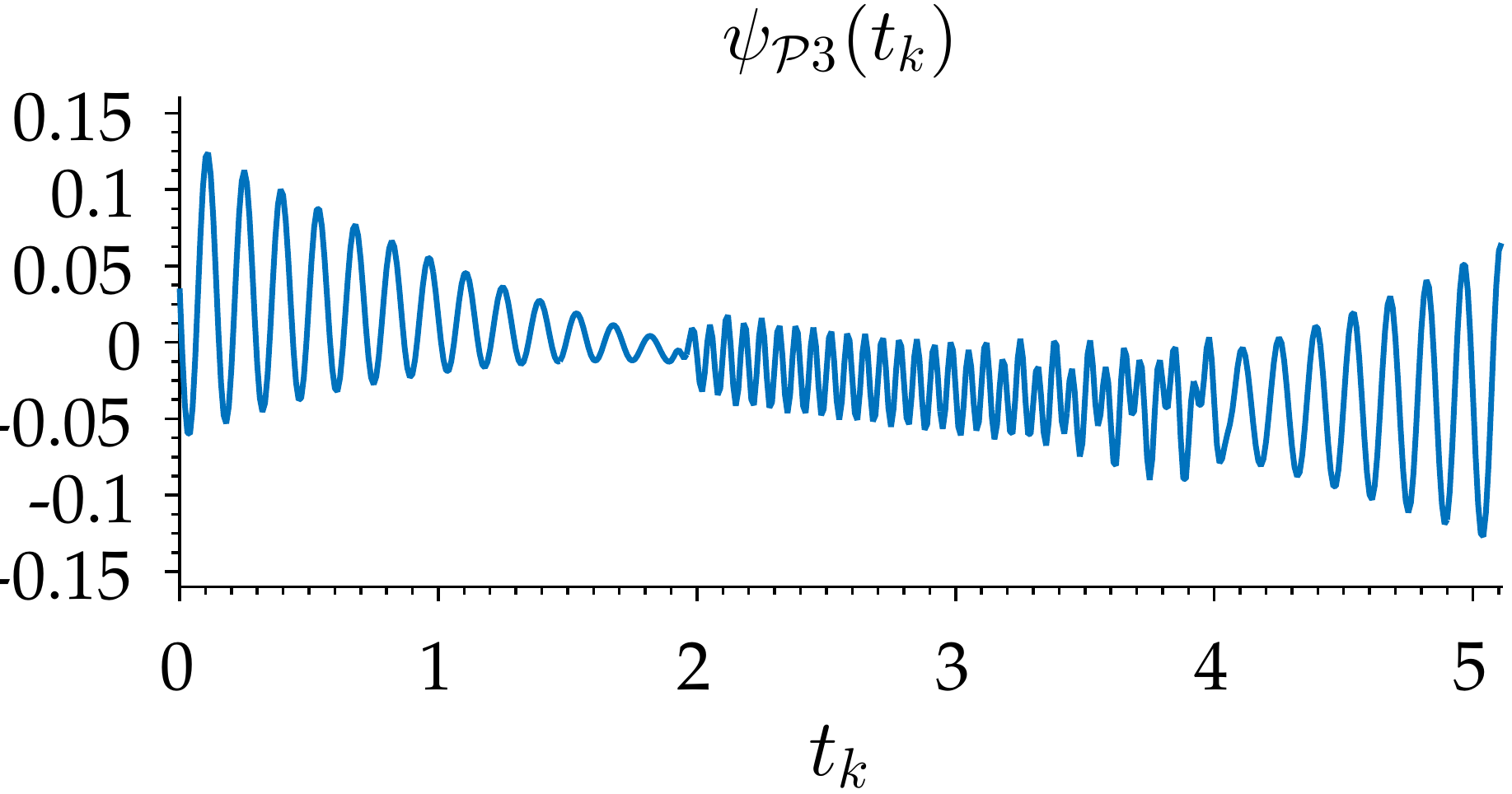}
\end{subfigure}
\begin{subfigure}{.08\textwidth}
\includegraphics[width=0.6cm]{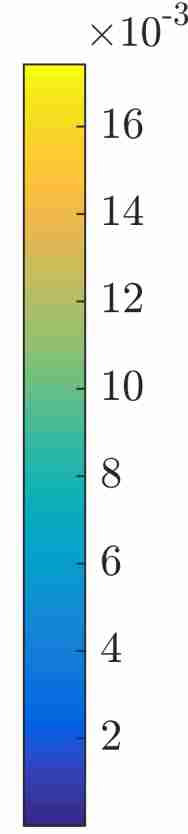}\\\vspace{18mm}
\end{subfigure}\\
\caption{Results from the POD of the synthetic test case. The first row shows the normalized temporal correlation matrix $K=D^\dag D/n_t$ (left) and corresponding DFT $\widehat{K}=\Psi_\mathcal{F}\,K\Psi_\mathcal{F}/n_t$ (right). The second row shows the three non-zero POD modes, including their spatial structures $\phi_{\mathcal{P}_r}$ and temporal structures  $\psi_{\mathcal{P}_r}$.}
\label{POD_RES}
\end{figure*}

By contrast, the POD avoids both the time localization and the convergence problems of DFT and DMD, but produces a severe spectral mixing between different modes. The results of the POD are collected in Figure \ref{POD_RES}, which includes the temporal correlation matrix $K$ and its DFT $\widehat{K}$, as well as the three identified modes. The first carries the information on the time localization of each event, as the superimposition of different patterns is visible (the onset of the temporal evolution $T_1$ is marked in the Figure). The second carries the information on the frequency localization, with the three dominant harmonics producing peaks mirrored along the center in both the diagonal and the anti-diagonal. The spectral leakage produces cross-like patterns centered in these peaks, with larger extension at the lowest frequencies (affected by the strongest windowing effect).

As this simple test case is not stationary, the correlation matrix is not Toeplitz circulant, as enforced in the SPOD proposed by \cite{SPOD}. On the contrary, its localized patterns are the footprint of the different phenomena in the dataset and give to the POD the time localization capabilities that are not achievable by DMD/DFT.

The temporal correlation matrix of this synthetic test case, however, is almost defective, because it is close to having repeated eigenvalues. The three detected singular values are $\Sigma_{\mathcal{P}}=diag(297,290,284)$: although the energy contribution of the introduced mode is identical, the non-orthogonality of the temporal structures requires a slight energy re-distribution. Yet, the singular values are sufficiently close to the limit $\sigma_1=\sigma_2=\sigma_3$ to approach the condition of wholly undetermined POD. 

\begin{figure}
\centering
\begin{subfigure}{.32\textwidth}
\centering
\includegraphics[width=4.35cm]{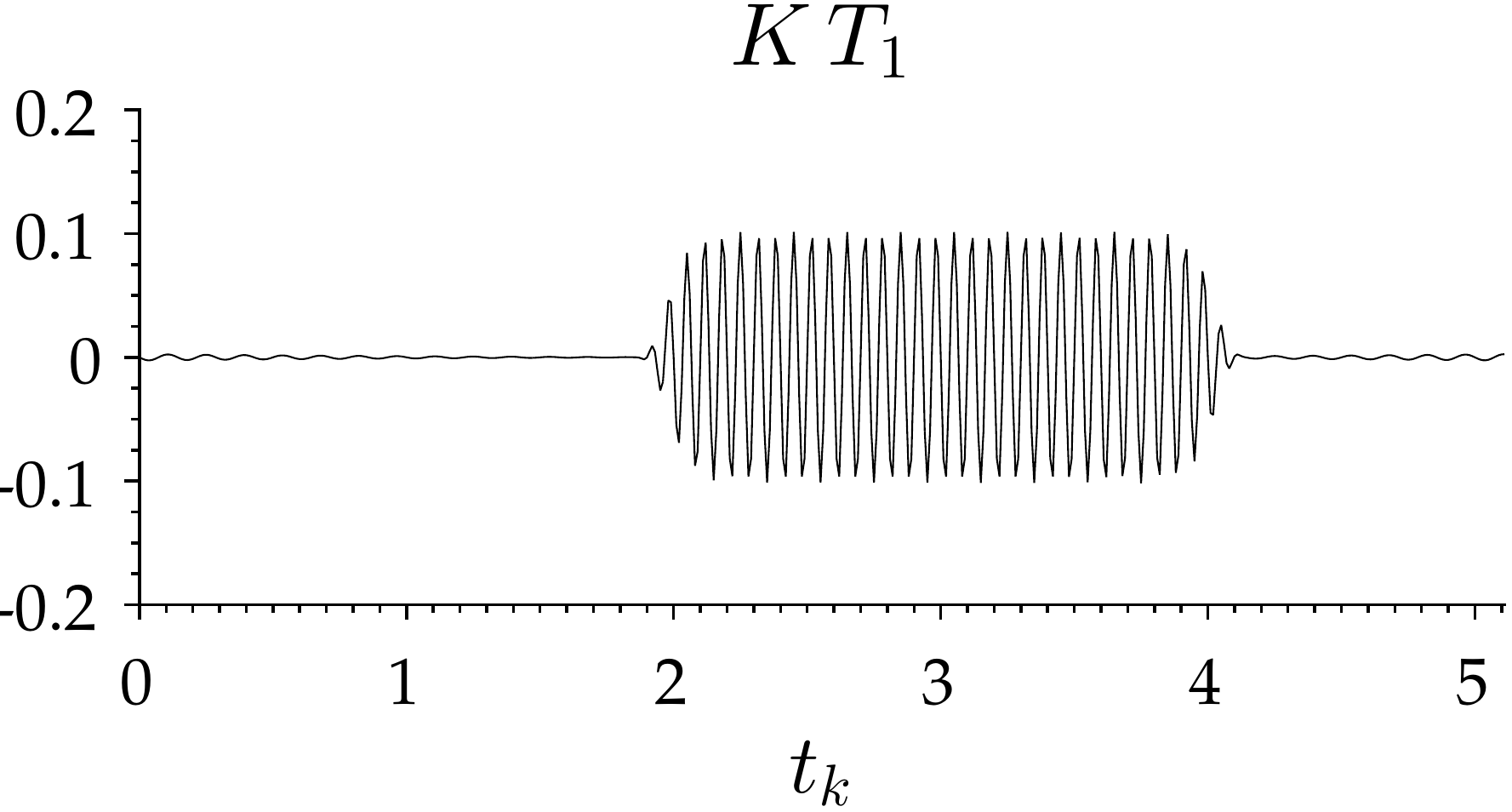}
\end{subfigure}
\begin{subfigure}{.32\textwidth}
\centering
\includegraphics[width=4.35cm]{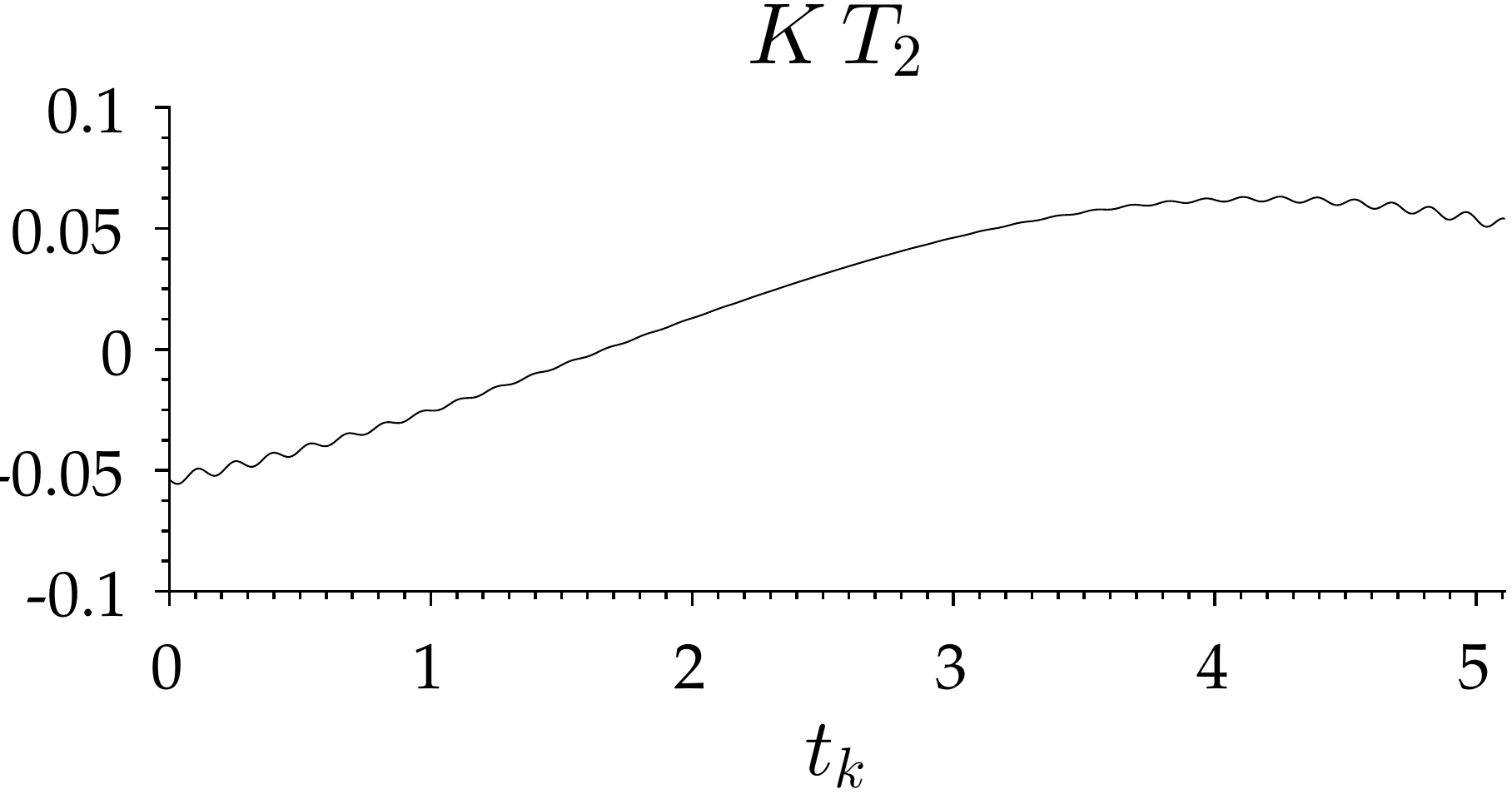}
\end{subfigure}
\begin{subfigure}{.32\textwidth}
\centering
\includegraphics[width=4.35cm]{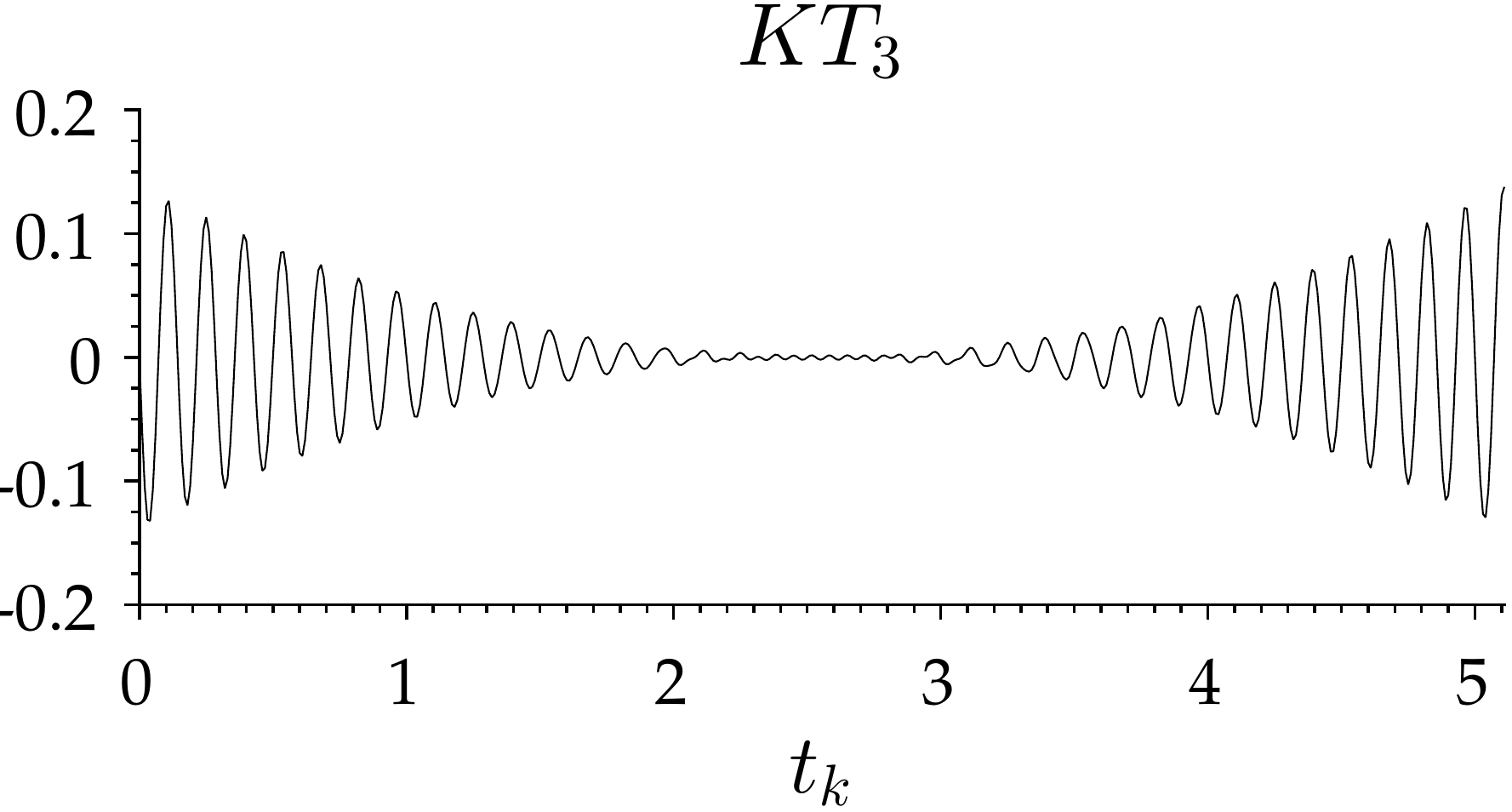}
\end{subfigure}
\caption{Test for POD uniqueness for the proposed test case: the temporal structures in Fig. \ref{Synthetic_Modes} are almost eigenvectors of the temporal correlation matrix, since $K\,T_k\sim \lambda_k \, T_k$. }
\label{Almost_Eigen}
\end{figure}

The eigenvectors of $K$, computed using the Matlab commands $\texttt{eig}$ are shown in figure \ref{POD_RES}. These modes are a linear combination of the introduced ones, with a severe mixing of different phenomena over different modes. To further highlight the uniqueness problem, Fig. \ref{Almost_Eigen} shows that the temporal evolutions in Fig. \ref{Synthetic_Modes} are \emph{almost} eigenvectors of $K$: the matrix multiplication $K T_k$ leads to vectors that are close to simple multiples of the temporal evolution $T_k$. The more $K$ is close to being defective, the larger the number of vectors that are \emph{almost} its eigenvectors and thus POD structures.

\begin{figure}
\centering
\vspace{2mm}
\begin{subfigure}{.32\textwidth}
  \centering
  \includegraphics[width=3.45cm]{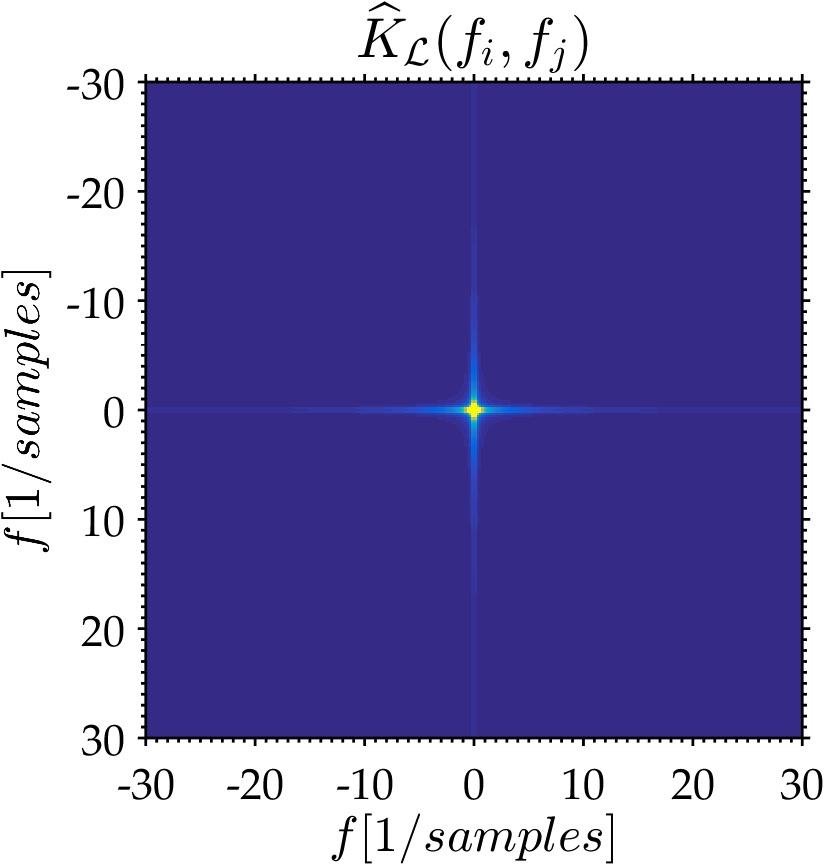}\\
\vspace{2mm}
\includegraphics[width=4cm]{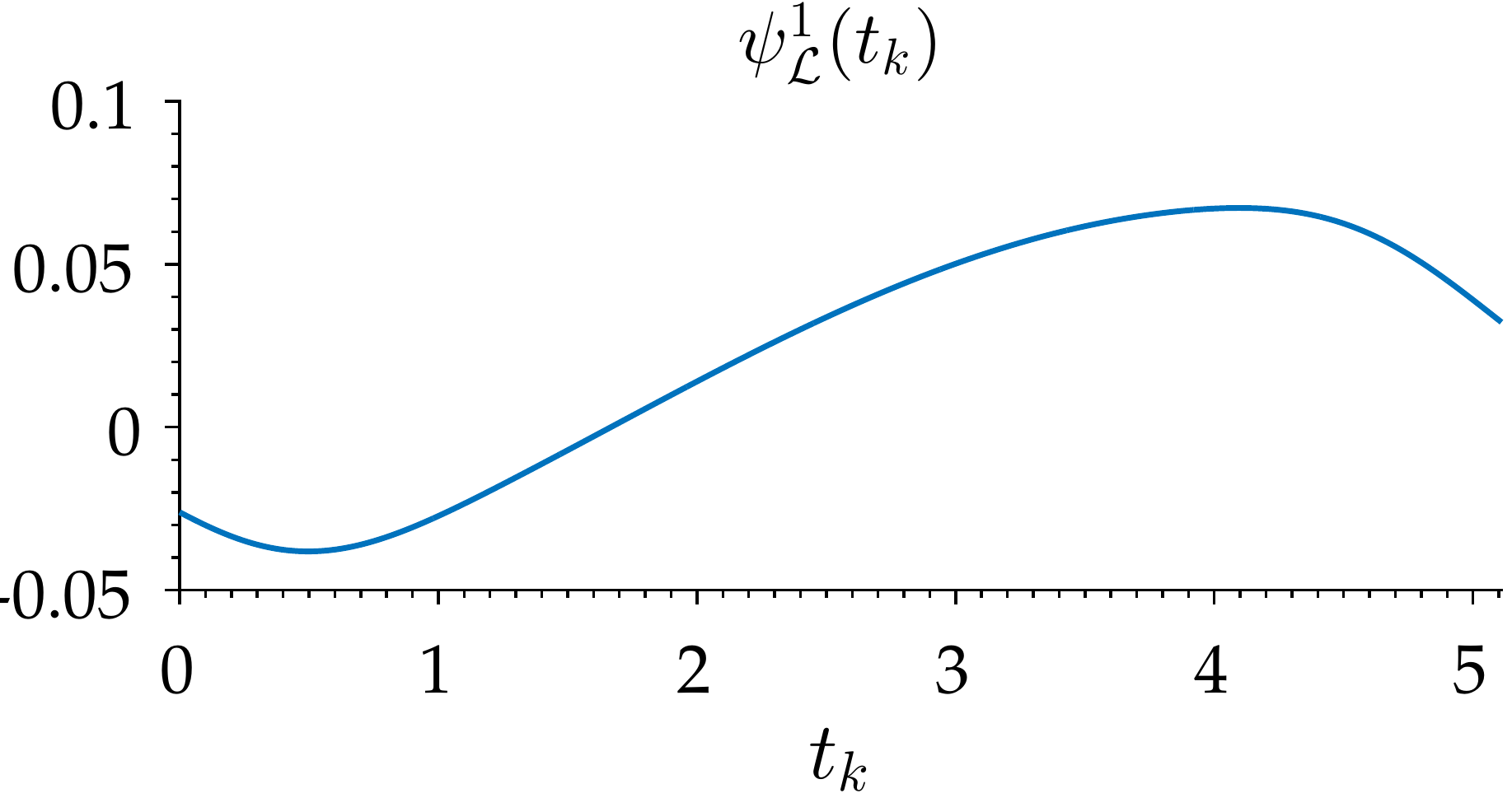}
\end{subfigure}
\begin{subfigure}{.32\textwidth}
  \centering
  \includegraphics[width=3.45cm]{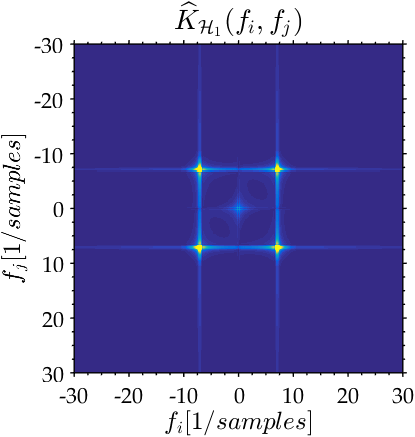}\\
\vspace{2mm}
\includegraphics[width=4cm]{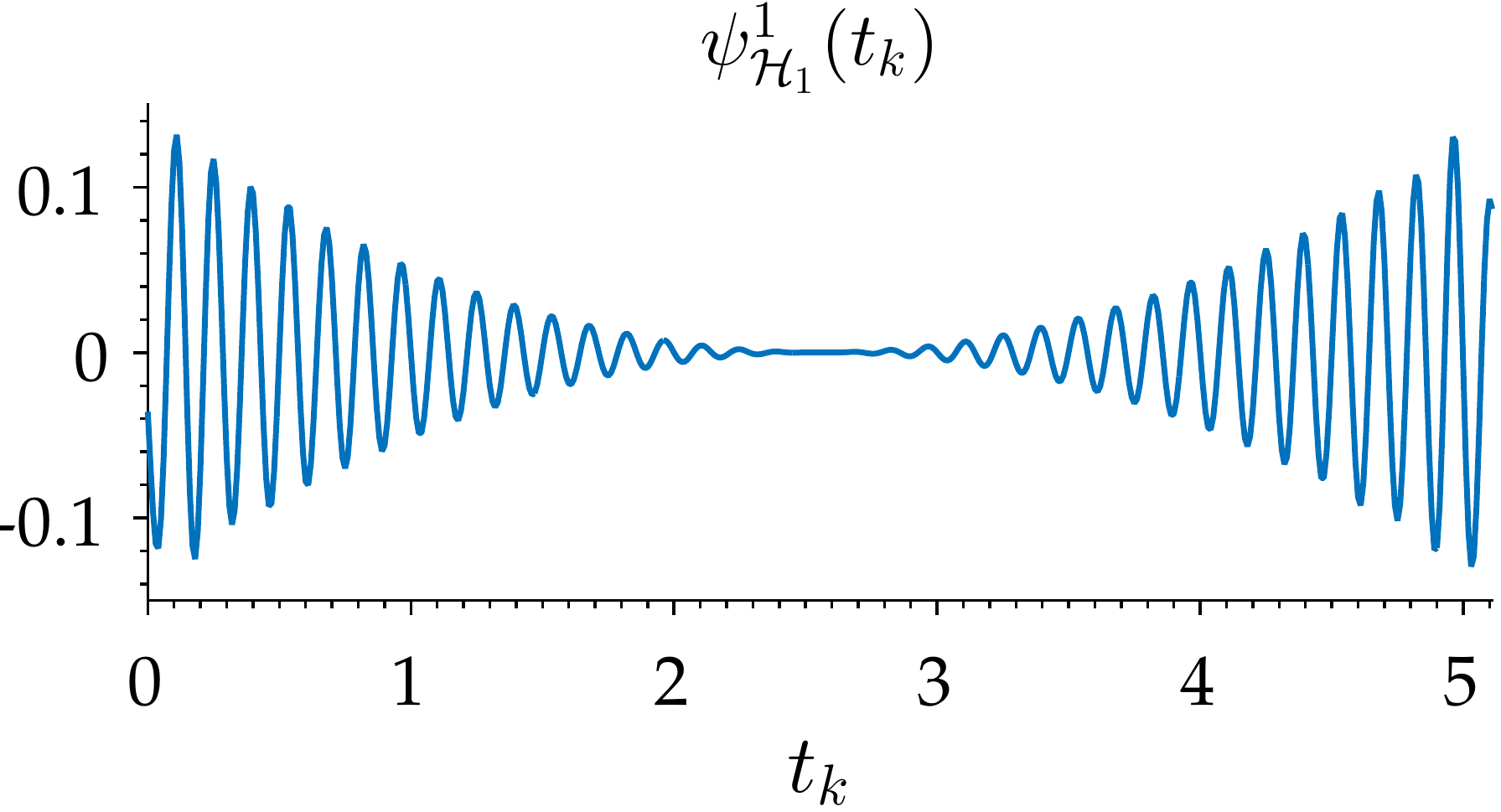}
\end{subfigure}
\begin{subfigure}{.32\textwidth}
  \centering
  \includegraphics[width=3.45cm]{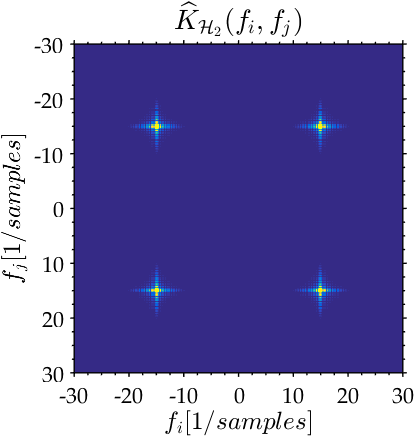}\\
\vspace{2mm}
\includegraphics[width=4cm]{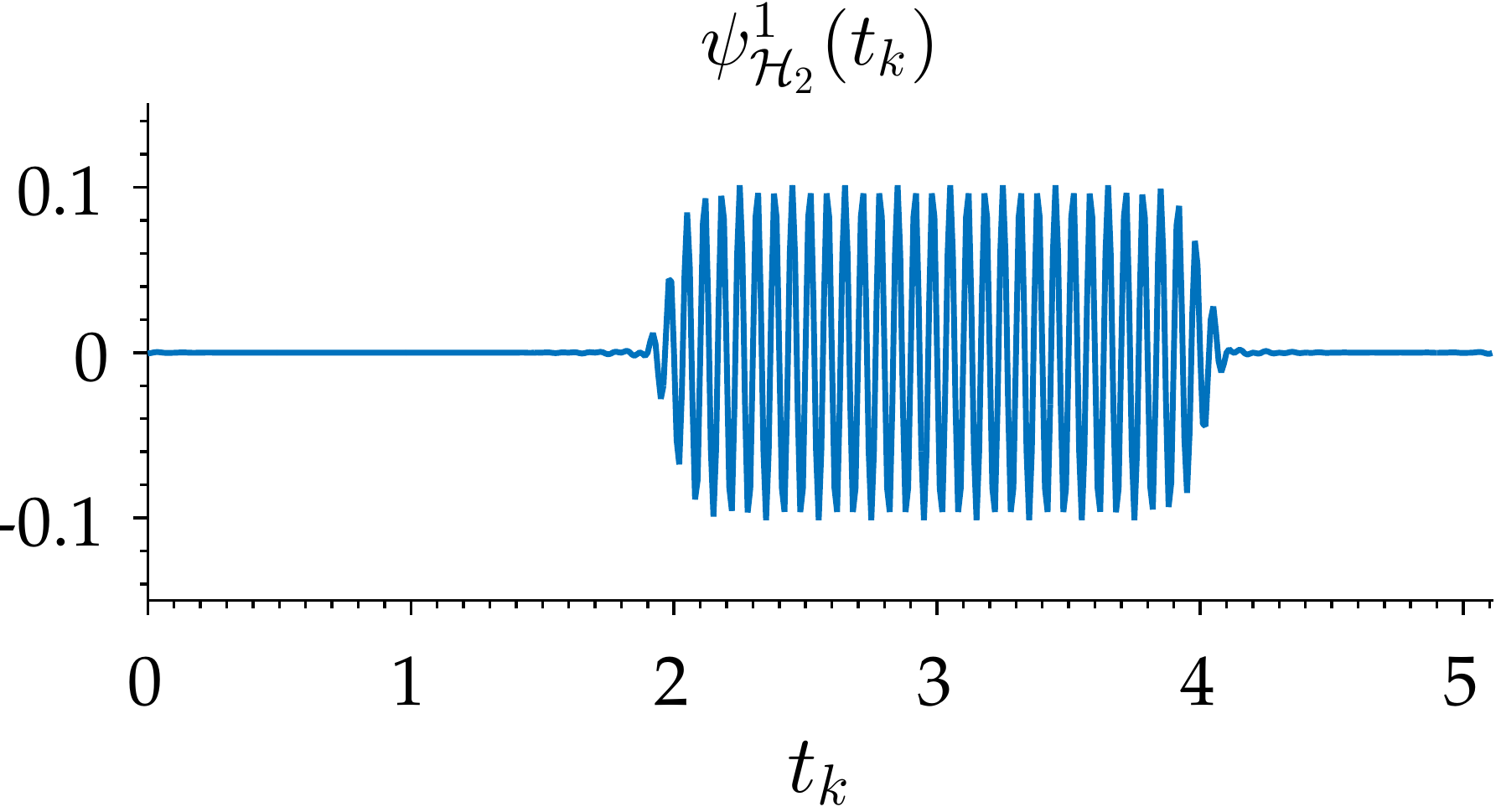}
\end{subfigure}
\caption{Fourier spectra of the correlation matrices $K_{\mathcal{L}_1}$, $K_{\mathcal{H}_1}$ and $K_{\mathcal{H}_2}$ for the three scales identified via filter bank in the synthetic test case. The bottom row shows the dominant eigenvector for each scale contribution. These are then assembled to construct the temporal structures  $\psi_{\mathcal{M}}$ as described in \S\ref{SUB3}.}
\label{mPOD_RES}
\end{figure}

 It is therefore clear that it is possible to obtain an orthonormal basis which is more representative of the coherent patterns in the data by constraining the POD within different frequency intervals (scales). In the proposed mPOD, this is achieved via the MRA of the correlation matrix, splitting the correlation matrix in $M=4$ contributions with $F_V=[0.4,10,20]$. The DFT of each scale is shown in the first row of Fig. \ref{mPOD_RES}; the dominant eigenvector for each scale is shown in the second row of Fig. \ref{mPOD_RES}. The spectral separation of the three contributions is achieved and the temporal structures recover the evolution of each mode. As a result, the spatial structures are identical to the introduced ones and are therefore not shown. At the cost of a minor loss in the decomposition convergence, if compared to the POD, the mPOD correctly identifies the coherent structures introduced and assigns a different mode to each of them.

\section{Example II: Source Detection in Nonlinear Vortex Dynamics}\label{VII}

\subsection{Dataset Description}

As a second test case, we propose a complex dataset which features the main physical mechanisms of complex fluid flows (i.e. nonlinear advection and diffusion), yet allowing for sufficient control over its dynamics so as to validate the decompositions. This test case is the numerical simulation of an advection-diffusion problem with random, pseudo-random and coherent sources. The problem is derived from the incompressible $2D$ Navier-Stokes (NS) equations in the vorticity-stream function formulation. This formulation is common in the study of large-scale geophysical systems \citep{Geo1,Geo2,Geo3}, and consists of a set of two equations. The first is the transport equation of the vorticity $\omega=\nabla\times \vec{u}$, where the velocity field $\vec{u}=(u,v)$ is described by a stream function $\zeta$, defined so that 
$\partial_y\zeta=u$ and $\partial_ x\,\zeta=-v$:

\begin{equation}
\label{vorticity_stream}
\partial_ t\omega=-\mathcal{J}\bigl(\omega,\zeta \bigr)+\frac{1}{Re}\,\nabla^2\omega+S_\omega\,\,,
\end{equation}

{\parindent0pt where} $\mathcal{J}\bigl(\omega,\zeta \bigr)=\partial_y \zeta\,\partial_x \omega-\partial_x \zeta\, \partial_y\omega$ is the Jacobian determinant of the solution vector $(\omega, \zeta)$; $Re$ is the Reynolds number weighting the importance of advection and diffusion; $S_\omega$ is the spatially and temporally varying source or sink of vorticity. The second equation is the Poisson problem linking vorticity and stream function:

\begin{equation}
\label{Omega_XI}
\omega=\nabla \times \vec{u}=\partial_x v-\partial_y u=-\nabla^2 \, \zeta\,\,.
\end{equation}

{Sources of vorticity $S_{\omega}$ in \eqref{vorticity_stream} are typically the vortex stretching/tilting due to the velocity gradient tensor, baroclinic effects due to the misalignment between pressure and density fields, and non-conservative body forces (e.g. the Coriolis effect). For the purpose of this work, we limit our interest to the nonlinear evolution of prescribed sources (regardless of their origin) and to the ability of data decomposition to correctly identify their spatial and temporal structures.}

The solution of \eqref{vorticity_stream}-\eqref{Omega_XI} is computed using a standard finite difference approach \citep{Num1,Kutz}, combining a fast Poisson solver for \eqref{Omega_XI} and a Runge-Kutta time marching scheme for \eqref{vorticity_stream}. The computational domain consists of a square Cartesian grid of $n_s=128\times128$ points over a domain $\mathbf{x}_i\in[-20,20]\times[-20,20]$, while the solution vectors is exported in a time domain $t_k\in[0, 20.48]$ with a total of $n_t=2048$ sample points. Periodic boundary conditions are set in all the boundaries.

The source term is treated explicitly, and consists of coherent ($S_{C\omega}$), pseudo-random $(S_{R\omega})$ and fully random $(S_{\mathcal{N}\omega})$ contributions:

\begin{equation}
\label{SOURCES_EQ}
S_\omega (\mathbf{x_i},t_k)=\overbrace{\sum^{4}_{l=1}\, A_{C_l}\,S_{C_{\phi_l}}(\mathbf{x_i})\, S_{C_{\psi_l}}(t_k)}^{S_{C\omega}}+\overbrace{A_R\,S_{R\phi}(\mathbf{x}_i)\,S_{R\psi}(t_k)}^{S_{R\omega}}+\overbrace{A_{\mathcal{N}}\mathcal{N}(\mathbf{x}_i,t_k)}^{S_{\mathcal{N}\omega}}\,.
\end{equation} 

\begin{figure*}
\centering
\includegraphics[width=3.3cm]{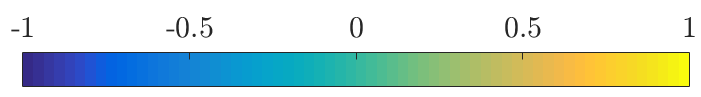}\\
\vspace{2mm}
\begin{minipage}[c][5.2cm][t]{.23\textwidth}
   \centering
   \includegraphics[width=2.9cm]{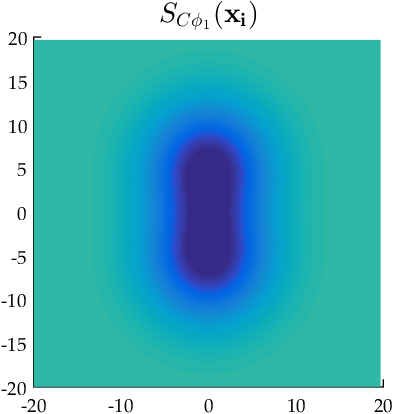}\\
\vspace{2mm}
  \includegraphics[width=3.2cm]{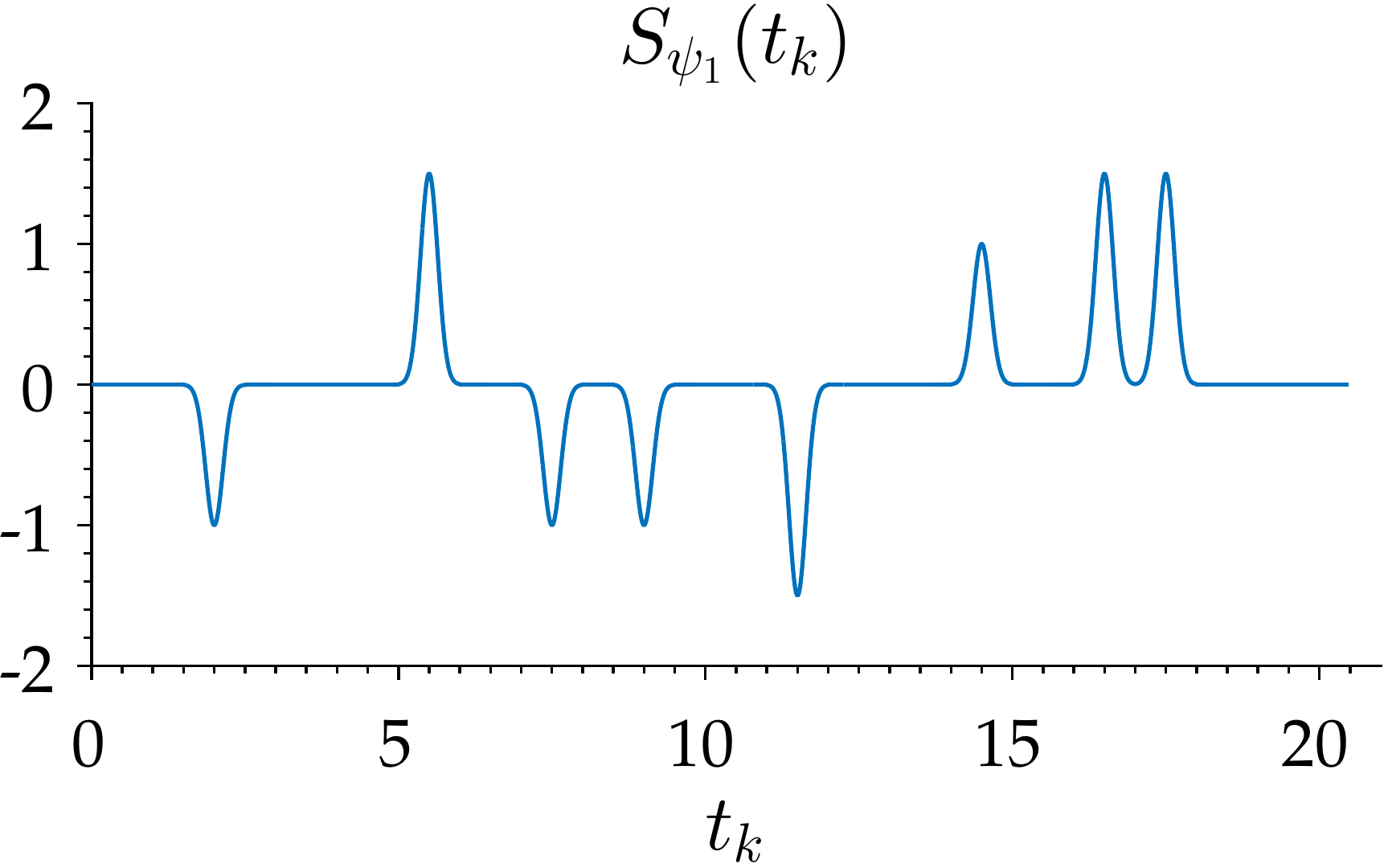}\\
\end{minipage}
\begin{minipage}[c][5.2cm][t]{.23\textwidth}
   \centering
   \includegraphics[width=2.9cm]{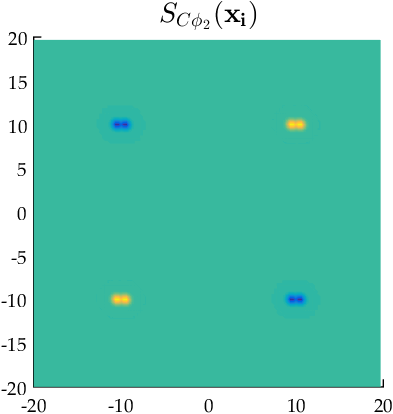}\\
\vspace{2mm}
  \includegraphics[width=3.2cm]{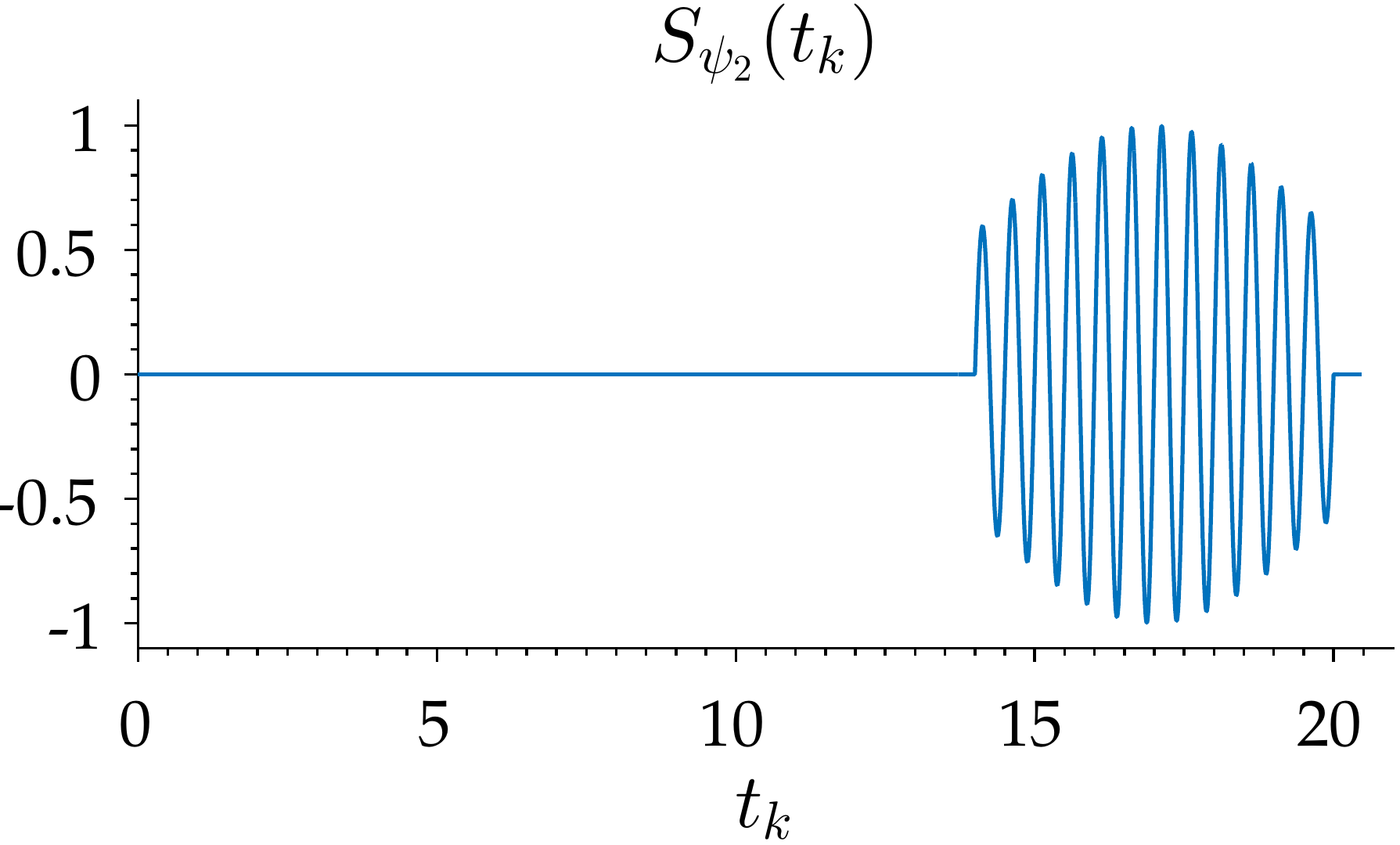}\\
 \end{minipage}
\begin{minipage}[c][5.2cm][t]{.23\textwidth}
   \centering
   \includegraphics[width=2.9cm]{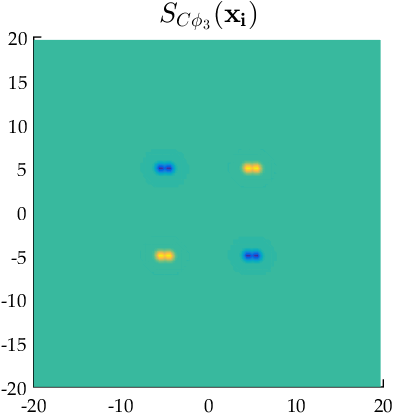}\\
\vspace{2mm}
  \includegraphics[width=3.2cm]{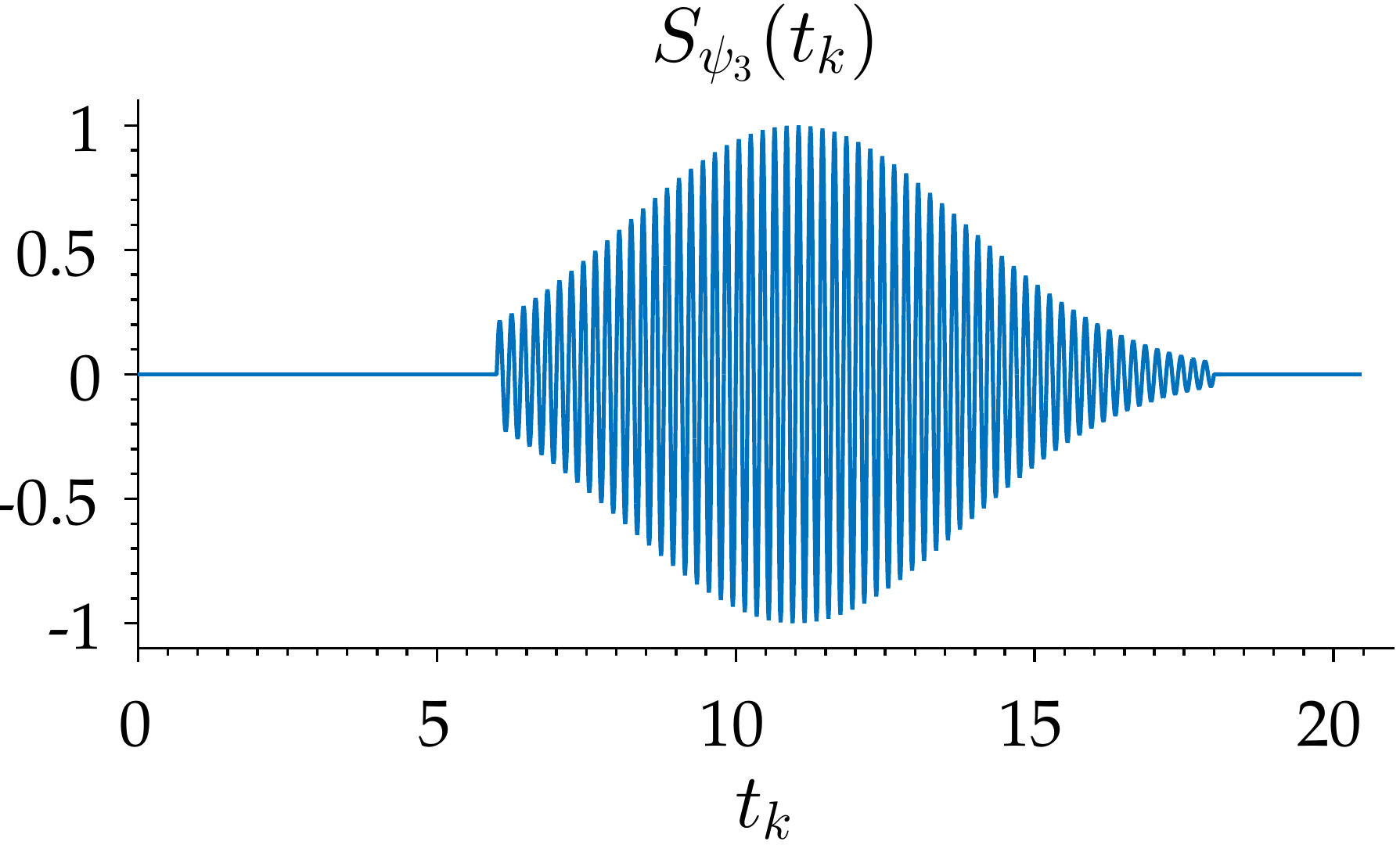}\\
\end{minipage}
\begin{minipage}[c][5.2cm][t]{.23\textwidth}
   \centering
   \includegraphics[width=2.9cm]{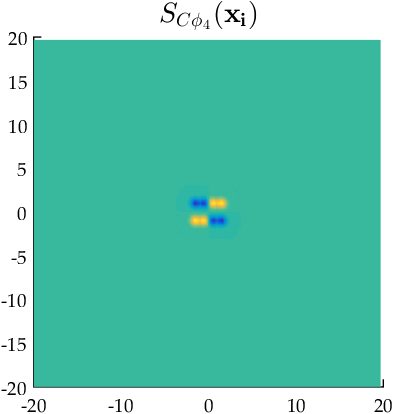}\\
\vspace{2mm}
  \includegraphics[width=3.2cm]{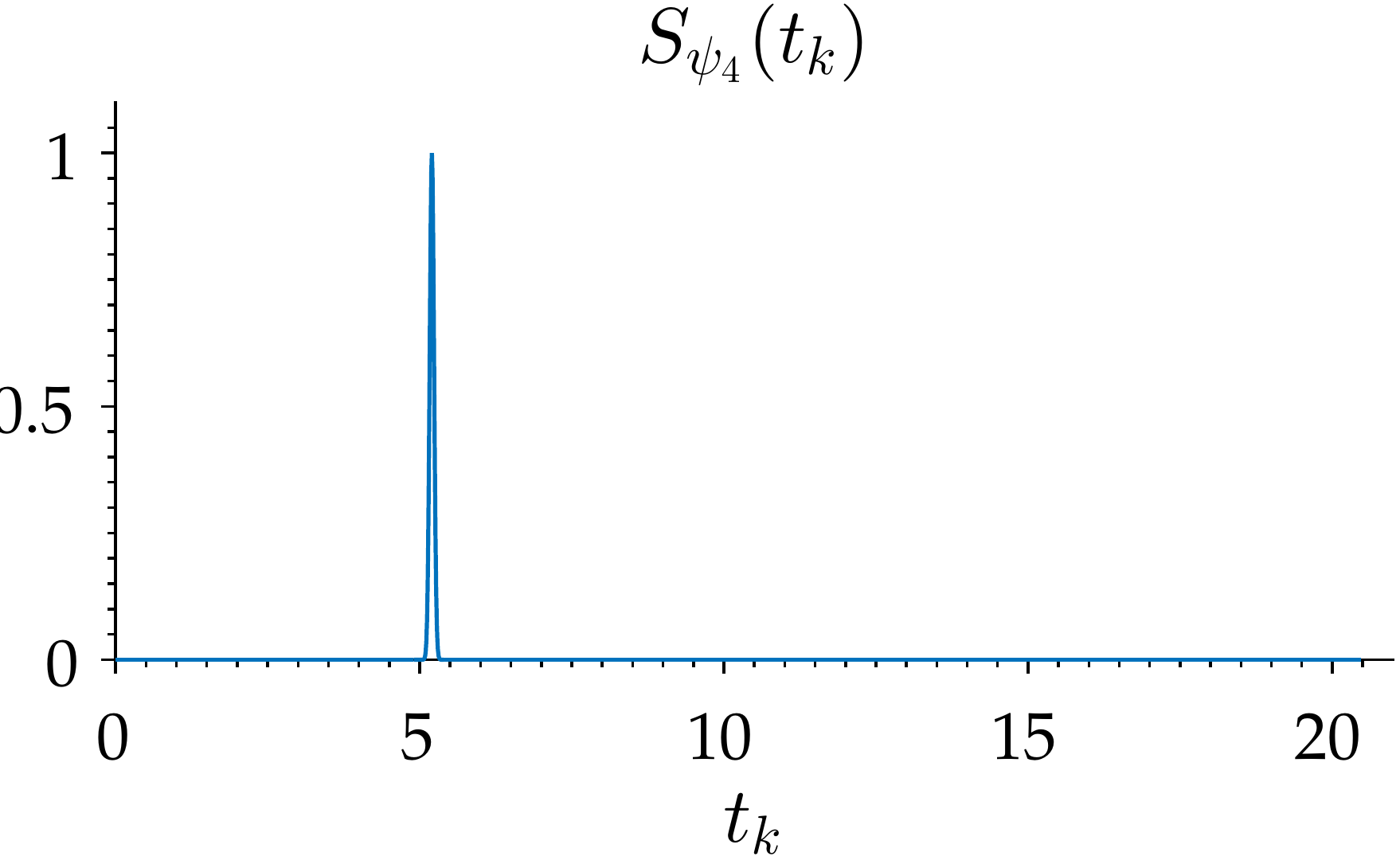}\\
\end{minipage}
\caption{Spatial ($S_{C_{\phi_l}}$, top) and temporal ($S_{C_{\psi_l}}$, bottom) structures of the coherent source term $S_{C\omega}$ in \eqref{SOURCES_EQ}.}
\label{SOURCES}
\end{figure*}

The four spatial $S_{C_{\phi_l}}(\mathbf{x_i})$ and temporal $S_{C_{\psi_l}}(t_k)$ structures of the coherent source $S_{C\omega}$ are shown in figure \ref{SOURCES}. The spatial structures are all constructed as sets of Gaussian vortex pairs of different size and position. The general form is 

\begin{equation}
S_{C_{\phi_l}}(\mathbf{x_i})=\pm\exp\Biggl (\frac{(\mathbf{x_i}\pm \mathbf{x}_{Cl}\pm \mathbf{d}_{l})^2}{\sigma_l^2}\Biggr )\,\,,
\end{equation}

{\parindent0pt
where the distance }between the vortices in each pair are 
$\mathbf{d}_{l}=(d_{xl},d_{yl})=([0,0.5,0.5,0.5],[5.5,0,0,0])$, the pair locations are $\mathbf{x}_{Cl}=(\mathbf{x}_{Cl},\mathbf{x}_{Cl})$ with $\mathbf{x}_{Cl}=[0,10,5,1]$ and the spreading factors are $\sigma_l=[49,0.5,0.5,0.5]$.

The first source is composed of a single vortex pair of equal sign, which naturally merges producing a large recirculation, with angular velocity controlled by its inertia and the sign of the associated temporal structure $S_{C_{\psi_1}}$. This temporal structure is composed of a set of Gaussian pulses $A_i\exp((t_k-t_i)^2/\sigma_{t1}^2)$ of amplitude varying between $\pm [1,1.5]$ and equal widths of $\sigma_{t1}=0.2$.
The large size of the vortex gives too much inertia to this structure to follow such sharp variations, but the change of sign in $S_{C_{\psi_1}}$ (for $t>12$) changes the large scale rotation from counter-clockwise to clockwise towards the end of the simulation. This process is much slower than all the other co-occurring.

The second and third coherent sources differ, in space, only in the distance between the vortex dipoles and have temporal structures consisting of Gaussian modulated harmonics:

\begin{equation}
\label{SPSIS}
S_{C_{\psi_{2,3}}}=\sin\Bigl [2\pi f_{2,3} \bigl( t_k-\alpha_{2,3}\Bigr)\Bigr ] \exp \Biggl[ \frac{-(t_k-\beta_{2,3})^2}{\sigma_{t{2,3}}^2}\Biggr]
\end{equation}

{\parindent0pt with frequencies} $f_{2,3}=[2,5]$, equal width $\sigma_{t2,3}=4$ but different phase delay for both the harmonic ($\alpha_{2,3}=[14,6]$) and the modulation terms ($\beta_{2,3}=[17,11]$). These differences in the phase delay create sharp variations that produce a regular vortex shedding. Finally, the fourth contribution is spatially similar to the second and third, with dipoles placed much closer to the center of the domain, and a temporal structure consisting of a single Gaussian impulse located a $t_k=5.2$ and width $\sigma_{t4}=0.05$.

\begin{figure}
\centering
\begin{subfigure}{.31\textwidth}
  \centering
  \includegraphics[height=3.2cm]{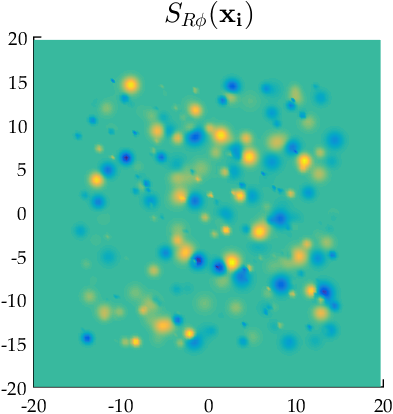}\\
\end{subfigure}
\begin{subfigure}{.45\textwidth}
  \centering
\vspace{2mm}
  \includegraphics[height=3.1cm]{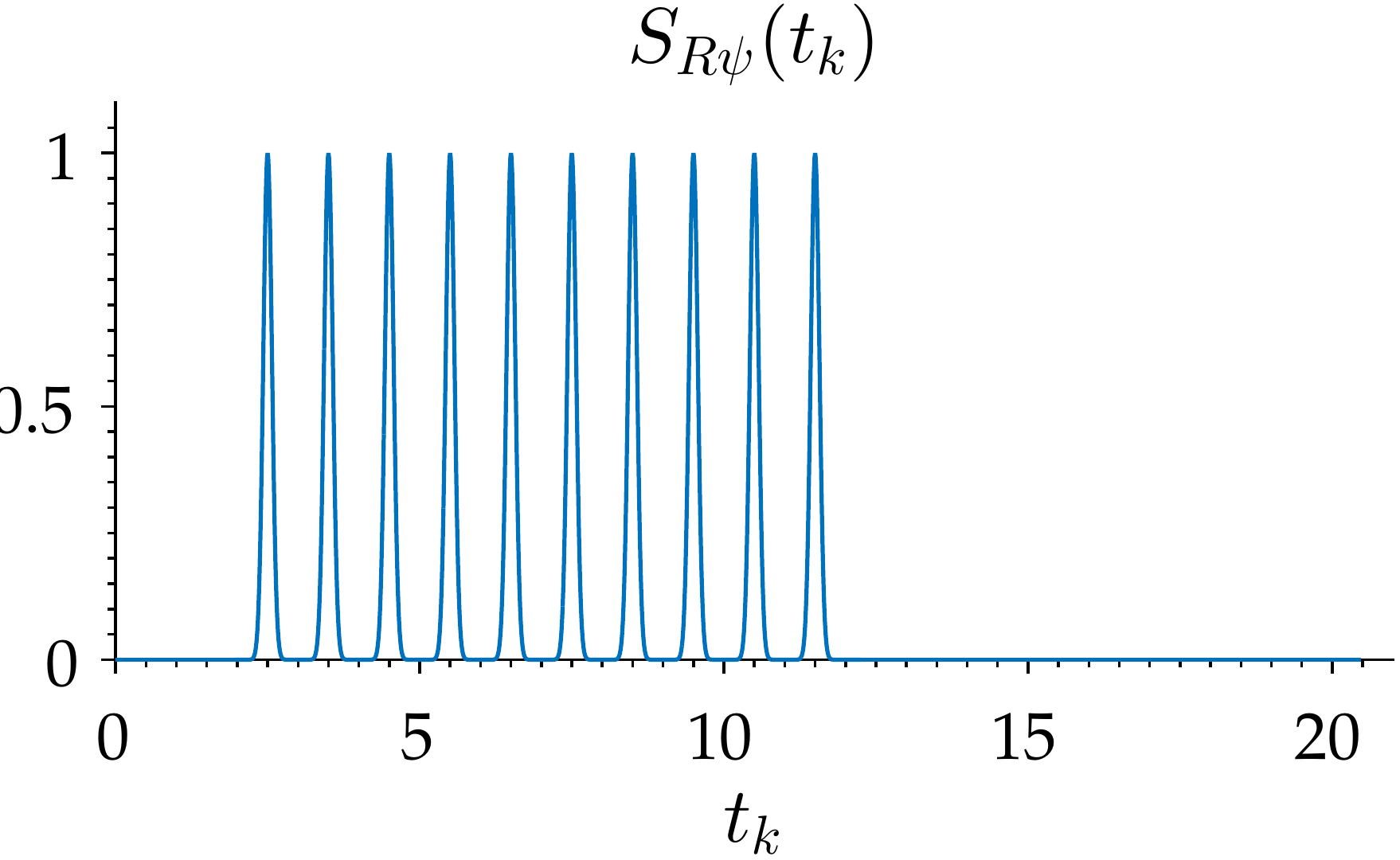}\\
\end{subfigure}
\caption{Exemplary spatial and temporal structures of the pseudo-random source term in \eqref{SOURCES_EQ}. At each of the time pulse, a random distribution of 300 Gaussian vortices is introduced.}
\label{SOURCES_RANDOM}
\end{figure}

The pseudo-random source term $S_{R\omega}$ in \eqref{SOURCES_EQ} produces a periodic injection of random vortices. The spatial structure $S_{R\phi}$ consist of a set of $300$ Gaussian vortices, randomly located with widths in the range $[0,1]$, and amplitudes in the range $[-1,1]$. An exemplary set of vortices representing one of these structures is shown in Figure \ref{SOURCES_RANDOM} (left). The temporal evolution $S_{R\psi}$, shown in Figure \ref{SOURCES_RANDOM} (right), consists of $10$ equal Gaussian impulses with width $0.1$ and unitary amplitude and constant time shift.
As the introduced spatial structures change randomly from one pulse to the other, this source contribution creates a chaotic background which has no spatial coherence but exhibits a temporal regularity. 
Finally, the white noise term $S_{\mathcal{N}_\omega}$ in \eqref{SOURCES_EQ} is introduced in the post-processing step, using a random number generator varying between $0$ and $1$ to generate the spatio-temporal noise $\mathcal{N}(\mathbf{x_i},t_k)$.

The coefficients $A_{C_l}$ ($l\in[1,4]$), $A_R$ and $A_\mathcal{N}$ in \eqref{SOURCES_EQ} set the relative importance of these three contributions and are taken as $A_l=[9,610,1100,2500]$, $A_R=300$ and $A_\mathcal{N}=7$. These coefficients were estimated, by trial and error, to give approximately the same importance to the four coherent sources, taking into account the strong damping effect produced by diffusion (especially on the smaller and faster scales), obtained by setting the Reynolds number in \eqref{vorticity_stream} to $Re=1$. The initial vorticity field is a random set of vortices similar to those introduced by the pseudo-random source term.

\begin{figure*}
\centering
           \includegraphics[width=2.45cm]{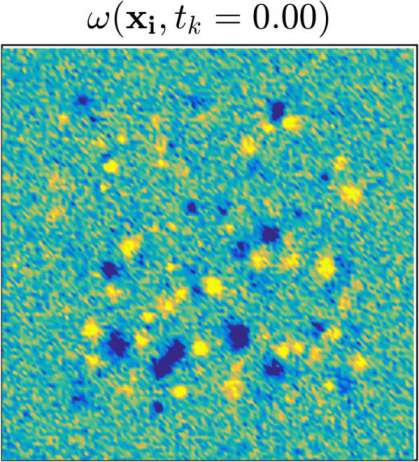} 
           \hspace{1mm}
           \includegraphics[width=2.45cm]{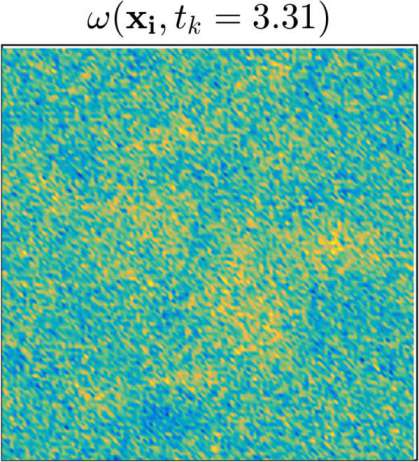}
                \hspace{1mm}
           \includegraphics[width=2.45cm]{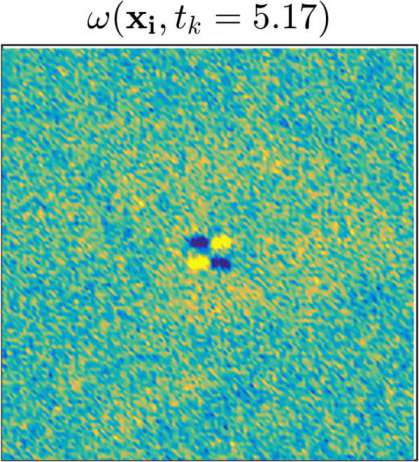}
                \hspace{1mm}
          \includegraphics[width=2.45cm]{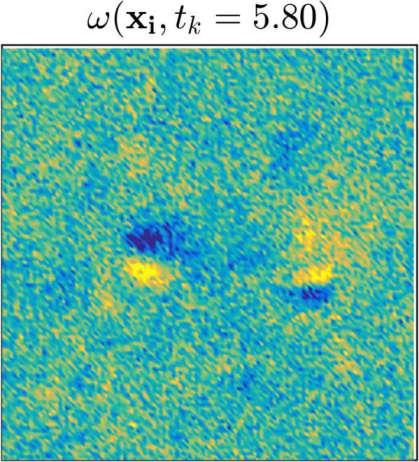}
           \hspace{1mm}
             \includegraphics[width=2.45cm]{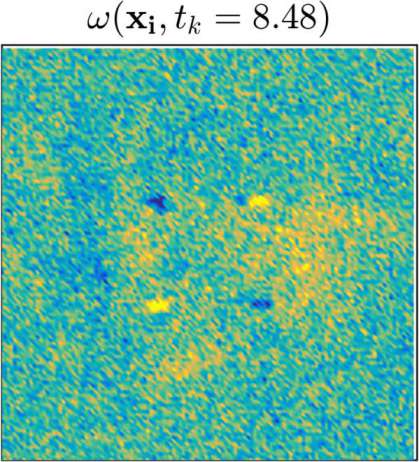} 
    \\  \vspace{1mm} 
            \includegraphics[width=2.45cm]{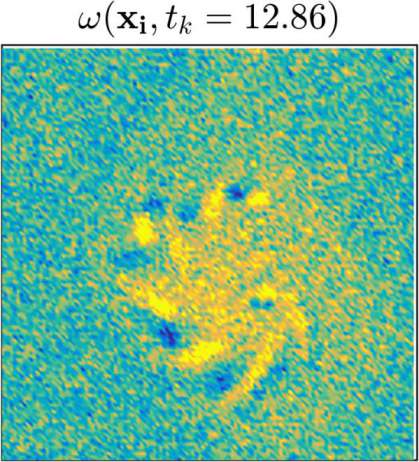}
\hspace{1mm}
             \includegraphics[width=2.45cm]{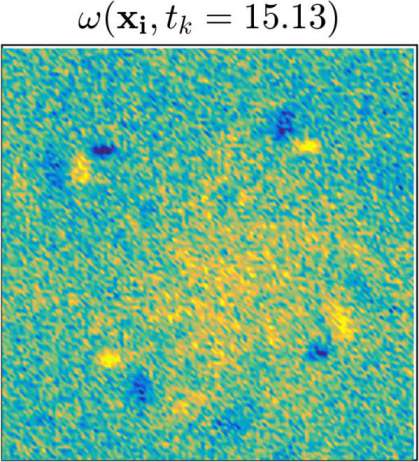} 
           \hspace{1mm}
           \includegraphics[width=2.45cm]{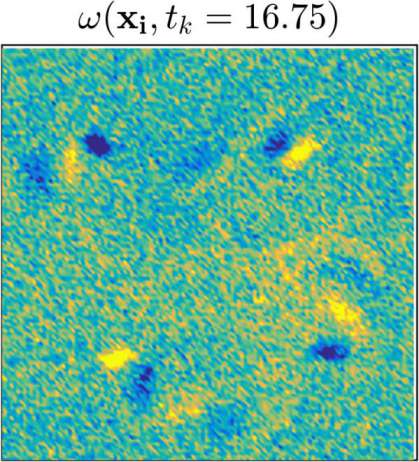}
                \hspace{1mm}
           \includegraphics[width=2.45cm]{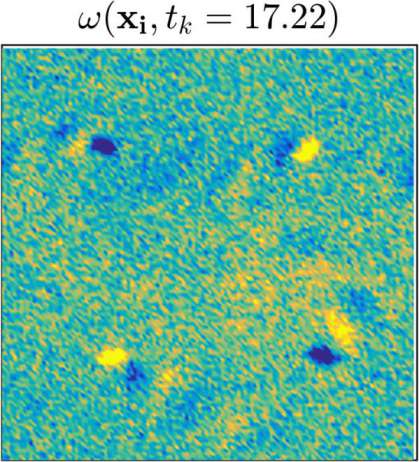}
                \hspace{1mm}
          \includegraphics[width=2.45cm]{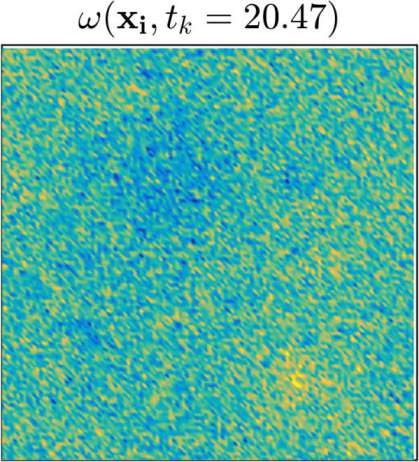}
\caption{Exemplary snapshots from the advection-diffusion problem in \eqref{vorticity_stream}. These selected snapshots describe some important instants in the evolution of the vorticity field considered in this test case. {An animation of this test case is available as supplemental Movie 2.} }
\label{TEST_1_SNAPSHOTS}
\end{figure*}

\begin{figure}
  \centering
  \includegraphics[width=9.6cm]{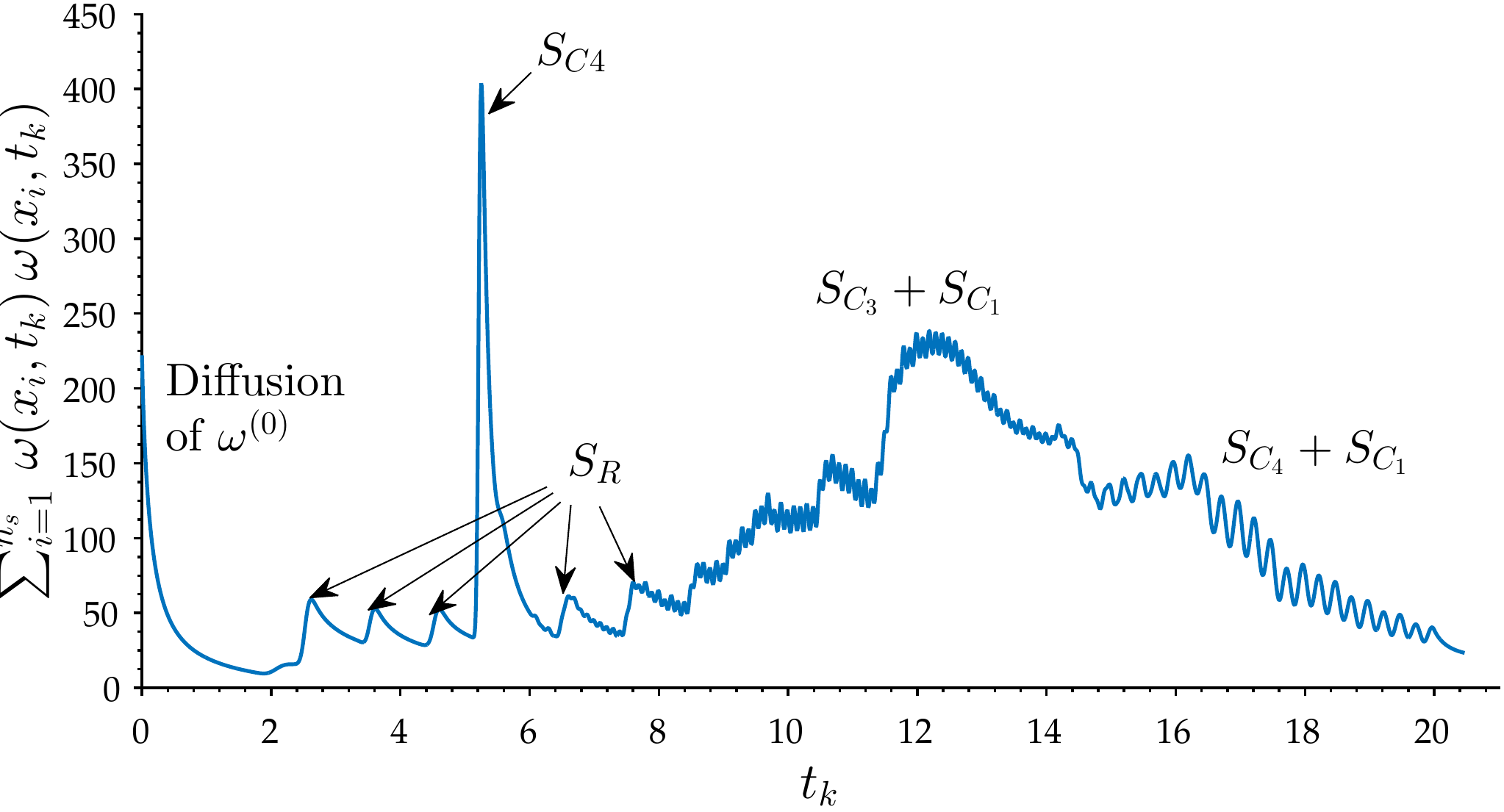}
\caption{Time evolution of the global energy content (norm of each vorticity realization), extracted from the diagonal of the temporal correlation matrix $K$.}
\label{K_DIAG_SIG}
\end{figure}

To visualize the complexity of the dataset obtained, Figure \ref{TEST_1_SNAPSHOTS} collects $10$ selected snapshots of the vorticity field, each corresponding to the time indicated in the title. The description of the vorticity evolution in these snapshots is well complemented by the evolution of the energy signal in time, that is the norm of each vorticity snapshot in the diagonal of the temporal correlation matrix $K$. This signal is shown in Figure \ref{K_DIAG_SIG}, and the occurrence of some key events are indicated with an annotation.

For $t_k<2$, the dataset describes the chaotic interaction of the initial field, which strongly diffuses to almost vanish (cf. Figure \ref{K_DIAG_SIG}). Before this occurs, the first coherent source gives its first pulse, producing a large-scale counter-clockwise rotation in the background. Until $t_k=5$, the three injections of random vortices produces three noticeable increase in the energy content against diffusion, but no coherent patterns can be identified at this stage.

At $t_k\approx5.15$, the injection of the source term $S_{C\psi_4}$ stands out from the chaotic motion, and the energy in the vorticity field reaches its highest peak. The four vortex pairs immediately merge into two identical dipoles of equal vorticity, which repel from the center of the domain in opposite direction and speed. However, due to the background rotation produced by the first source and the chaotic interaction with the other random injections from the pseudo-random term $S_{R\omega}$, these two dipoles do not travel the same distance and are broken and wholly dissipated by $t\approx 6.5$.

At $t_k>6$, the third coherent source begins its pulsation. At $t_k\approx7$ this generates four wakes which shed at a constant frequency. The overall vorticity in the domain increases and the large-scale clock-wise rotation involves the four wakes in a closed circular pattern. Before this large-scale rotation vanishes, at $t_k\approx 14$, the second coherent source starts to pulse at a lower frequency, producing other four trains of vortices, initially also rotating counterclockwise. Until approximately $t_k=17$ the second and third source coexist and produces patterns of similar energy content but different frequencies.

From $t_k>17$, the second source vanishes and the vortex pair introduced by the last three pulses of the first coherent source slowly changes the direction of the rotation. This breaks the circular shedding into two vertical ones before a new circular shedding rotating in the opposite direction is produced. The change of direction and the strong diffusion reduce the energy content of the flow until only a large scale vortex, rotating clockwise, slowly dissipates.

\subsection{Modal Analysis}

\begin{figure}
  \centering
  \includegraphics[width=7.5cm]{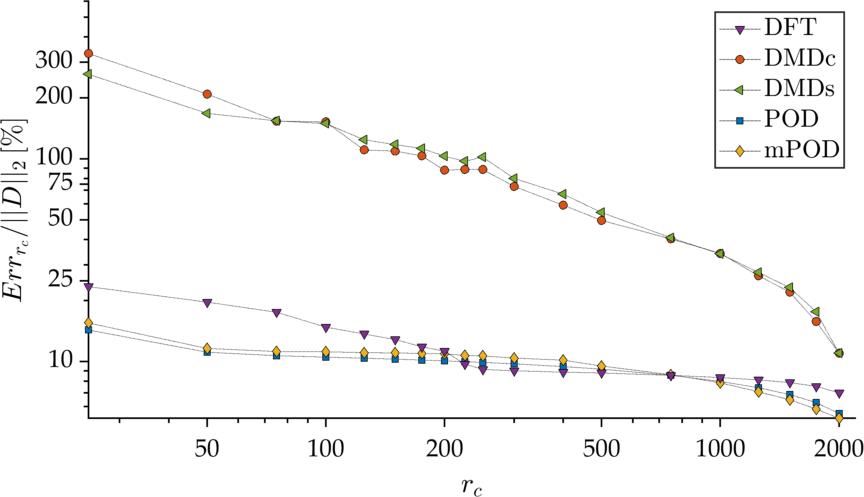}
\caption{Decomposition convergence for DFT, cDMD, sDMD, POD  and mPOD for the vorticity field in the nonlinear advection diffusion problem considered in this section.}
\label{Deco_TOT} 
\end{figure}

The relative $L^2$ error convergence (\ref{L2}) for the DFT, cDMD, sDMD, POD and mPOD are shown in Figure \ref{Deco_TOT}. The cDMD and the sDMD formulations yield very similar (yet not identical) results if the full POD basis is used in the construction of the reduced propagator in \eqref{PROPA_Schmidt}.

This test case strongly highlights the convergence limits of the DMD for a noisy data with nonlinear evolution.
As shown in \S\ref{DMD}, the amplitude coefficients $\sigma_{\mathcal{D}r}$ in the DMD are projections of the initial conditions onto the DMD basis: large amplitudes are assigned to modes which are more strongly linked to the initial conditions, even is these vanishes immediately afterwards -- in this case due to the strong diffusion. As a result, both algorithms produce DMD eigenvalues lying well inside the unit circle and exponentially decaying. Since decaying modes with real frequencies are not orthogonal, the DMD convergence is drastically compromised and yields, for $r_c<200$, an $L^2$ approximation error which is larger than the norm of the dataset itself.

\begin{figure}
\centering
\begin{subfigure}[c]{.32\textwidth}
  \centering
\includegraphics[width=3.6cm]{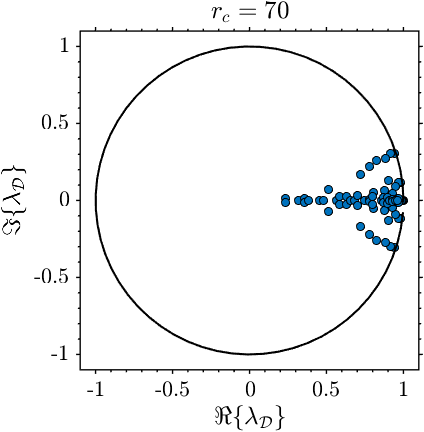}
 \end{subfigure}
 \begin{subfigure}[c]{.32\textwidth}
  \centering
\includegraphics[width=3.6cm]{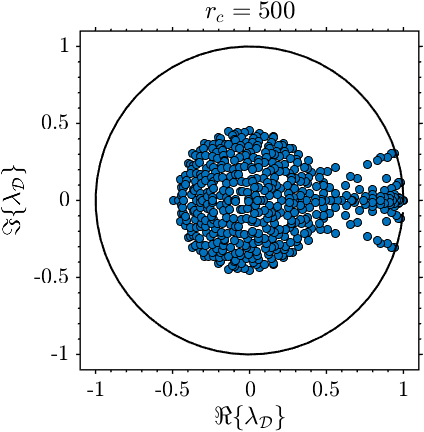}
 \end{subfigure}
 \begin{subfigure}[c]{.32\textwidth}
  \centering
\includegraphics[width=3.6cm]{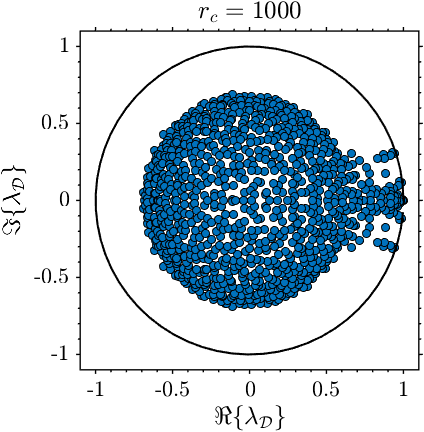}
 \end{subfigure}\\
\vspace{2mm}
 \begin{subfigure}[c]{.32\textwidth}
  \centering
\includegraphics[width=3.6cm]{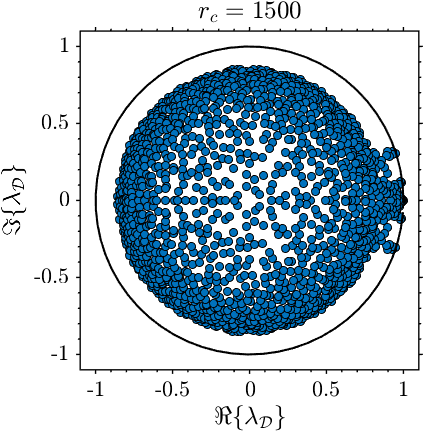}
 \end{subfigure}
 \begin{subfigure}[c]{.32\textwidth}
  \centering
\includegraphics[width=3.6cm]{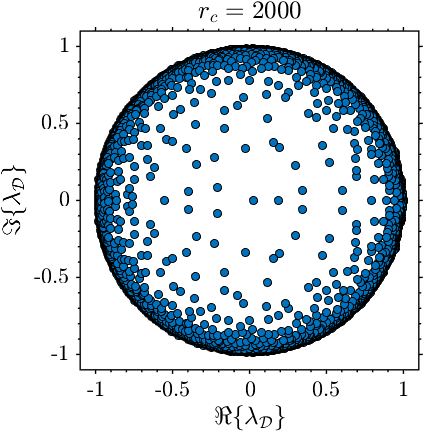}
 \end{subfigure}
 \begin{subfigure}[c]{.32\textwidth}
  \centering
\includegraphics[width=3.6cm]{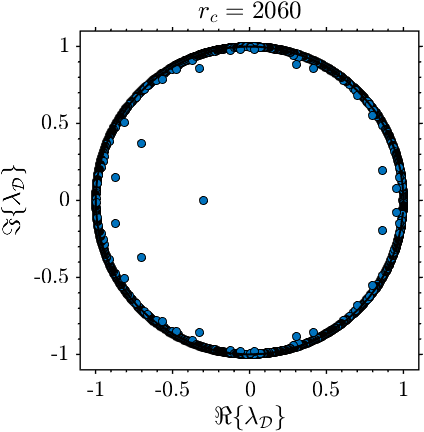}
 \end{subfigure}
 \caption{Eigenvalue distribution for reduced propagators $\mathcal{S}\in \mathbb{C}^{r_c\times r_c}$ in \eqref{PROPA_Schmidt} obtained by projecting the full propagator $P$ onto the first $r_c$ POD modes. The smaller the propagator, the more the decomposition convergence is limited within a smaller number of time steps from the initial data. At $r_c=n_t-1$, the sDMD approaches the cDMD.}
\label{eigensDMD}
\end{figure}

The problem is particularly evident for the sDMD with the reduced propagator $\mathcal{S}$ expressed in \eqref{PROPA_Schmidt}. Figure \ref{eigensDMD} shows the eigenvalues distribution in the complex plane for reduced propagators $\mathcal{S}$ of different dimensions $r_c\times r_c$, obtained by projecting the full propagator onto the first $r_c$ POD modes. For $r_c\ll n_t$, most of the eigenvalues are well inside the unit circle, and their contribution to the decomposition is limited to the first few time steps. Since each of these modes is normalized, the shortest is their duration, the largest the associated amplitude. As $r_c$ approaches $n_t$, and the DMD basis approaches completeness, the number of modes with exponential decays is reduced, and the sDMD approaches the cDMD.

\begin{figure}
\centering
\begin{subfigure}[c]{.45\textwidth}
  \centering
\includegraphics[width=5.5cm]{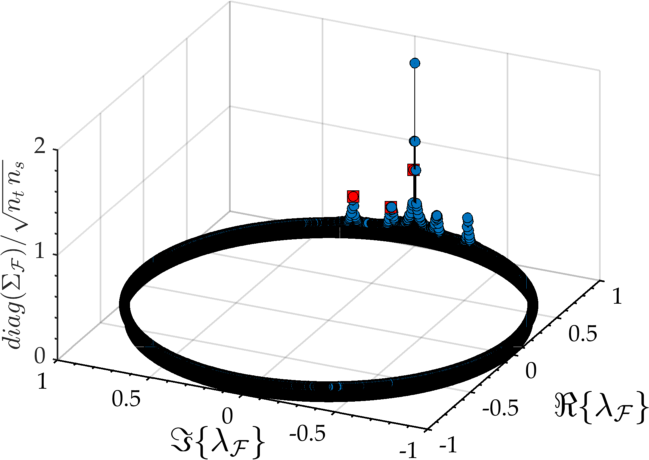}
 \end{subfigure}
 \begin{subfigure}[c]{.45\textwidth}
\includegraphics[width=5.5cm]{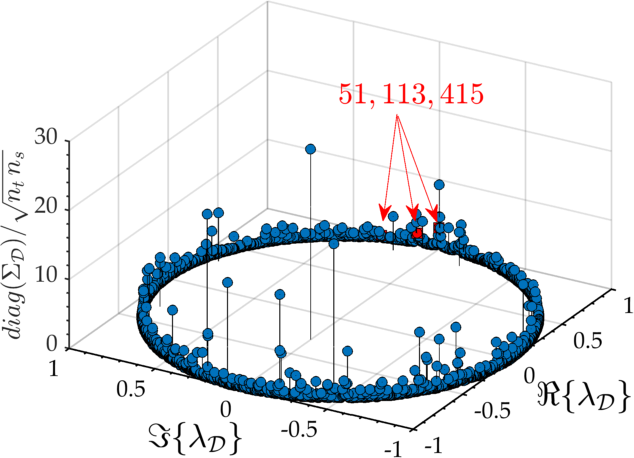}
 \end{subfigure}
 \caption{Amplitude spectra of the DFT (left), and cDMD (right) mapped in the complex plane. While both frequency based decompositions are symmetric along the imaginary axis, the DMD is sensitive to the high noise level and the strong diffusion of the initial dataset, which results in all the leading modes lying inside the unit circle.}
\label{Vort_AMP}
\end{figure}

The amplitudes of the modes for the DFT and the cDMD are mapped in the complex plane in Figure \ref{Vort_AMP}. The DFT spectrum shows that the energy contribution of the mean flow is stronger than in the previous test case, and highlights three dominant frequencies together with a wide range of modes with nearly identical energy content. These are produced by the high random noise level in the dataset and the presence of impulsive events.
For the cDMD, the random noise makes the calculation of the Companion matrix more difficult, producing an error of $92\%$ of the $L^2$ norm of the last realization in the minimization $||d_{n_t}-D_1\,\mathbf{c}||_2$.

\begin{figure*}
\begin{subfigure}{.29\textwidth}
  \centering
  \includegraphics[width=2.9cm]{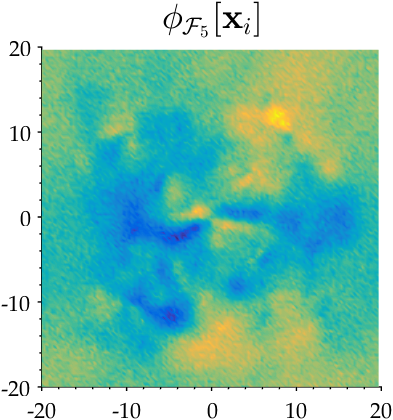}
\end{subfigure}
\begin{subfigure}{.29\textwidth}
  \centering
  \includegraphics[width=2.9cm]{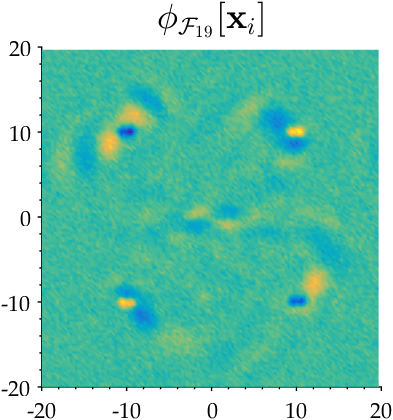}\\
\end{subfigure}
\begin{subfigure}{.29\textwidth}
  \centering
  \includegraphics[width=2.9cm]{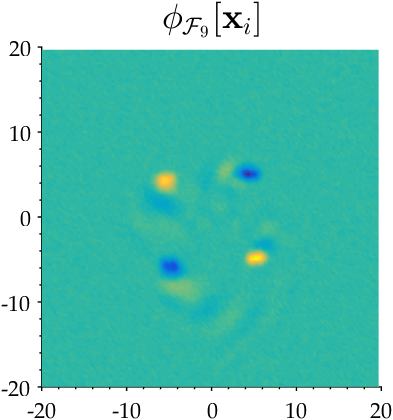}\\
\end{subfigure}
\begin{subfigure}{.08\textwidth}
\hspace{1mm}
\includegraphics[width=0.9cm]{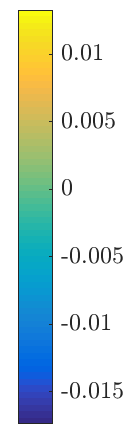}
\end{subfigure}\\
\vspace{2mm}
\begin{subfigure}{.29\textwidth}
  \centering
  \includegraphics[width=2.9cm]{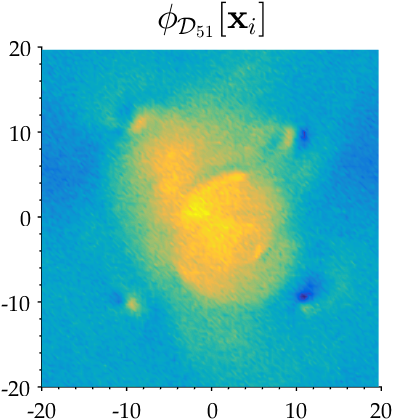}
\end{subfigure}
\begin{subfigure}{.29\textwidth}
  \centering
  \includegraphics[width=2.9cm]{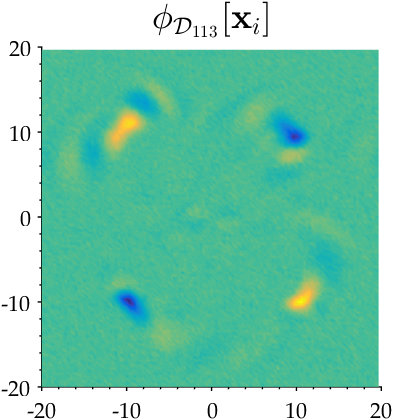}\\
\end{subfigure}
\begin{subfigure}{.29\textwidth}
  \centering
  \includegraphics[width=2.9cm]{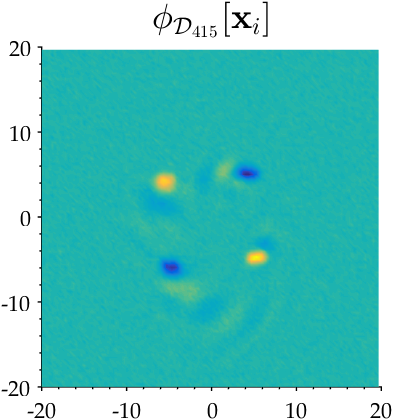}\\
\end{subfigure}
\begin{subfigure}{.08\textwidth}
\includegraphics[width=0.9cm]{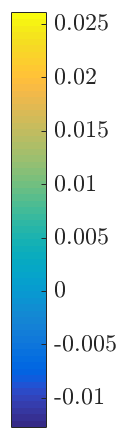}
\end{subfigure}\\
\caption{First row: spatial structures $\phi_\mathcal{F}[\mathbf{x}_i]$ of three dominant DFT modes, pulsing, from left to right, at frequencies of $f_n=[0.12,2.01,5.02]$. These corresponds to leading frequencies of the coherent source term. Second row: spatial structures $\phi_\mathcal{D}[\mathbf{x}_i]$ of the dominant sDMD modes with frequencies $[0.08,1.97,4.99]$, that is the ones closer to those dominating the DFT spectra.}
\label{DFT_DMD_SPATIAL_VORTI}
\end{figure*}

The three dominant frequencies in the DFT are $f_n=[0.12,2.01,5.02]$; the first is linked to the large-scale motion produced by the first coherent source, the second and third are related to the frequencies of the second and third sources pulsing at $f_{2,3}=[2,5]$. Figure \ref{DFT_DMD_SPATIAL_VORTI} (first row) shows the corresponding spatial structures, which are well related to the introduced ones. The three cDMD modes with unitary modulus having frequencies closer to these DFT modes -- also captured by all the sDMD in Figure \ref{eigensDMD}-- are shown in the second row of Figure \ref{DFT_DMD_SPATIAL_VORTI}. The amplitude of these modes is labeled with a red marker in the complex plane of Figure \ref{Vort_AMP}, and is much lower than the dominant ones, associated to the vanishing initial conditions.

As for the synthetic test case in \S\ref{VI}, both the DFT and the DMD have sufficient frequency resolution to correctly localize periodic phenomena in the frequency domain and --thanks to the projection in \eqref{PHI_CALC}-- in the space domain. However, as no temporal localization is possible, the convergence of these decompositions is remarkably poorer than the POD and results in a high redundancy of the spatial basis. This is particularly evident in the description of the impulsive event produced by the fourth coherent source, the footprint of which appears in a large number of DFT and DMD modes.

%
%

\begin{figure*}
 \centering
  \includegraphics[height=3.7cm]{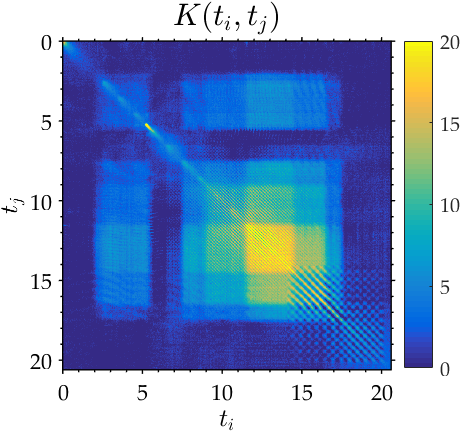}
\hspace{2mm}
  \includegraphics[height=3.7cm]{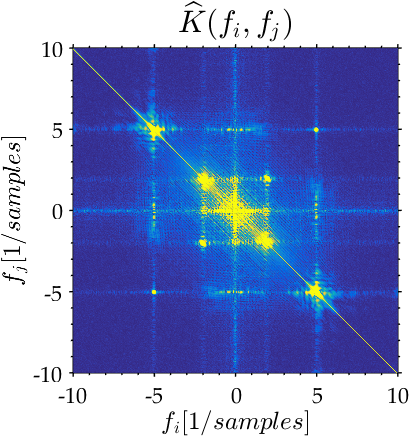}\\
\begin{subfigure}{.29\textwidth}
  \centering
  \includegraphics[width=2.9cm]{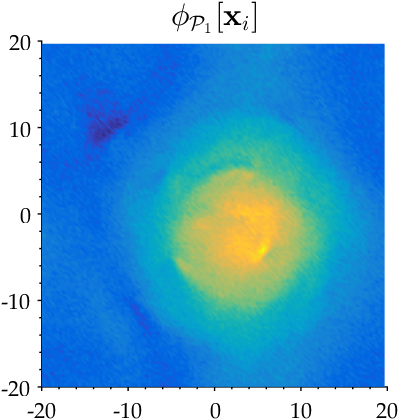}\\
\vspace{2mm}
\includegraphics[width=3.6cm]{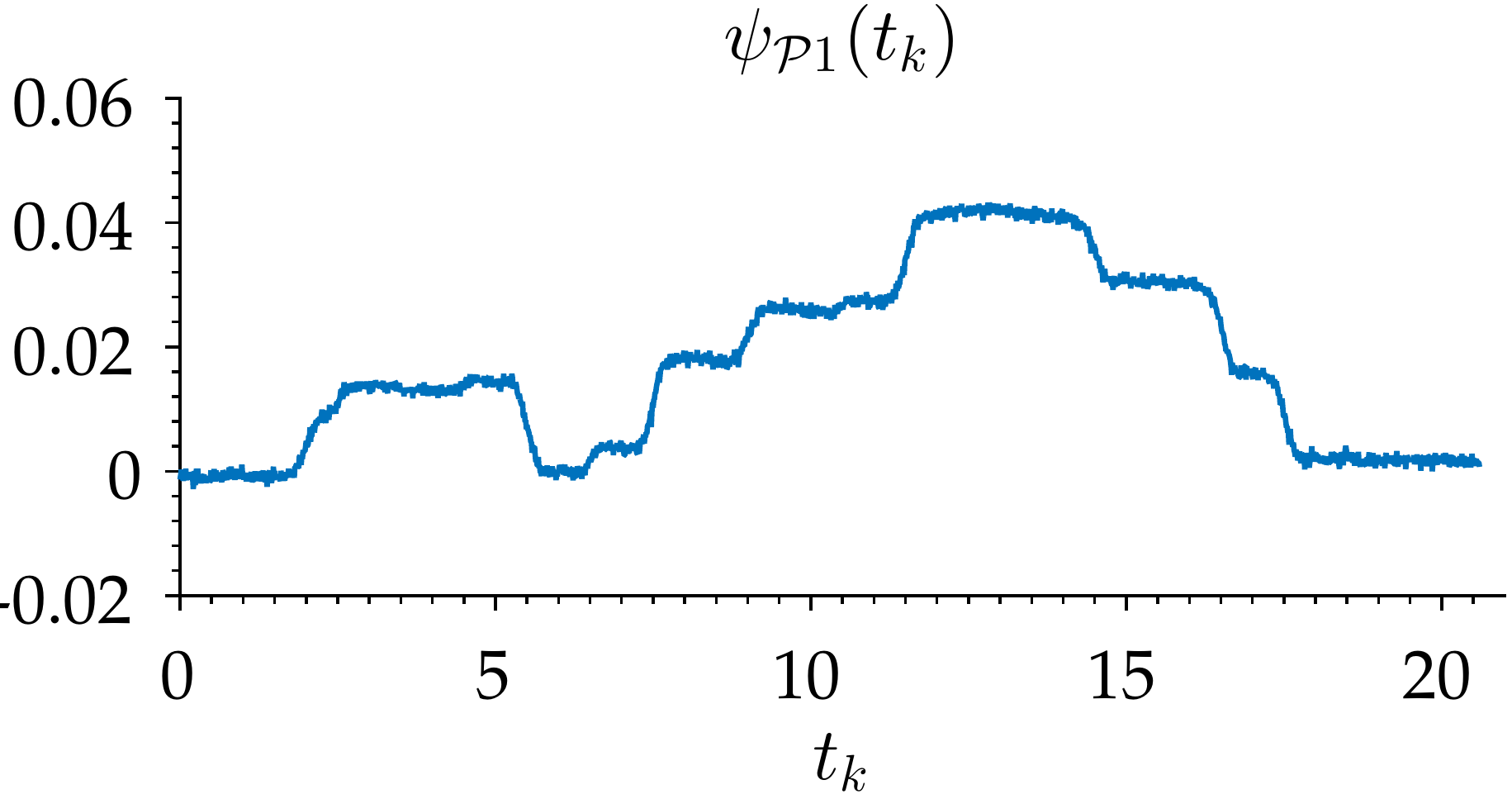}
\end{subfigure}
\begin{subfigure}{.29\textwidth}
  \centering
  \includegraphics[width=2.9cm]{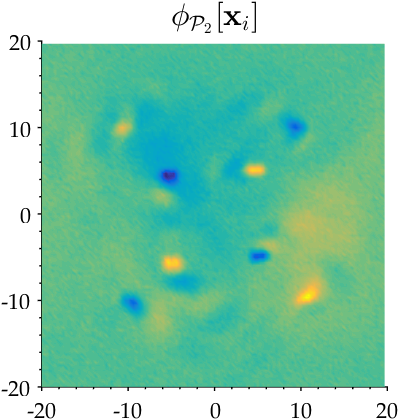}\\
\vspace{2mm}
\includegraphics[width=3.6cm]{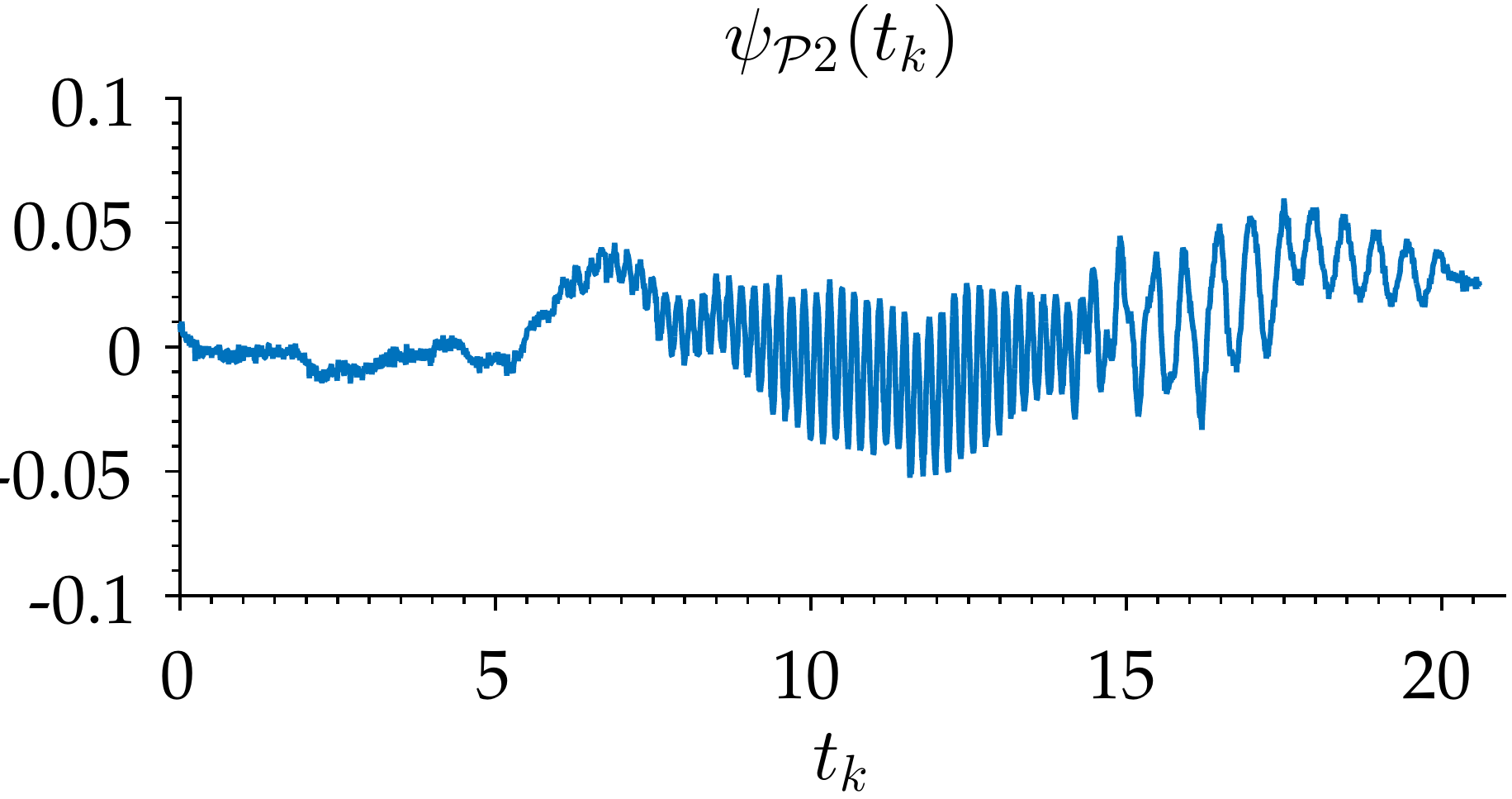}
\end{subfigure}
\begin{subfigure}{.29\textwidth}
  \centering
  \includegraphics[width=2.9cm]{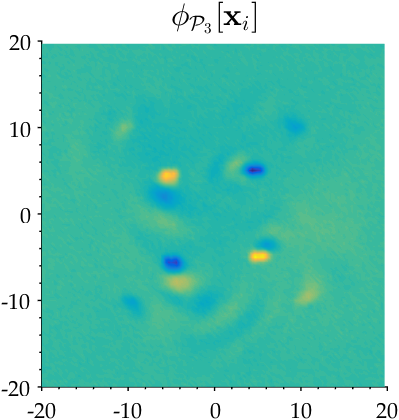}\\
\vspace{2mm}
\includegraphics[width=3.6cm]{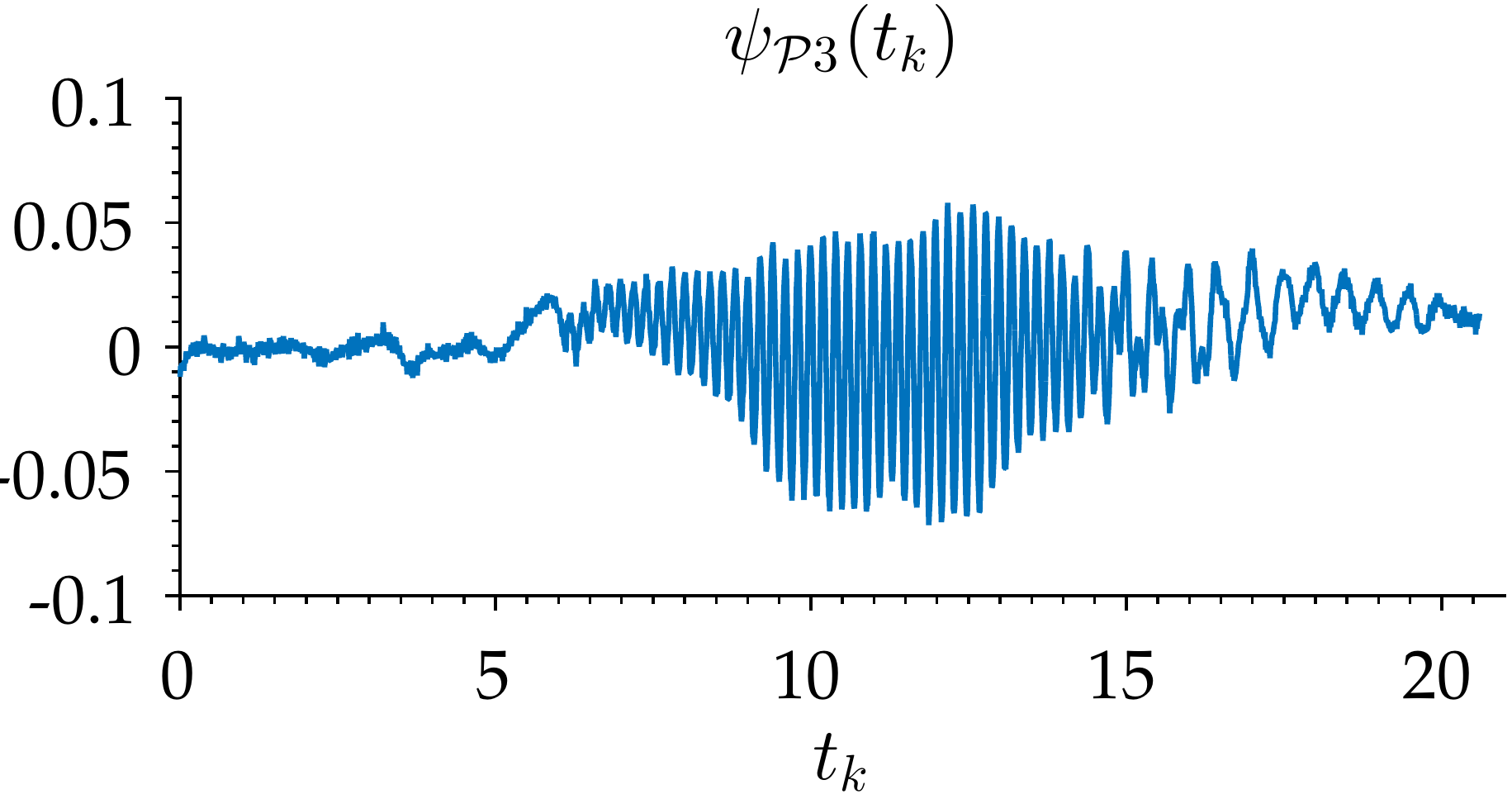}
\end{subfigure}
\begin{subfigure}{.08\textwidth}
\includegraphics[width=0.8cm]{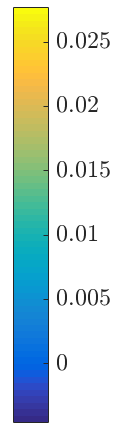}\\\vspace{18mm}
\end{subfigure}\\
\begin{subfigure}{.29\textwidth}
  \centering
  \includegraphics[width=2.9cm]{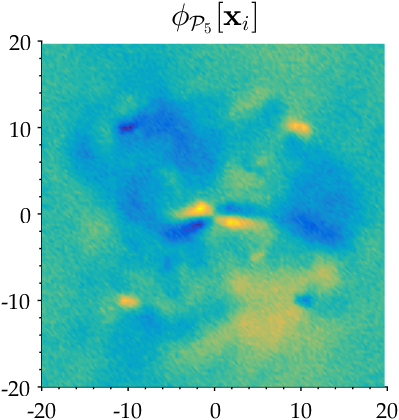}\\
\vspace{2mm}
\includegraphics[width=3.6cm]{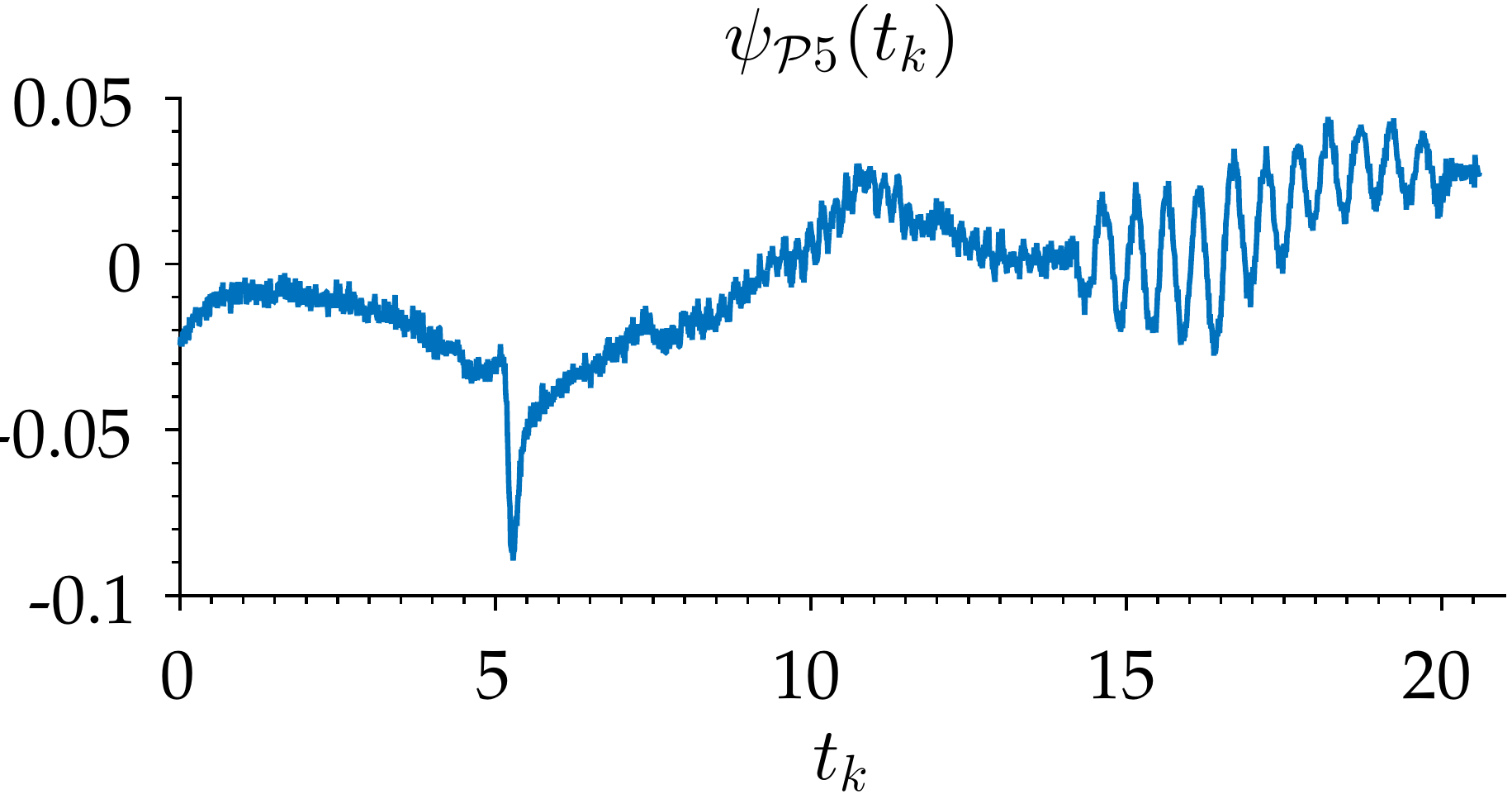}
\end{subfigure}
\begin{subfigure}{.29\textwidth}
  \centering
  \includegraphics[width=2.9cm]{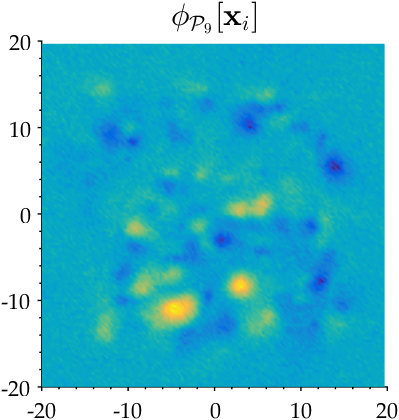}\\
\vspace{2mm}
\includegraphics[width=3.6cm]{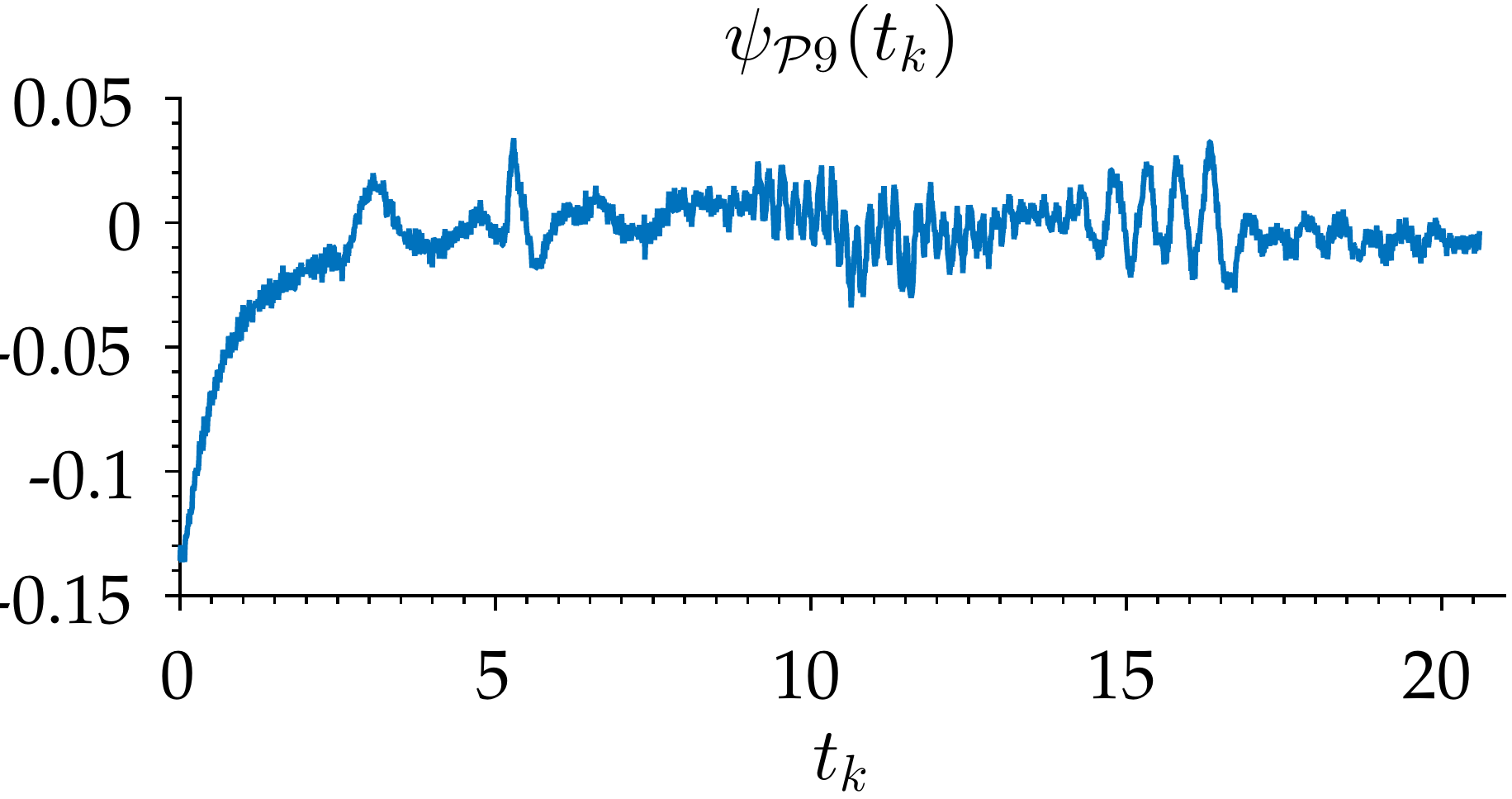}
\end{subfigure}
\begin{subfigure}{.29\textwidth}
  \centering
  \includegraphics[width=2.9cm]{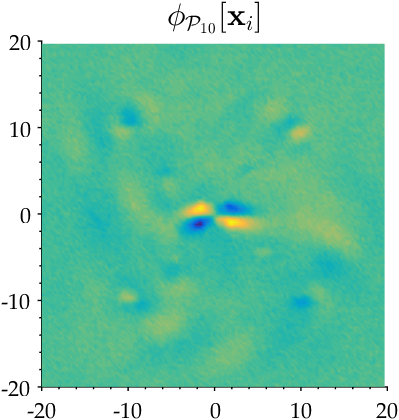}\\
\vspace{2mm}
\includegraphics[width=3.6cm]{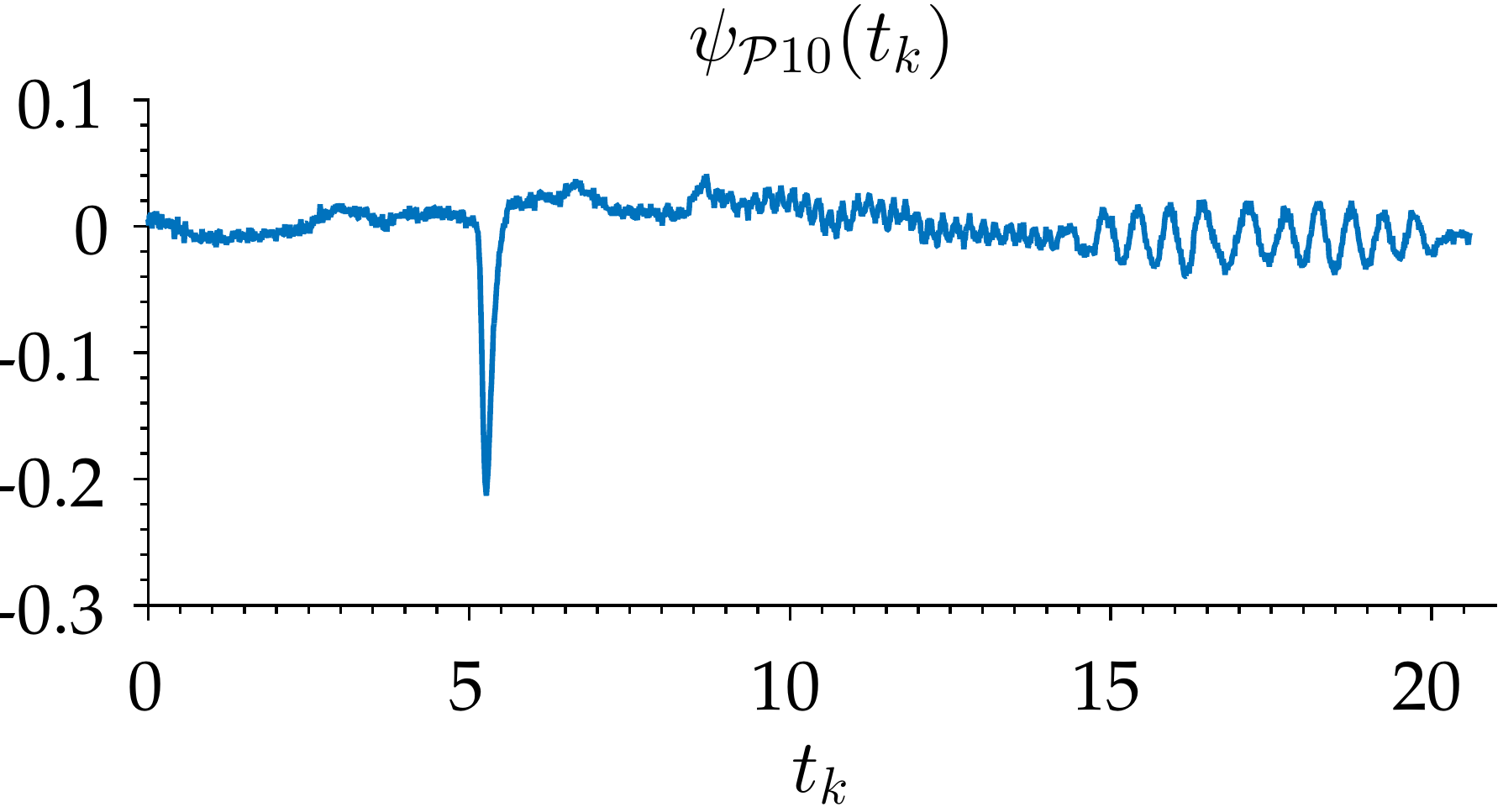}
\end{subfigure}
\begin{subfigure}{.08\textwidth}
\includegraphics[width=0.8cm]{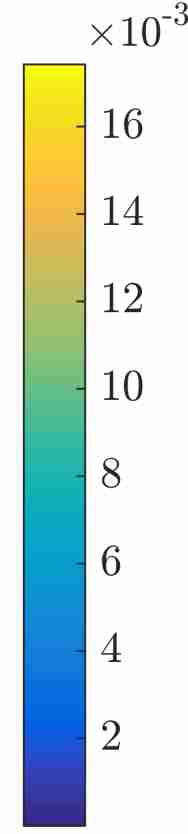}\\\vspace{18mm}
\end{subfigure}
\caption{Same as Fig. \ref{POD_RES} but for the second test case discussed in this section.}
\label{POD_RES_VOR}
\end{figure*}

The POD has complementary advantages and limitations. 
Figure \ref{POD_RES_VOR} shows the results of the POD analysis, including the temporal correlation matrix $K$ and its DFT spectra $\widehat{K}$, together with the spatial and the temporal structures of six representative POD modes.

The temporal correlation matrix $K$ gives information on the time localization of different events, featuring two significant chessboard patterns corresponding to the harmonic contributions of the second and third sources shown in Figure \ref{SOURCES}, and rectangular areas corresponding to the time interval of major activity of the first large-scale coherent source. In the diagonal, a clear peak is located at $t_k=5.2$, corresponding to the introduction of the fourth impulsive source, while the random noise gives a pedestal correlation level along the whole diagonal.

\begin{figure*}
 \begin{subfigure}{.24\textwidth}
  \centering
  \includegraphics[height=3.4cm]{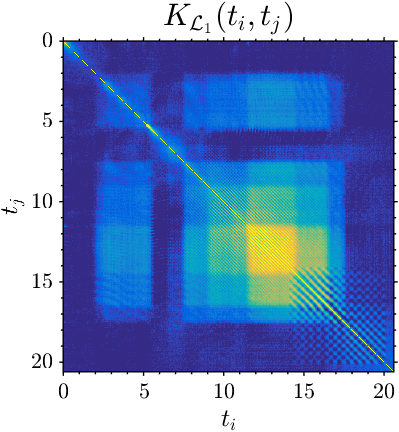}\\
\vspace{1mm}
\includegraphics[height=3.4cm]{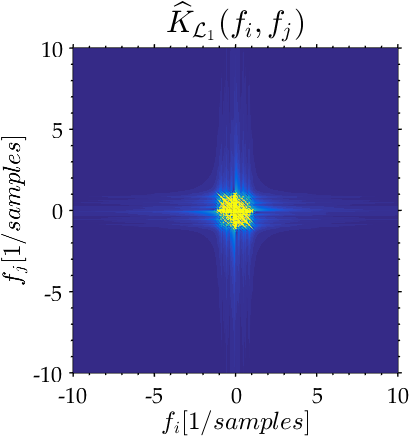}
\end{subfigure}
\begin{subfigure}{.24\textwidth}
  \centering
  \includegraphics[height=3.4cm]{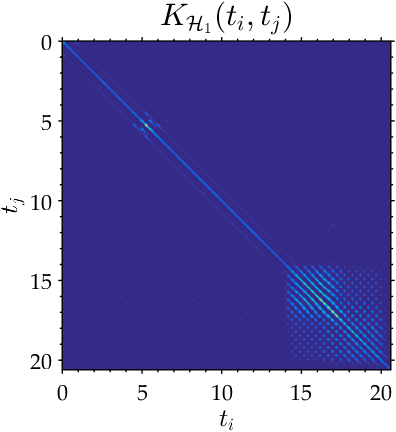}\\
\vspace{1mm}
\includegraphics[height=3.4cm]{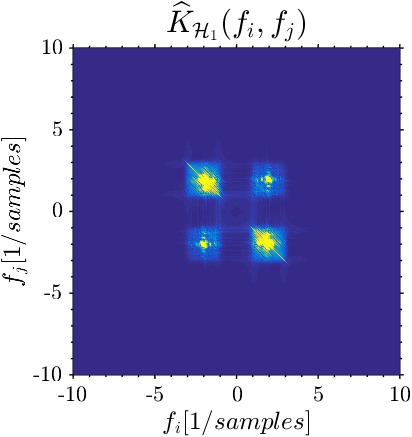}
\end{subfigure}
\begin{subfigure}{.24\textwidth}
  \centering
  \includegraphics[height=3.4cm]{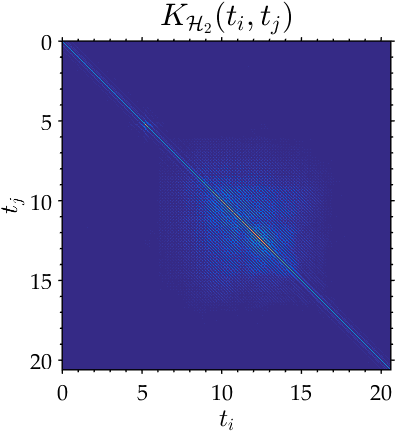}\\
\vspace{1mm}
\includegraphics[height=3.4cm]{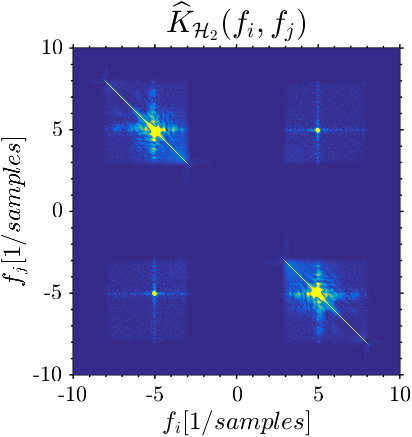}
\end{subfigure}
\begin{subfigure}{.24\textwidth}
  \centering
  \includegraphics[height=3.4cm]{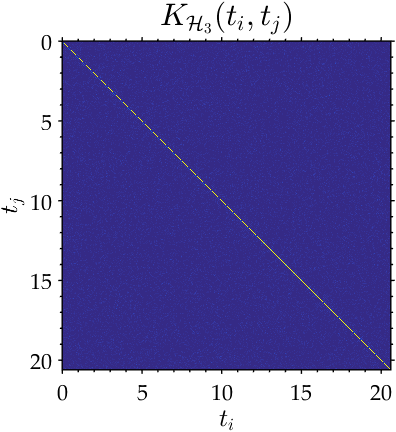}\\
\vspace{1mm}
\includegraphics[height=3.4cm]{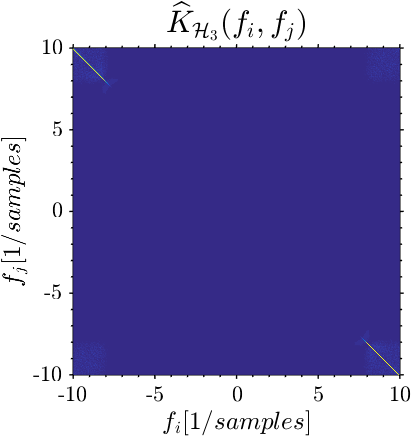}
\end{subfigure}
\caption{Results from the MRA of the correlation matrix $K$, split into four contributions by a filter bank with frequency splitting vector $F_V=[1,3,8]$. The first row shows the temporal correlations; the bottom row shows their frequency content.}
\label{mPOD_CORRs}
\end{figure*}

\begin{figure*}
\begin{subfigure}{.32\textwidth}
  \centering
  \includegraphics[width=2.9cm]{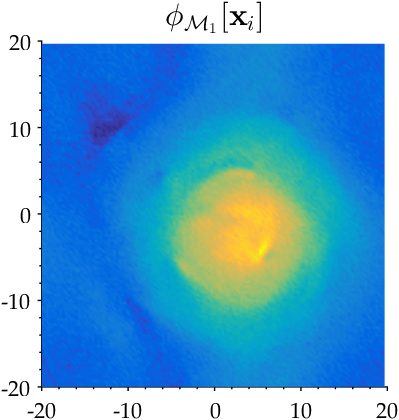}\\
\vspace{2mm}
\includegraphics[width=3.7cm]{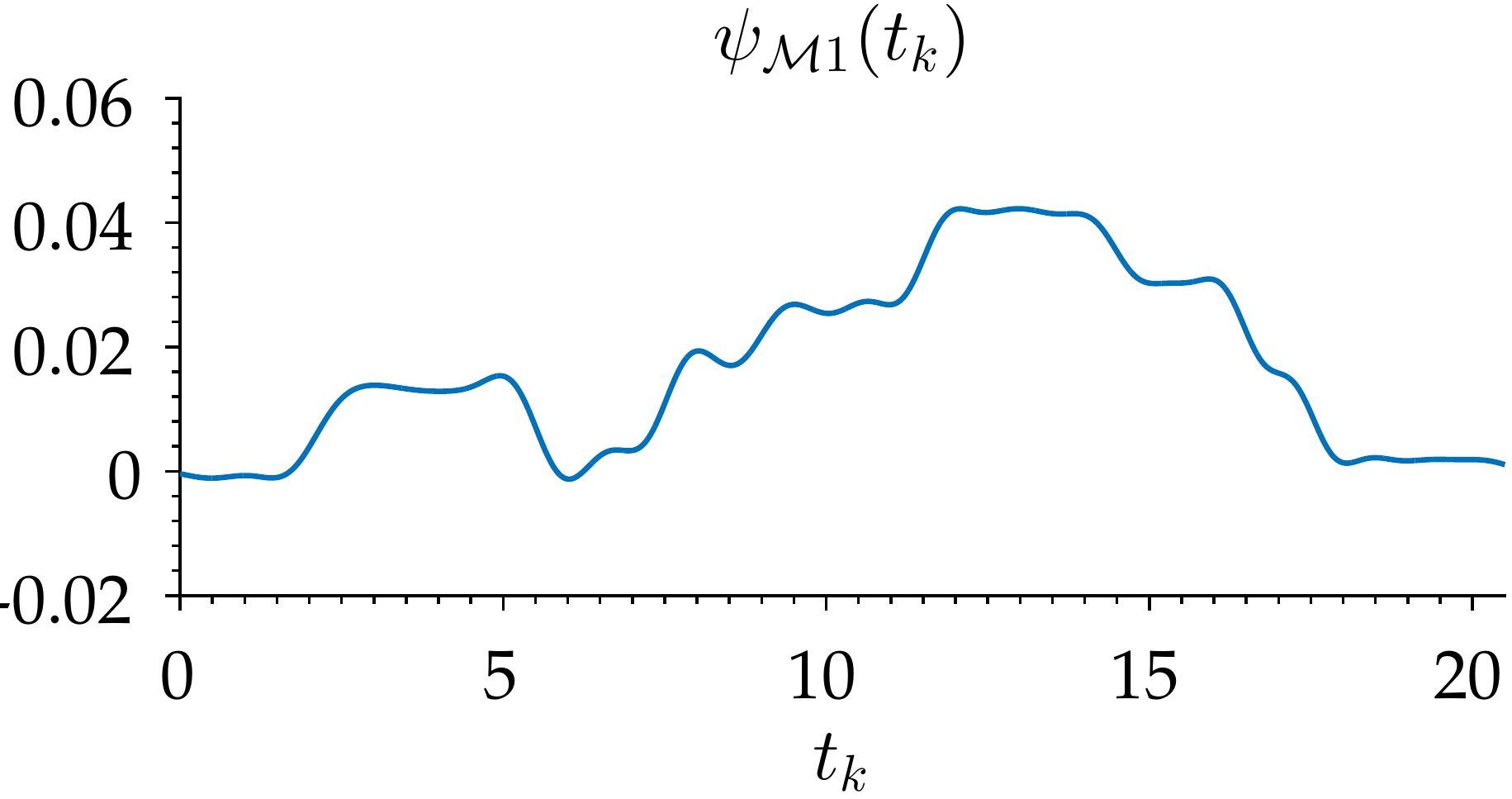}
\end{subfigure}
\begin{subfigure}{.32\textwidth}
  \centering
  \includegraphics[width=2.9cm]{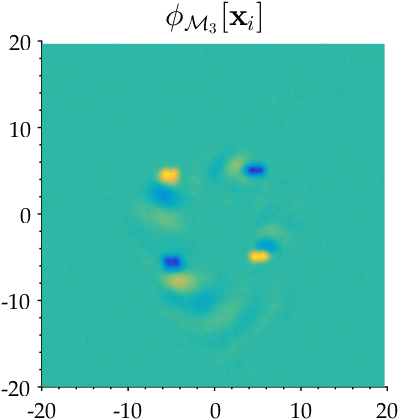}\\
\vspace{2mm}
\includegraphics[width=3.7cm]{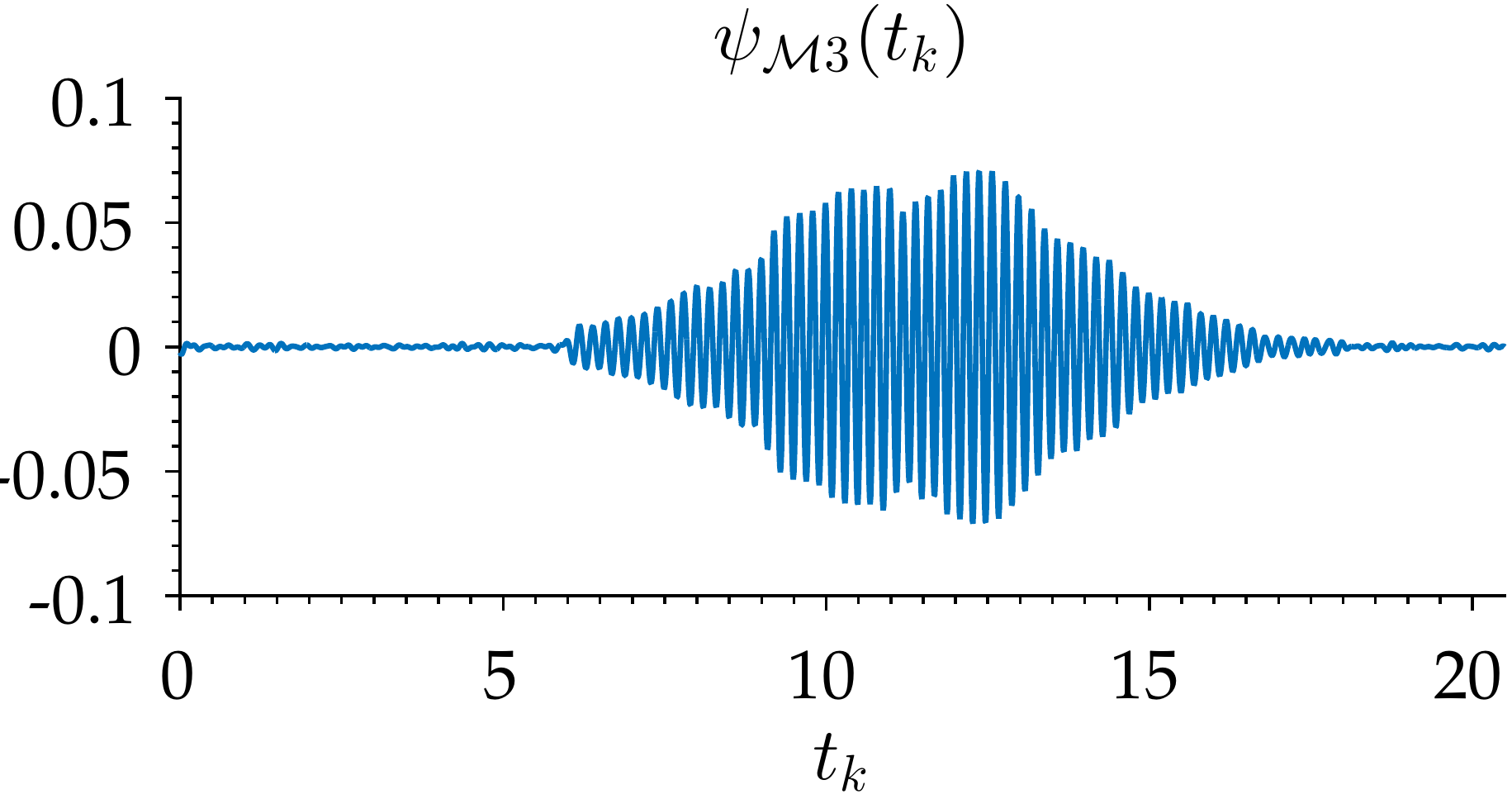}
\end{subfigure}
\begin{subfigure}{.32\textwidth}
  \centering
  \includegraphics[width=2.9cm]{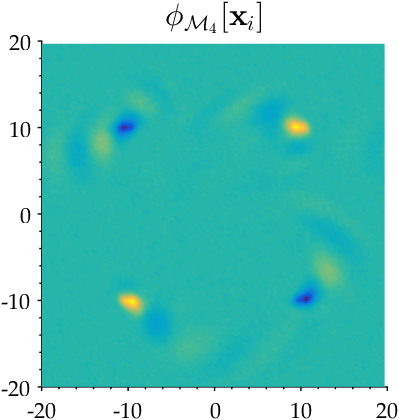}\\
\vspace{2mm}
\includegraphics[width=3.7cm]{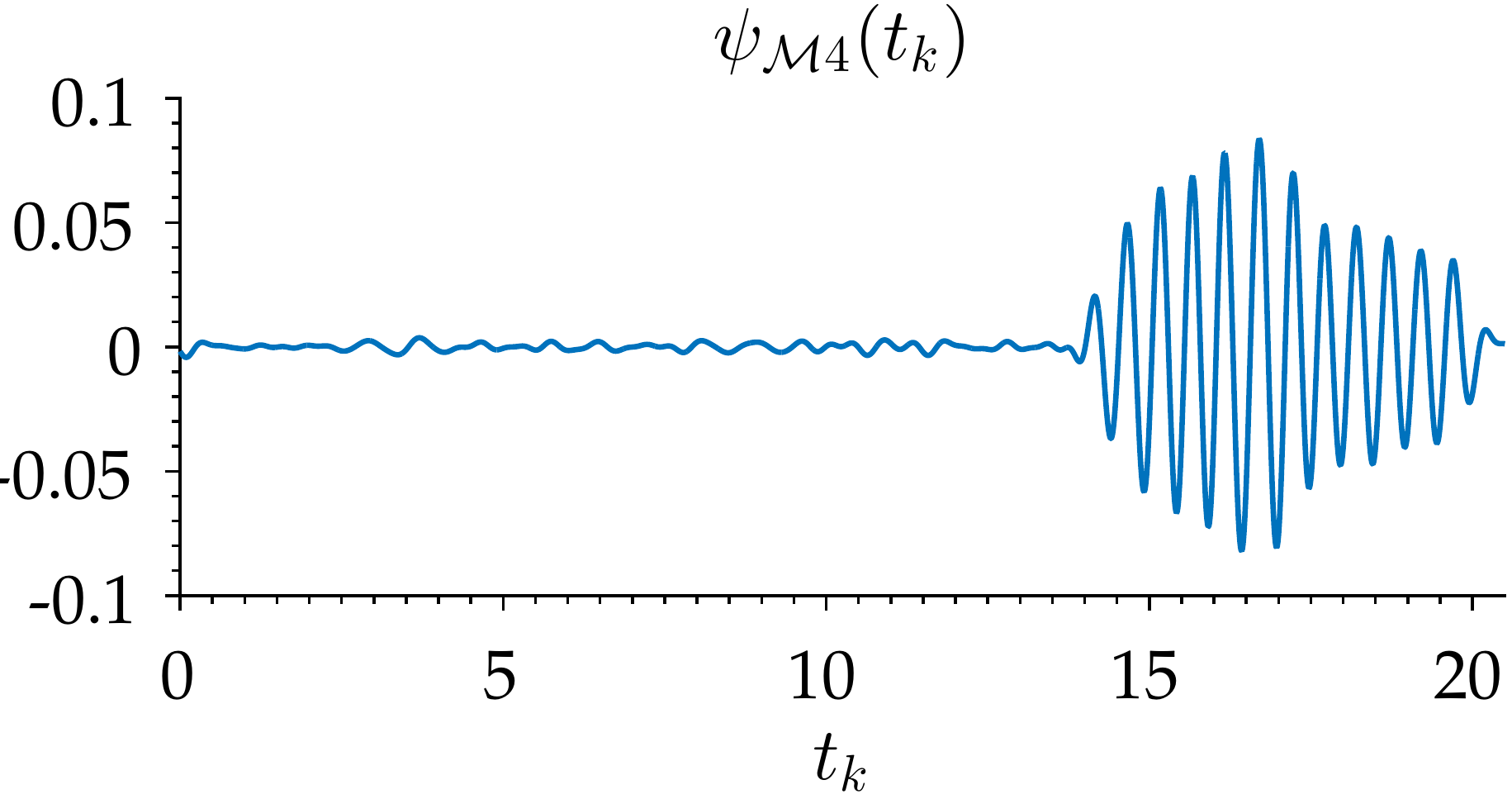}
\end{subfigure}\\
\begin{subfigure}{.32\textwidth}
  \centering
  \includegraphics[width=2.9cm]{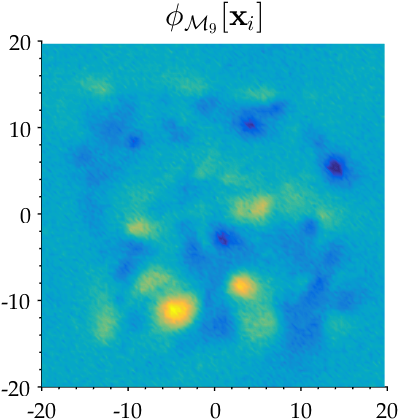}\\
\vspace{2mm}
\includegraphics[width=3.7cm]{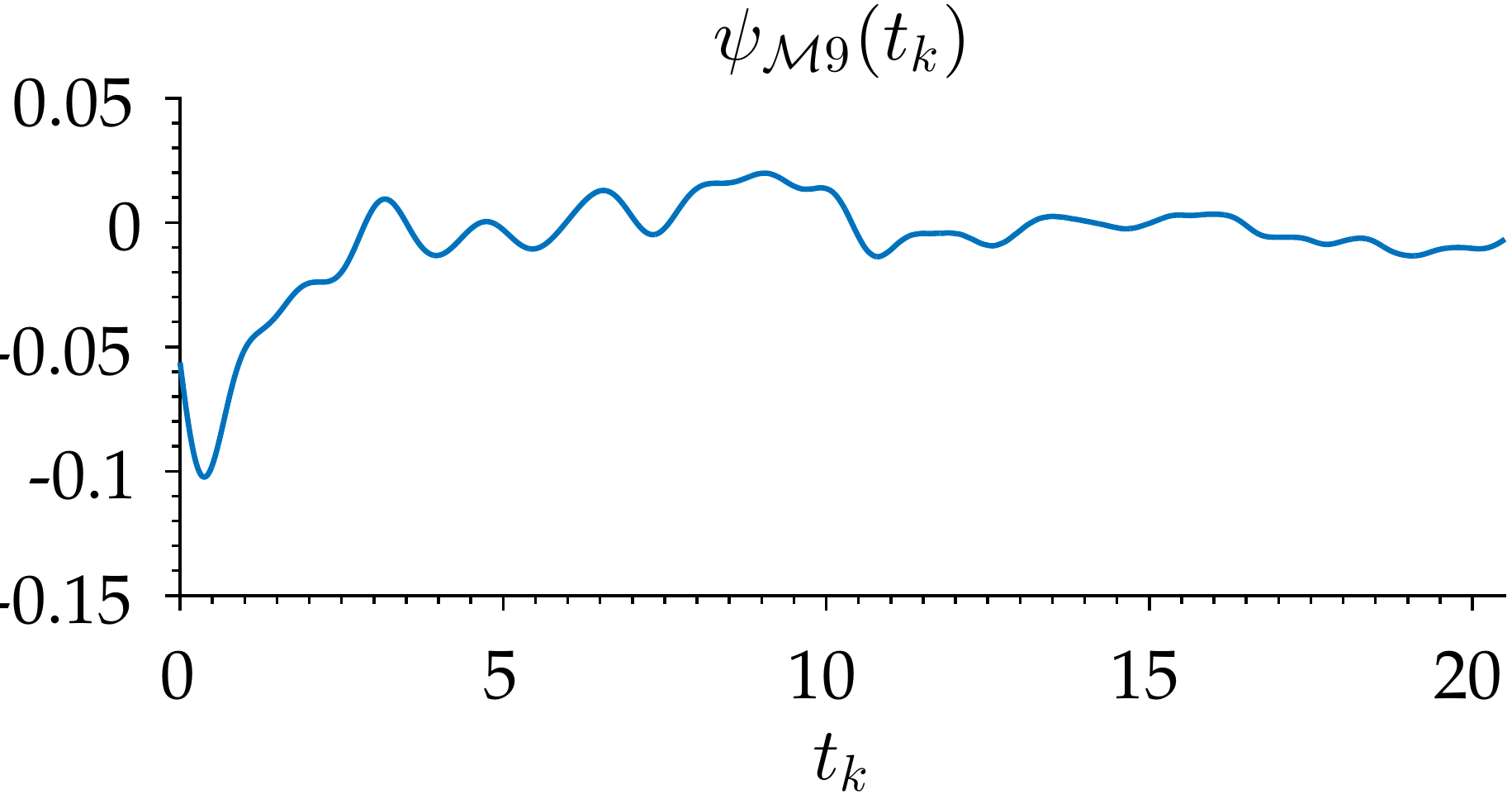}
\end{subfigure}
\begin{subfigure}{.32\textwidth}
  \centering
  \includegraphics[width=2.9cm]{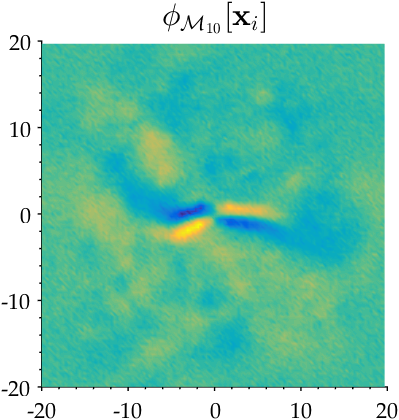}\\
\vspace{2mm}
\includegraphics[width=3.7cm]{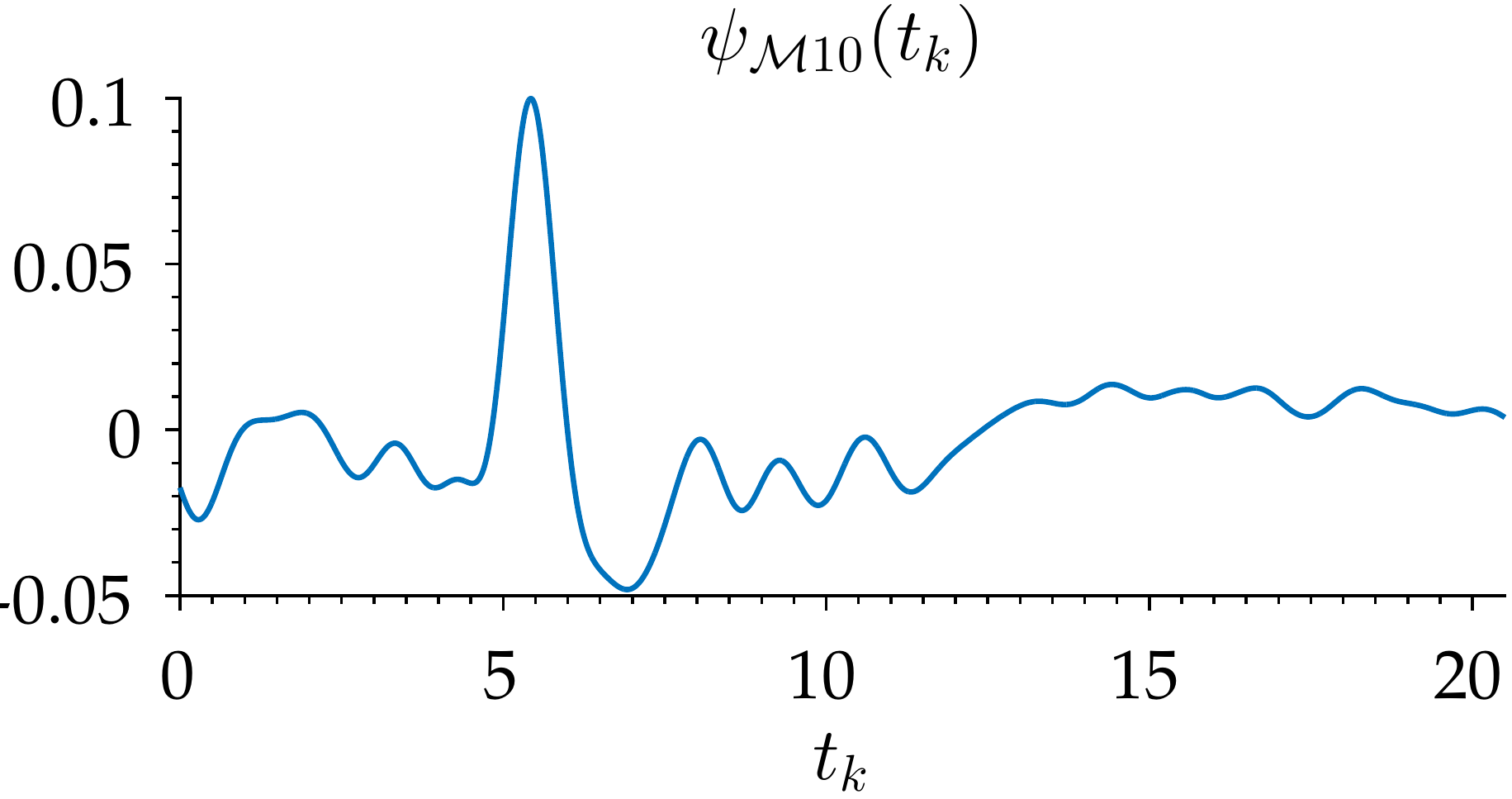}
\end{subfigure}
\begin{subfigure}{.32\textwidth}
  \centering
  \includegraphics[width=2.9cm]{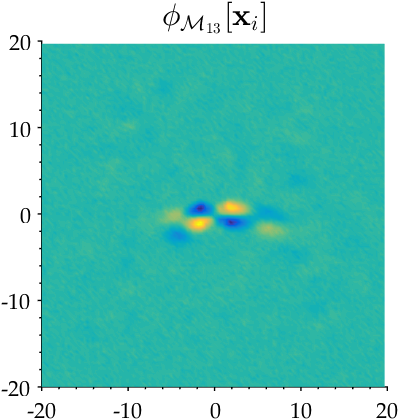}\\
\vspace{2mm}
\includegraphics[width=3.7cm]{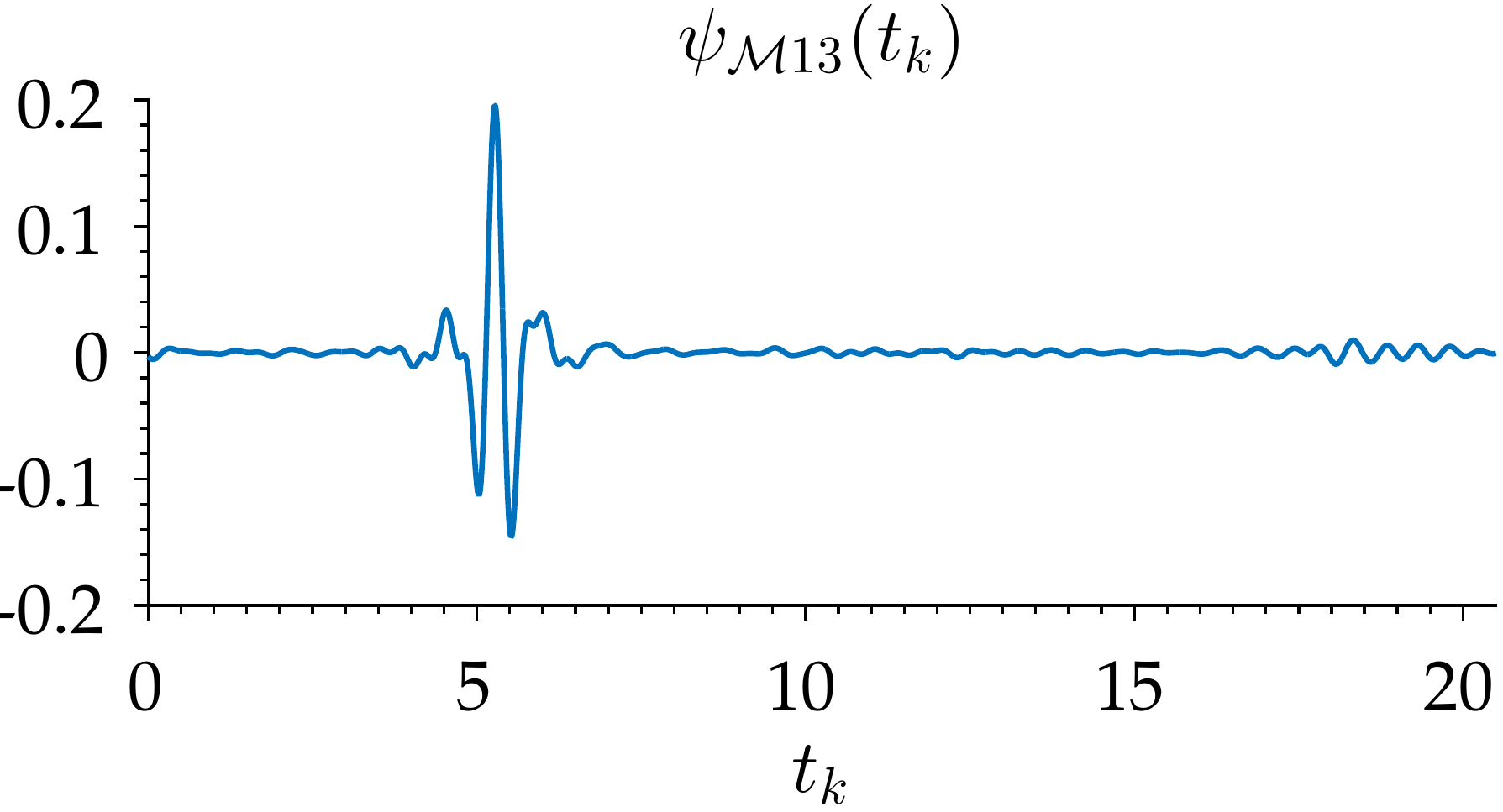}
\end{subfigure}
\caption{Spatial and temporal structures for six representative mPOD modes for the nonlinear vortex test case presented in this section.}
\label{mPOD_RES_VOR}
\end{figure*}

The first mode captures the large-scale motion produced by the first coherent source, with a temporal structure slightly polluted by white noise. Uncorrelated noise is uniformly distributed over the entire POD spectra \citep{Mendez_J_1} and the low noise level present in this mode is due to the large dataset available. The second and third modes show a severe spectral mixing, as their energy contribution is similar ($\sigma_{\mathcal{P}_2}=4272$ and $\sigma_{\mathcal{P}_2}=4433$). This problem, also illustrated on the simple synthetic test case in \S\ref{VI}, results in a combination of the features from the second and third coherent sources. Such mixing extends to other POD modes with different phase delays, to cancel with the summation of multiple modes. The fifth mode mixes three phenomena: a large-scale contribution to the rotation produced by the first source, a harmonic contribution due to the second source, and a sharp peak due to the impulsive fourth source. The same occurs with the ninth mode, which mostly describes the diffusion of the initial data. Interestingly, the tenth mode well captures the impulsive event, although both the harmonic contribution of the second and third sources are present to a minor extent.

Figure \ref{mPOD_CORRs} shows the results of the MRA analysis of the temporal correlation matrix $K$, using $F_V=[1,3,8]$ and computing the filter orders from \eqref{N_f}. The first row displays the four contributions of the identified scales and the second row their DFT. No significant phenomenon appears in the last contributions, which mostly contains a substantial portion of the noise energy, whereas the other three scales isolate the contributions of different sources. Being the overlapping of frequencies negligible, the eigenvectors of these matrices are almost orthogonal complements for $\mathbb{R}^{n_t}$, and the resulting full basis $\Psi^0_\mathcal{M}$ (\ref{mPOD_0}) requires a minor re-orthogonalization via QR factorization.

The spatial and temporal structures of six representative mPOD modes are shown in Figure \ref{mPOD_RES_VOR}. These modes are similar to the POD modes in Figure \ref{POD_RES_VOR}, but with no frequency overlapping. The first mode is a cleaner version of the first POD mode, with negligible random noise; the third and fourth captures the contributions of the second and third coherent sources with no mixing. The sixth mode describes the diffusion of the initial data while the tenth and the thirteenth are related to the impulsive event. The frequency bounding makes the decomposition stronger than the standard POD versus the random noise since only a minor portion of its contribution remains in each scale. Yet, the capabilities of capturing impulsive phenomena (also distributed over the entire frequency spectra) are reasonably preserved: traces of this impulse event are present in all the scales (the tenth mode is eigenvector of $K_{\mathcal{L}}$; the thirteenth is eigenvector of  $K_{\mathcal{H}_3}$), but no significant overlapping between this event and the other harmonics is produced.


\section{Example III: PIV of an Impinging Jet Flow}\label{VIII}
\subsection{Dataset Description}

As a third test case, we propose an experimental dataset obtained via time-resolved Particle Image Velocimetry (TR-PIV) on a classical, statistically stationary flow configuration: a plane jet impinging normally onto a flat plate. Impinging jet flows have received considerable attention for their application in drying, cooling or heating processes \citep{heat1,dry,heat2} and the POD of PIV data has been successfully used by many authors to identify the resulting coherent structures \citep{Kim,Charmiyan,Pieris,Hammad}.

These flows are characterized by largely different time scales in three main regions: the free jet, close to the nozzle exit; the stagnation region, close to the flow impact; the wall jets, released parallel to the wall from impact \citep{Gutmark}.

\begin{figure}
\centering
\vspace{2mm}
\includegraphics[width=6cm]{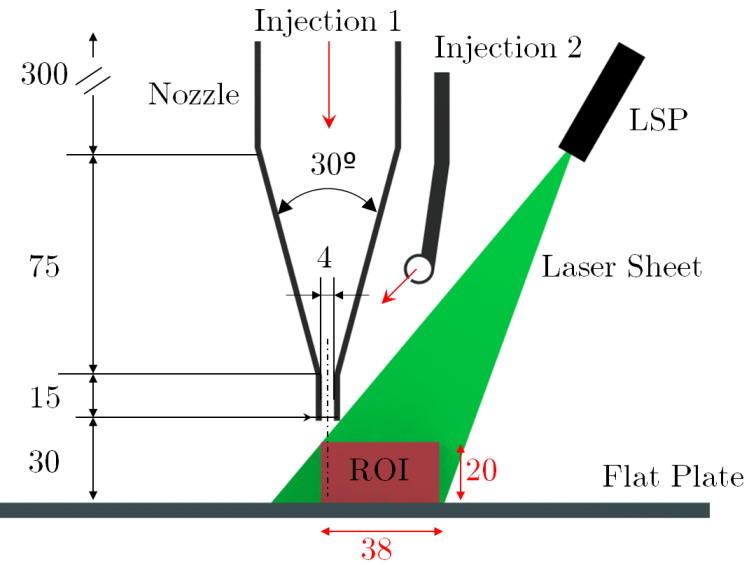}
 \caption{Schematic of the experimental set up for TR-PIV analysis of an impinging plane jet, showing the region of interest (ROI). The dimensions are in mm. }
\label{TR_SETUP}
\end{figure}

The experimental set up is shown in Figure \ref{TR_SETUP}. The rectangular outlet section of the nozzle has an opening of $H=4mm$ and an aspect ratio of $H/W=62.5$. The chamber gauge pressure $P_N$ is monitored by a piezoresistive pressure transducer AMS5812 connected to a pressure tap and the flat surface consist of a polished aluminum plate placed at a distance of $Z=30\pm 1\%mm$. The experiment is carried out with $P_N=30\pm1\%\,Pa$ leading to an average jet flow velocity of  $U_0= 6.5\pm2\% m/s$, that is a Reynolds number of $Re\approx U_0\,H/\nu=1733$ considering a kinematic viscosity of $\nu=15cSt$. 

The seeding tracers for the PIV measurements are microdroplets of mineral oil (Ondina Shell 917), produced by a Laskin nozzle (PIVTEC45-M). These are introduced both in the stagnation chamber and in a second injection on one side of the nozzle to seed the entrainement flow. The laser source is a dual diode-pumped ND:YLF laser (Quantronix Darwin Duo 527), reaching the test section via a light sheet probe (LSP, from ILA). The videos are acquired by a CMOS Photron FASTCAM SA1, positioned at about $30cm$ and equipped with a zoom objective Tamron CZ-735 $70-300mm$ to provide an image scaling factor of $24.1\pm0.1 pixel/mm$.

The investigated Region of Interest (ROI) is a rectangular area of $20mm\times38mm$ at a distance of $10mm$ from the nozzle outlet (cf. Figure \ref{TR_SETUP}). The video sequence is made of $n_t=2728$ images pairs of $1000\times 740$ pixels, acquired in frame-straddling mode at $f_s=2kHz$. The separation time between the frames is set to $\Delta t_p=70\mu s$ to produce a maximum displacement, in the jet flow, of about $8$ pixels. The PIV evaluation is carried out with the Matlab package PIVlab \citep{PIVLAB}, using standard iterative multi-step interrogation (four passes from $128\times128$ to $8\times8$ windows) with spline window deformation, 2D Gaussian sub-pixel interpolation, and vector validation via median test. The resulting vector field consists of $n_s=60\times114$ spatial realizations sampled on a uniform Cartesian grid.

\begin{figure}
\centering
\begin{subfigure}[c]{.52\textwidth}
  \centering
\vspace{1mm}
\includegraphics[height=3.95cm]{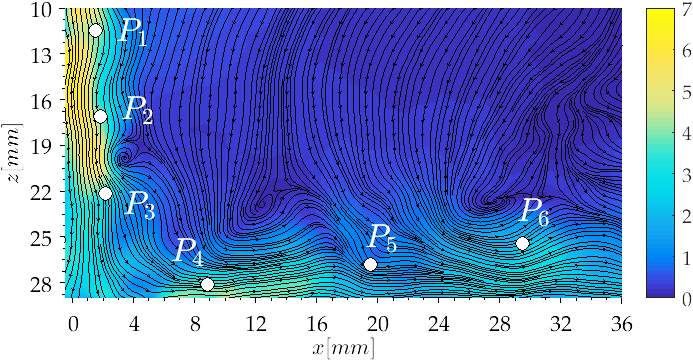}\\a)
 \end{subfigure}
 \hspace{2mm}
 \begin{subfigure}[c]{.45\textwidth}
\centering
\includegraphics[height=4.25cm]{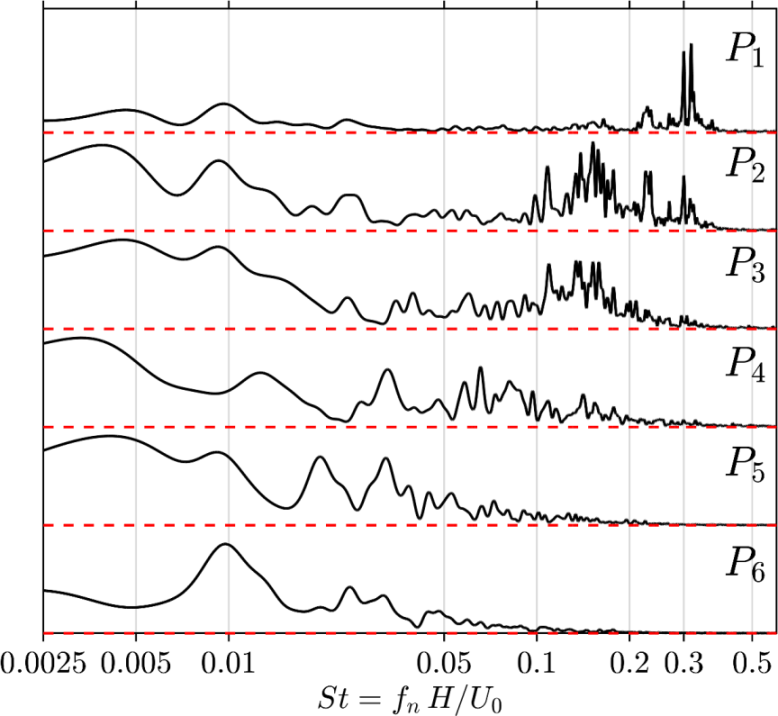}\\\hspace{2mm}b)
 \end{subfigure}
\caption{a) Exemplary snapshot of velocity field from TR-PIV of an impinging jet flow, with contour-plot displaying the magnitude of the velocity in $m/s$, and location of six representative probes. {An animation of this test case is available as supplemental Movie 3.} b) Normalized power spectral densities in the probes in a), versus Strouhal number.}
\label{Sample_Spectra}
\end{figure}

An exemplary instantaneous flow field is shown in Figure \ref{Sample_Spectra}, together with the location of six probes. The normalized power spectral density computed from the velocity magnitude in these locations is shown in Figure \ref{Sample_Spectra}b as a function of the dimensionless frequency $St=f_n\,H/U_0$ and normalized by the corresponding maximum value. These spectra are computed using the Welch's method \citep{Welch} with Hanning windows of $n_t/4$ width and $25\%$ overlapping. The first probe $P_1$ is located in the free-jet flow, where the shear layer instabilities develop unaffected by the presence of the impinged wall. In this location, most of the energy is centered around a peak at $St\approx 0.3$, in agreement with other studies on the free plane jet in transitional regime \citep{Thomas_Ch5,Suresh}. This corresponds to the formation of large vortices (primary vortices) which originates from a Kelvin-Helmholtz (K-H) instability and evolve into roller-like structures \citep{Mumford}.

As the flow approaches the stagnation region (probes $P_2$, $P_3$), the frequency spectra broaden and moves towards lower frequencies due to the vortex merging and the velocity decay, as also reported by \cite{Pieris} for impinging jet flows at higher Reynolds numbers. Far from the stagnation region, the wall jet flow ($P_4$, $P_5$, $P_6$) is governed by a much lower frequency and the unsteady separation due to the intermittent passage of the primary vortices \citep{Didden,Hanj}.

\subsection{Modal Analysis}

\begin{figure}
\centering
\includegraphics[width=7.5cm]{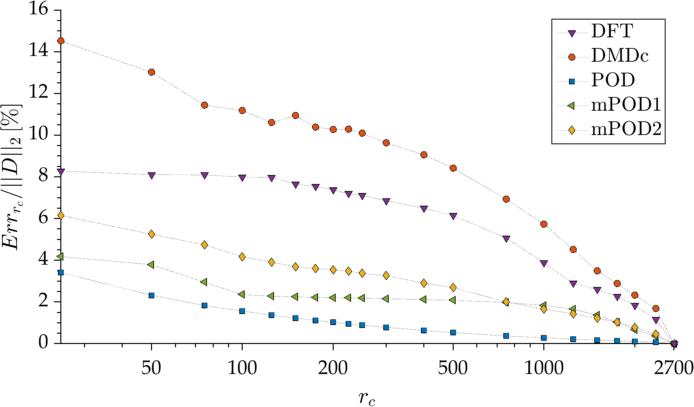}
\caption{$L^2$ convergence (\ref{L2}) of the POD, cDMD, DFT and two mPOD with different frequency splitting vectors for the TR-PIV field of the impinging jet flow.}
\label{Conv_3}
\end{figure}

Figure \ref{Conv_3} shows the $L^2$ convergence in (\ref{L2}) for POD, cDMD, DFT and two mPOD with frequency splitting vectors. The first, in terms of Strouhal numbers, consists of $M=4$ with $F_{V1}(m)=[0.1, 0.2, 0.5]$; the second consist of $M=50$ with $F_{V2}(m)=0.01+m \Delta f$ with equal bandwidth $\Delta f=0.02$. As discussed in \S\ref{SUB2}, the finer the frequency splitting, the more the mPOD approaches a DFT.

\begin{figure}
\centering
\begin{subfigure}[c]{.45\textwidth}
  \centering
\includegraphics[width=5.2cm]{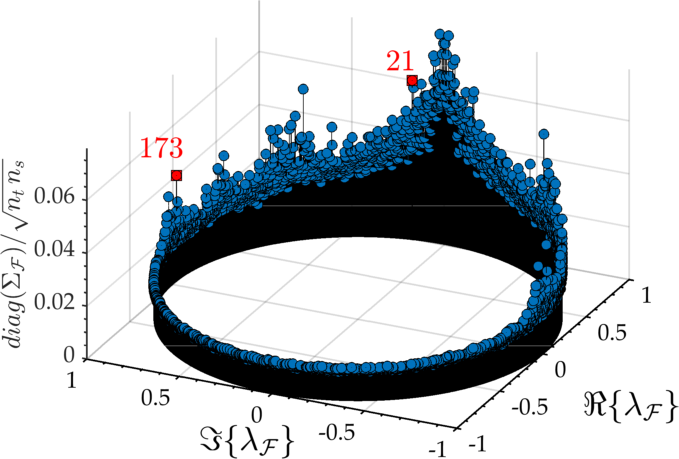}
 \end{subfigure}
 \begin{subfigure}[c]{.45\textwidth}
\includegraphics[width=5.2cm]{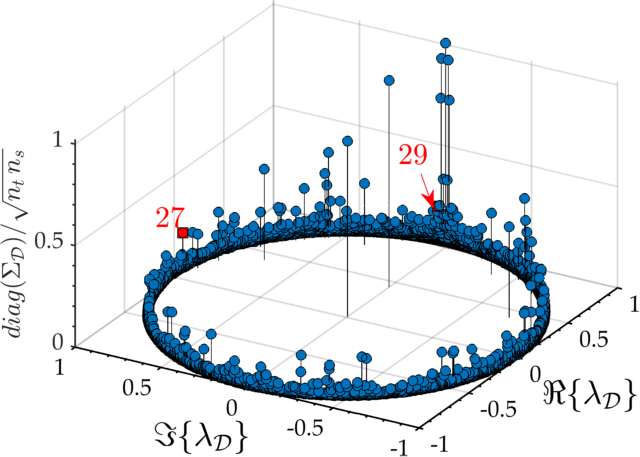}
 \end{subfigure}
 \caption{Same as Figures \ref{COM_1} and \ref{Vort_AMP}, but for the third test case presented in this section.}
\label{EXP_AMP}
\end{figure}

The frequency spectra for DFT and cDMD are mapped in Figure \ref{EXP_AMP}. The investigated jet flow is in statistically stationary conditions and the mean flow accounts for a significant portion of the dataset. For plotting purposes, therefore, the entry corresponding to the zero frequency is removed. The spectra of the DFT shows that the investigated test case is rather broadband, with few frequencies reasonably standing over the others. Despite the significant random noise and turbulence contribution, the presence of a statistically converged time average results in a relative error $||d_{n_t}-D_1\,\mathbf{c}||_2/||d_{n_t}||_2\approx 6\%$ in the minimization defining the Companion matrix for the cDMD. The resulting DMD dominant modes identify the relevant frequencies more sharply than the DFT. Nevertheless, several eigenvalues lie inside the unit circle and the convergence of the cDMD, although not dramatically poor as in the previous test case, is below the DFT.

\begin{figure*}
\begin{subfigure}{.49\textwidth}
  \centering
  \includegraphics[width=5.5cm]{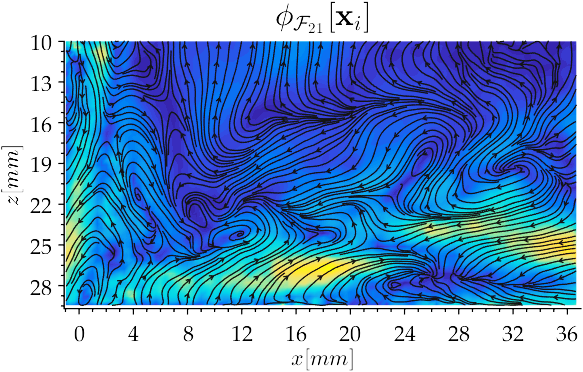}
\end{subfigure}
\begin{subfigure}{.49\textwidth}
  \centering
  \includegraphics[width=5.5cm]{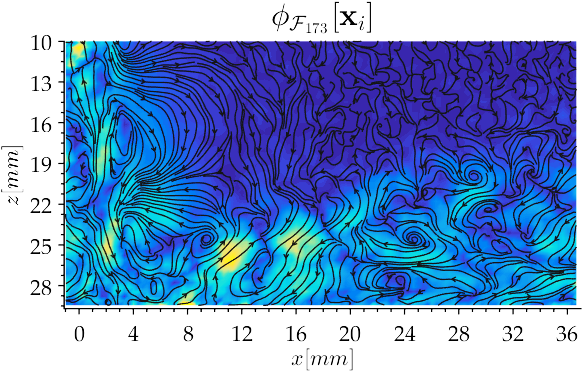}\\
\end{subfigure}
\vspace{2mm}\\
\begin{subfigure}{.49\textwidth}
  \centering
  \includegraphics[width=5.5cm]{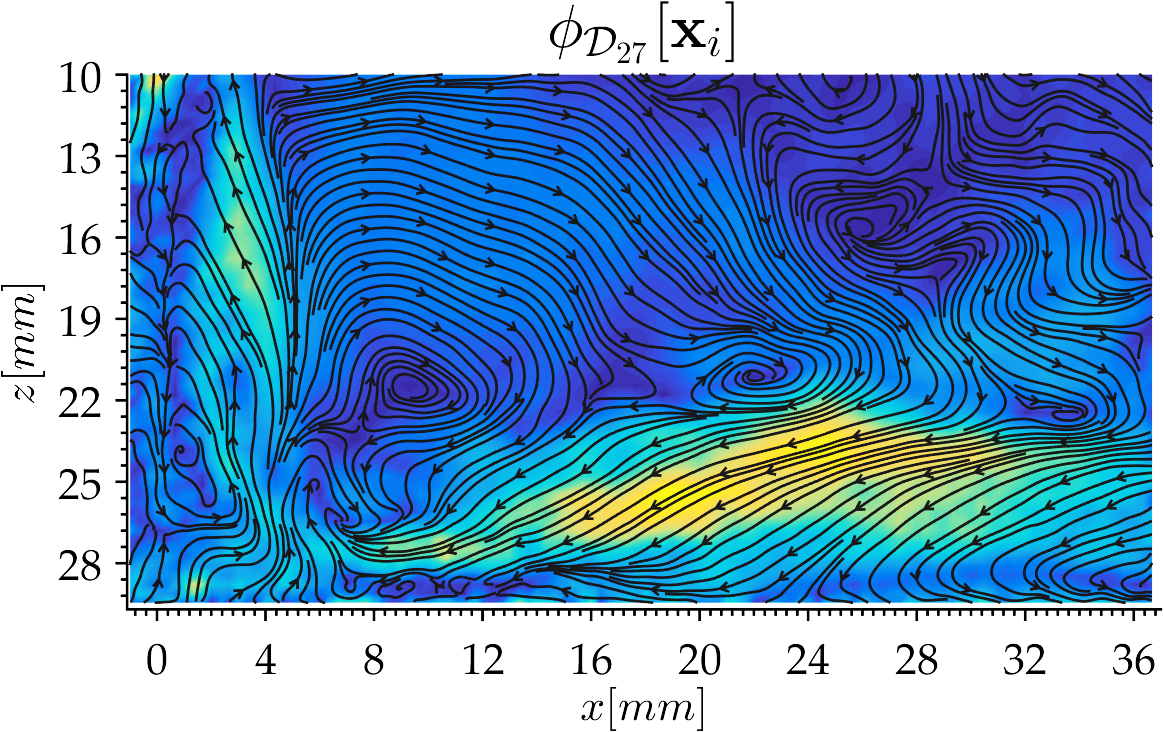}
\end{subfigure}
\begin{subfigure}{.49\textwidth}
  \centering
  \includegraphics[width=5.5cm]{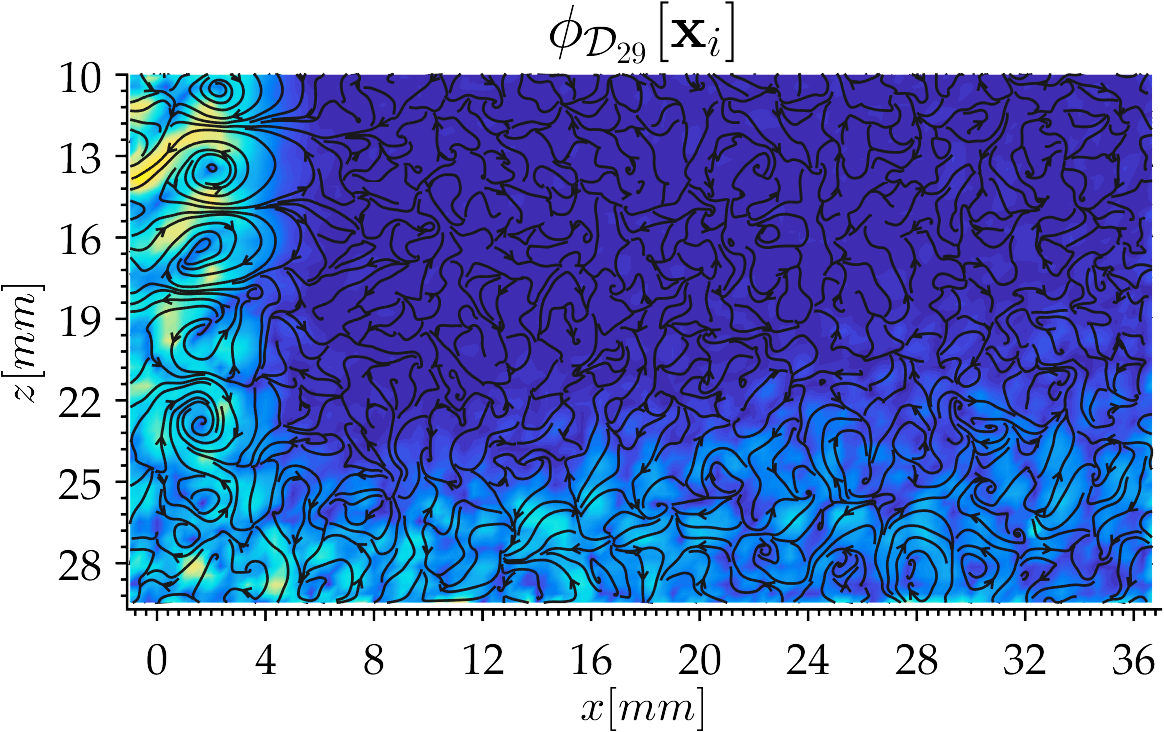}\\
\end{subfigure}
\caption{Spatial structures of two representative DFT (top row) and DMD (bottom row) modes, with amplitudes shown in Figure \ref{EXP_AMP} with a red marker. The DFT modes are associated to dimensionless frequencies of $St=[0.033,0.299]$; for the DMD these are $St=[0.009,0.293]$.}
\label{DFT_DMD_SPATIAL_EXP}
\end{figure*}

The spatial structures of two representative DFT and DMD modes are shown in Figure \ref{DFT_DMD_SPATIAL_EXP}. For the DFT, these structures are associated to dimensionless frequencies of $St=[0.033,0.299]$ and are thus expected to be linked to the large-scale oscillations of the wall jet and K-H structures produced in the free jet region respectively (see Figure \ref{Sample_Spectra}). The constraint of single harmonic is nevertheless too restrictive to isolate these mechanisms, and the corresponding spatial modes are not capable of fully identifying the related coherent patterns. The DMD spatial structures, pulsing at dimensionless frequencies of $St=[0.009,0.293]$, appear less noisy than the DFT ones, thanks to the better frequency selection. The mode $\Phi_{\mathcal{D}_{27}}$ describes the large-scale lifting of the wall jet, due to the interaction with the roller structures that in this region lose most of their momentum and mix with the entrainment flow. As expected, the mode $\Phi_{\mathcal{D}_{29}}$ captures the K-H structures, vortical structures of about $2mm$ in diameter that arise from the primary shear layer instability of the free jet flow.

\begin{figure*}
 \centering
  \includegraphics[height=3.8cm]{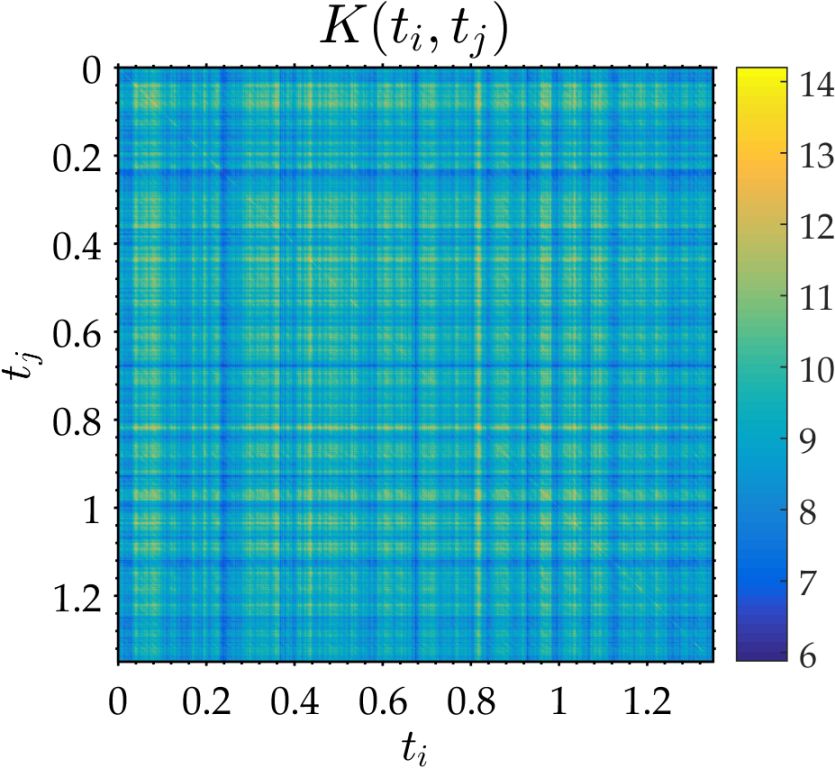}
\hspace{2mm}
  \includegraphics[height=3.8cm]{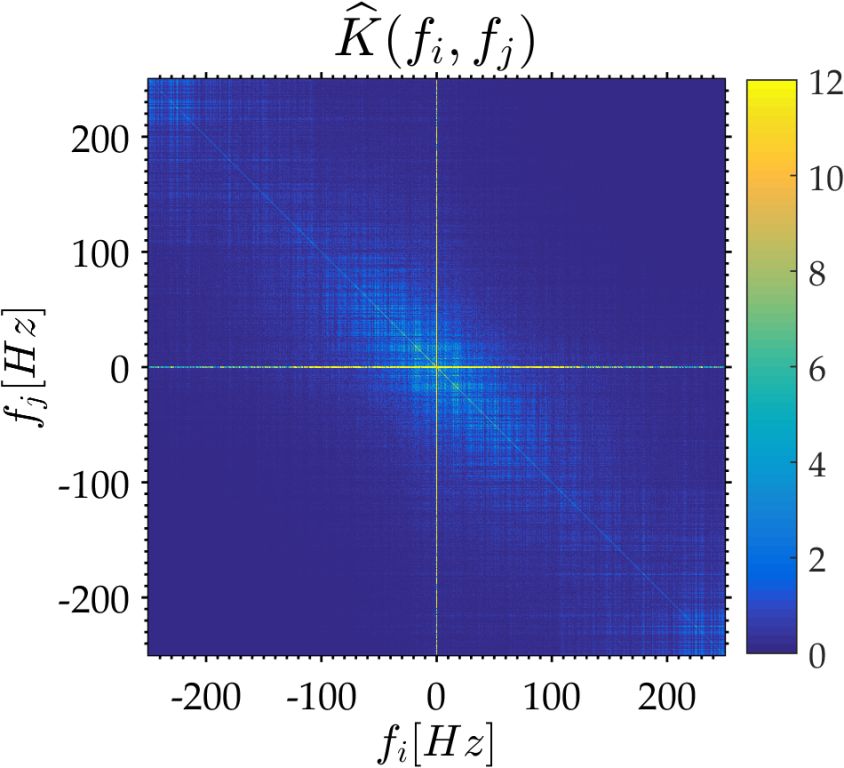}\\
\vspace{1mm}
\begin{subfigure}{.49\textwidth}
  \centering
  \includegraphics[width=5.5cm]{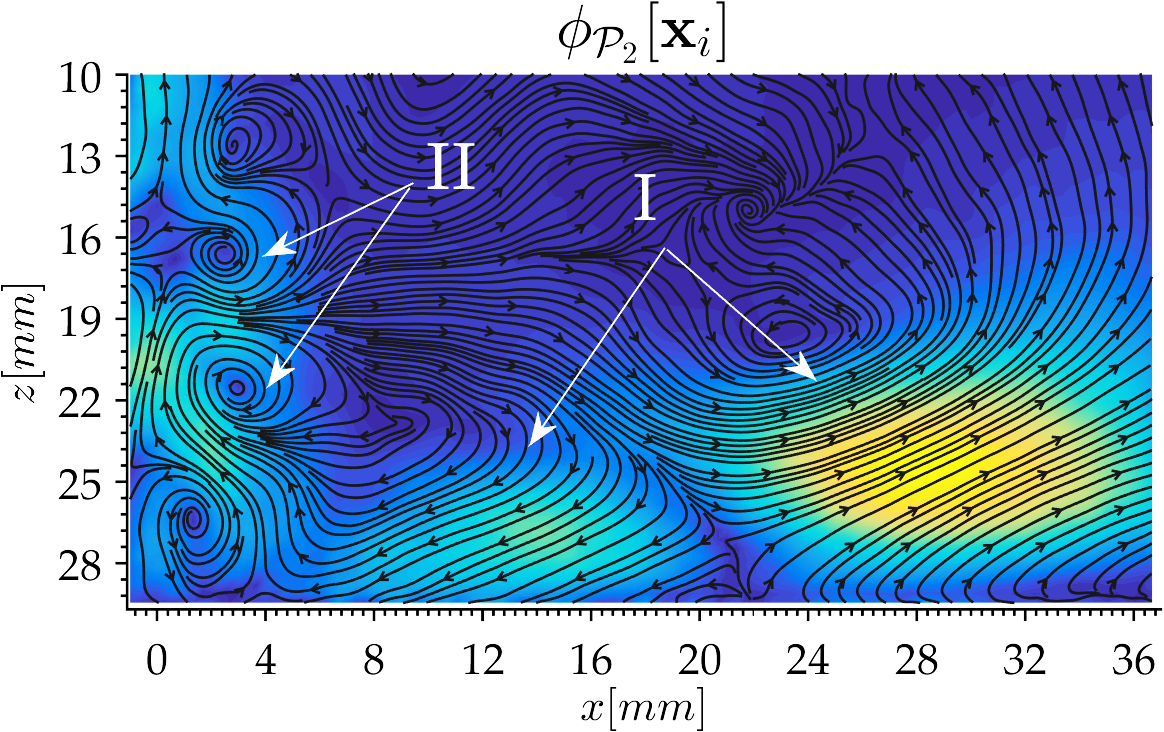}\\
\vspace{1mm}
\includegraphics[width=4.7cm]{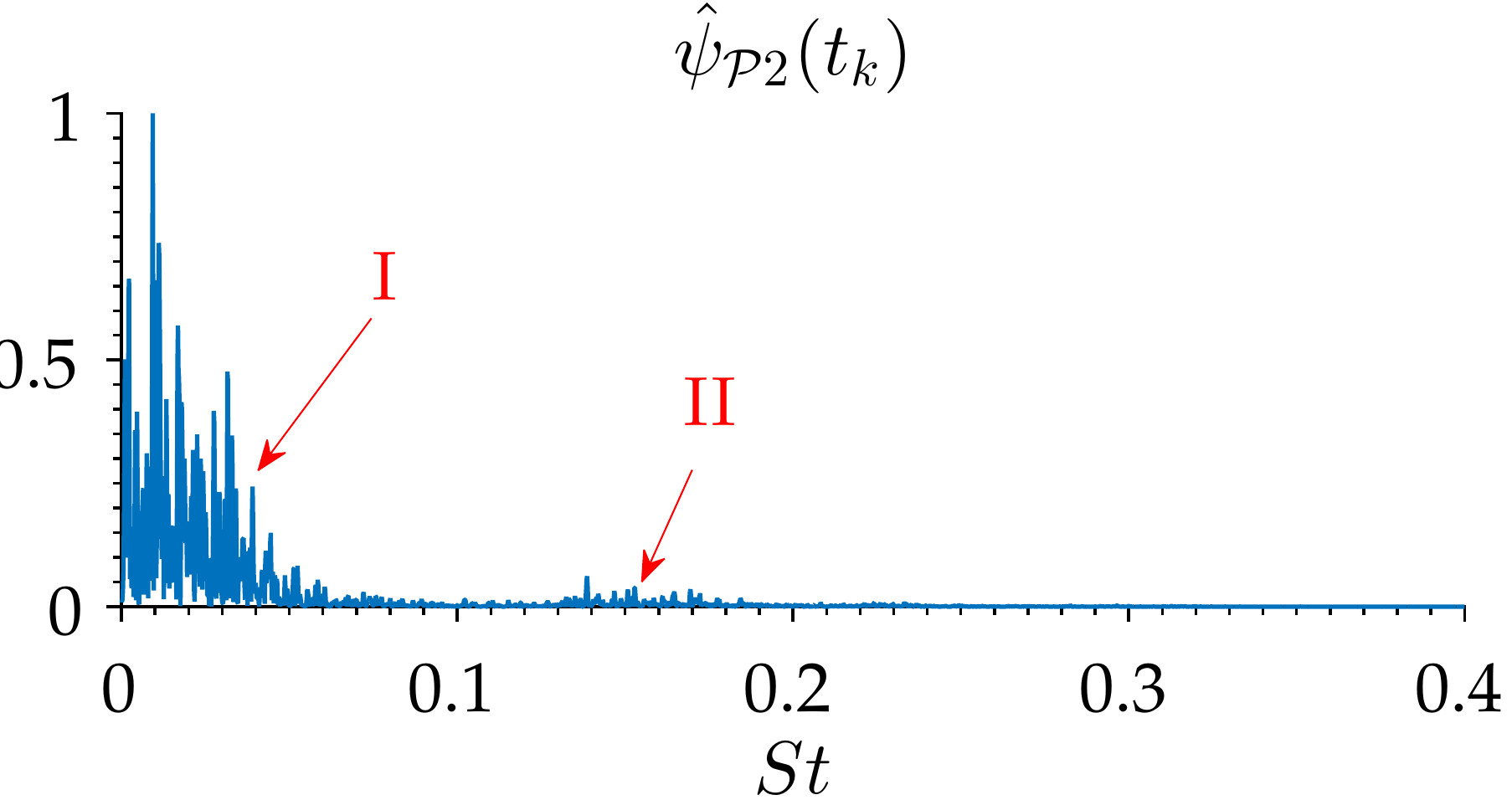}
\end{subfigure}
\begin{subfigure}{.49\textwidth}
  \centering
  \includegraphics[width=5.5cm]{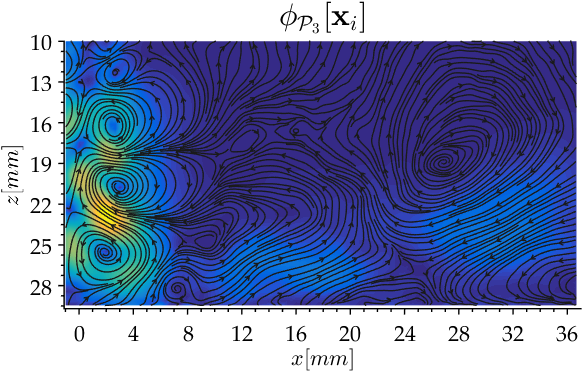}\\
\vspace{1mm}
\includegraphics[width=4.7cm]{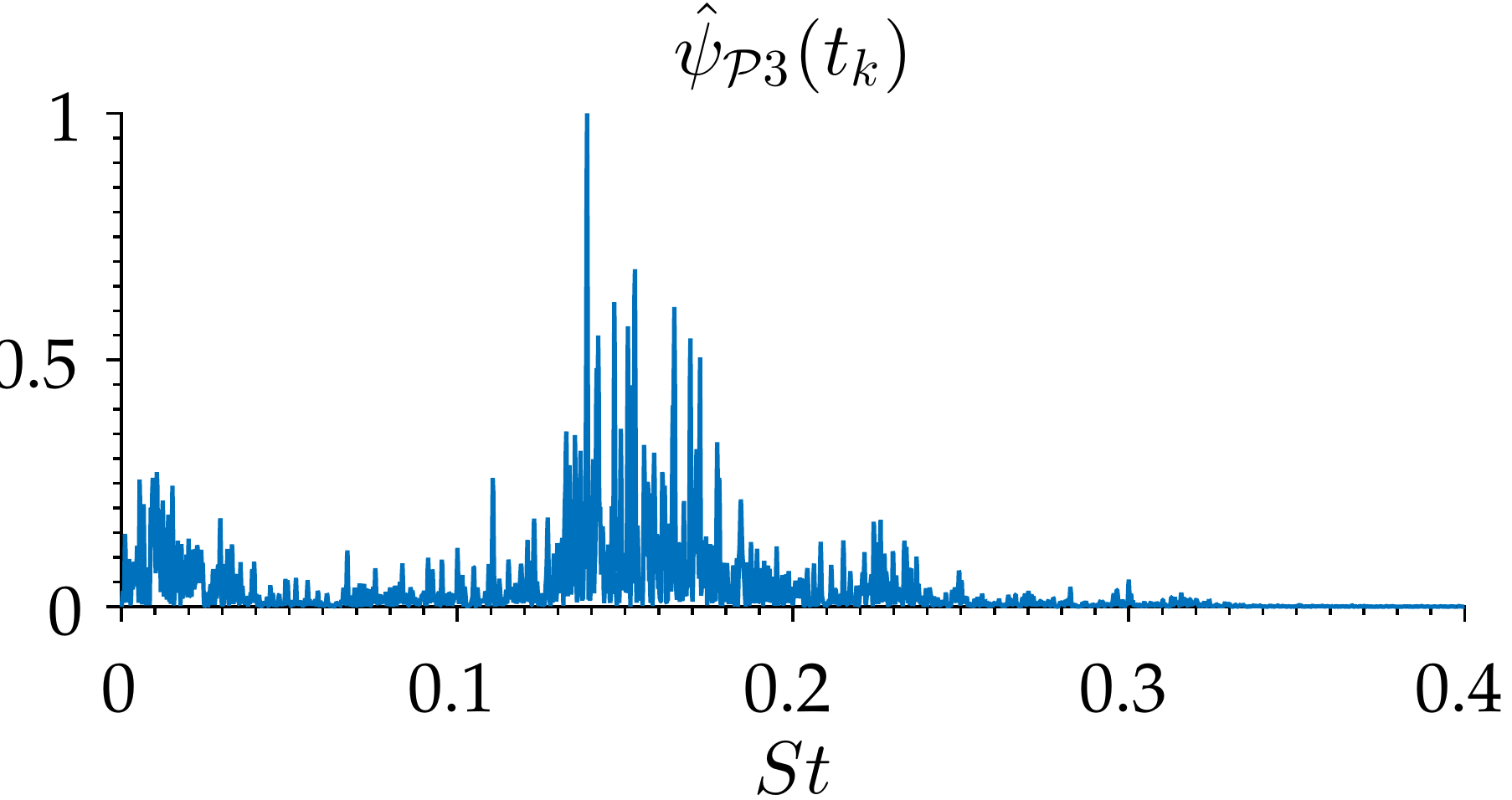}
\end{subfigure}
\vspace{1mm}\\
\begin{subfigure}{.49\textwidth}
  \centering
  \includegraphics[width=5.5cm]{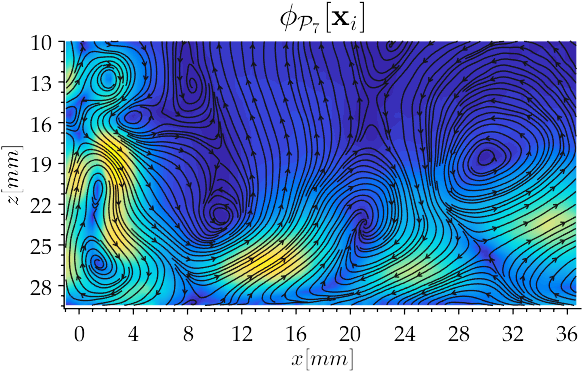}\\
\vspace{1mm}
\includegraphics[width=4.7cm]{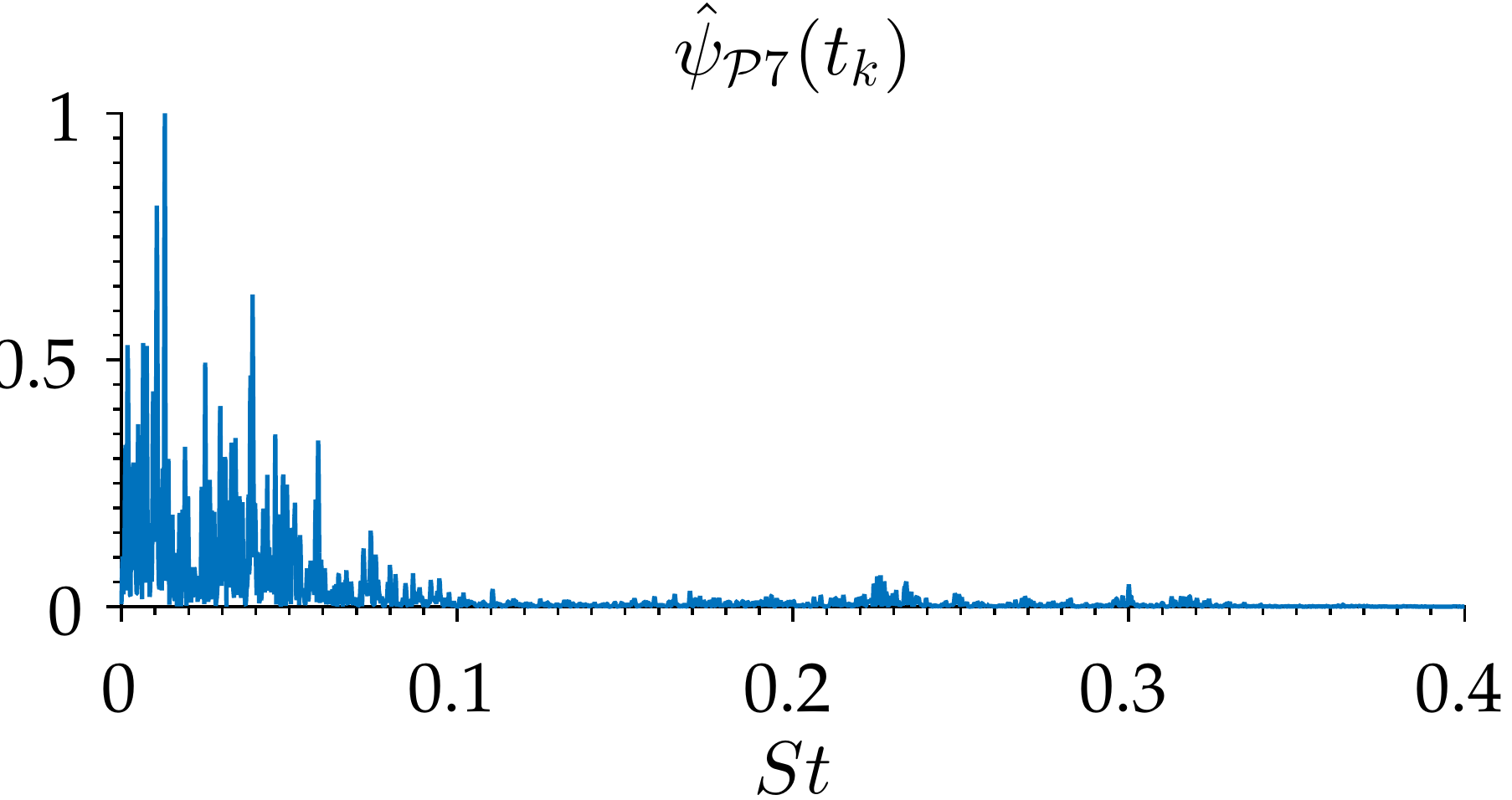}
\end{subfigure}
\begin{subfigure}{.49\textwidth}
  \centering
  \includegraphics[width=5.5cm]{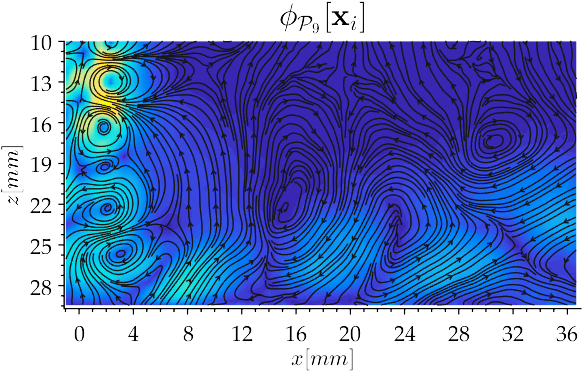}\\
\vspace{1mm}
\includegraphics[width=4.7cm]{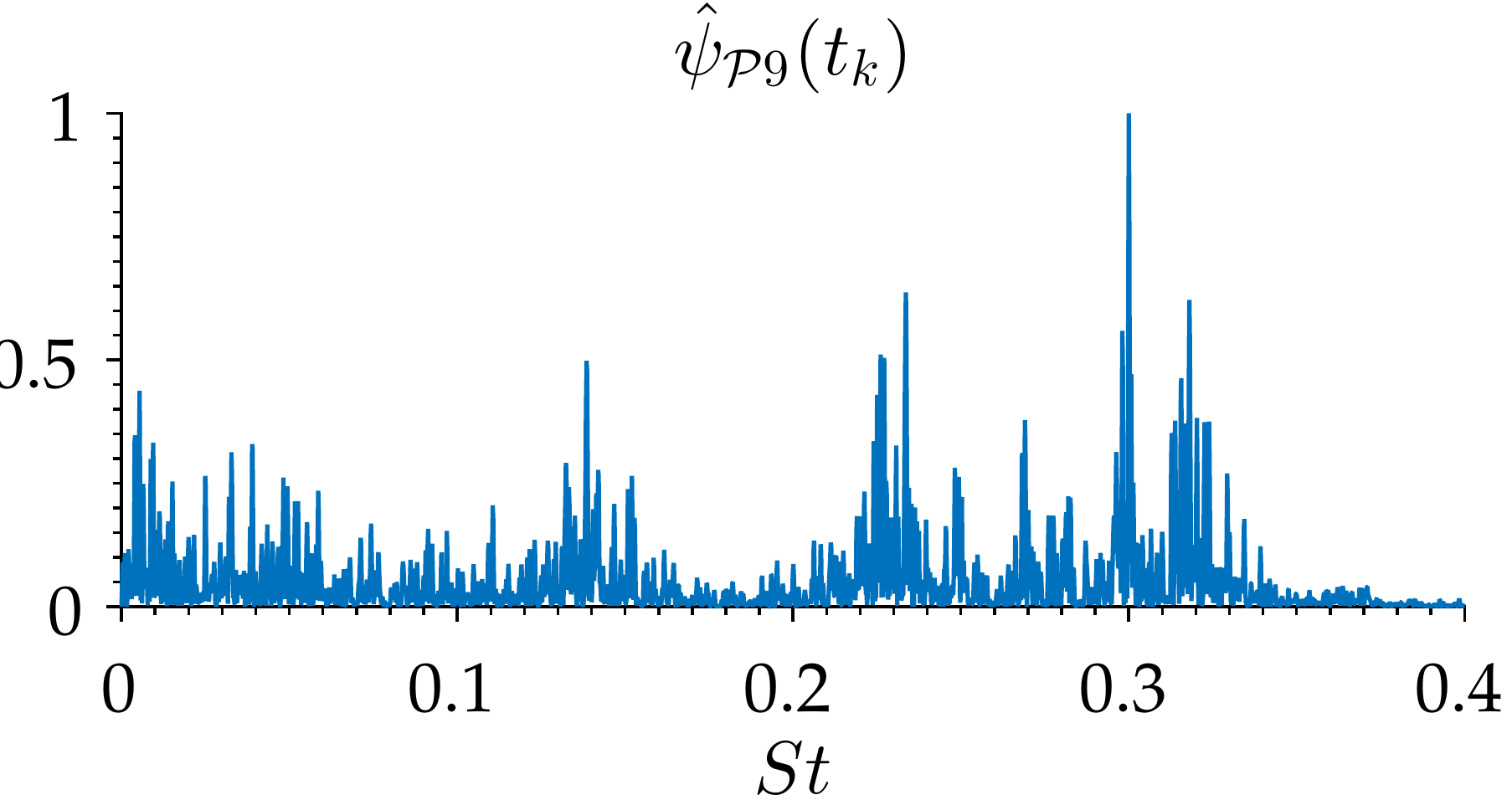}
\end{subfigure}

\caption{Same as \ref{POD_RES} and \ref{POD_RES_VOR}, but for the third test case discussed in this section and replacing the temporal structures ${\psi}_{\mathcal{P}}$ with their normalized frequency content $|\widehat{\psi}_{\mathcal{P}}|/max\{\widehat{\psi}_{\mathcal{P}}\}$.}
\label{POD_RES_EXP}
\end{figure*}

The results of a standard POD analysis are collected in Figure \ref{POD_RES_EXP}. Since the POD modes are stationary, it is interesting to show the frequency content $\widehat{\psi}_{\mathcal{P}r}$ associated to each spatial structure, rather than their temporal evolution ${\psi}_{\mathcal{P}r}$. All the modes in the range $r=4$ to $r=11$ are related to traveling structures and arise therefore in pairs that have $\pi/2$ phase delay in space and time. Only the four representative modes are shown.

The temporal correlation matrix $K$ displays a regular pattern composed of a broad range of frequencies and $\widehat{K}$ displays a band-like structure. Yet, a dominant frequency around $12 Hz$ ($St=0.0074$) is visible. This corresponds to the large-scale flow motion in the wall jet flow, which covers a larger portion of the domain and, therefore, gives a more significant contribution to the overall correlation level. Being the data statistically stationary, the constant vector $\psi_{\mu}=\underline{1}/\sqrt{n_t}$ is almost an eigenvector of the correlation matrix, and thus the first POD mode reproduces closely the mean flow (not shown). For plotting purposes, the mean correlation is removed from $\widehat{K}$.

The spatial structures identified by the POD are considerably less influenced by the uncorrelated portion of the dataset and highlight coherent features in the flow. Besides the mean flow, the dominant contribution $\phi_{\mathcal{P}_2}$ in Figure \ref{POD_RES_EXP} is mostly related to the large-scale pulsation in the wall jet (labeled with $I$) and its periodic separation, as expected from the spectra of the correlation matrix $\widehat{K}$. However, this spatial structure also captures, to a minor extent, coherent structures in the free jet region (labeled with $II$). These are expected to occur at much higher frequency (in the range $St=0.1-0.2$) as shown by the probes $P_2$-$P_3$ in Figure \ref{Sample_Spectra}.

The frequency spectra $\widehat{\psi}_{\mathcal{P}_2}$ gives the final proof of spectral mixing for this POD mode: a minor energetic content is also present in the frequency range belonging to the flow structures of the free jet flow, and the projection of the data along this temporal evolution propagates this mixing to the spatial domain. The same occurs in the mode $\phi_{\mathcal{P}_3}$, which is mostly linked to the traveling vortices in the jet, but -- to a minor extent-- also to the wall jet separation. All the other modes suffer from the same problem, which reaches its worse case in the mode $\phi_{\mathcal{P}_9}$. In addition to the two mechanisms previously described, this mode also captures the K-H structures produced further upstream, where they originate at $St\approx 0.3$ as clearly shown by the probe $P_1$ in Figure \ref{Sample_Spectra}. This test case shows that the spectral mixing problems of the POD, highlighted by the two test cases of \S\ref{VI} and \S\ref{VII}, can easily occur also in statistically stationary flows if these are governed by largely different scales with similar energy content.

\begin{figure*}
\vspace{1mm}
\begin{subfigure}{.49\textwidth}
  \centering
  \includegraphics[width=5.5cm]{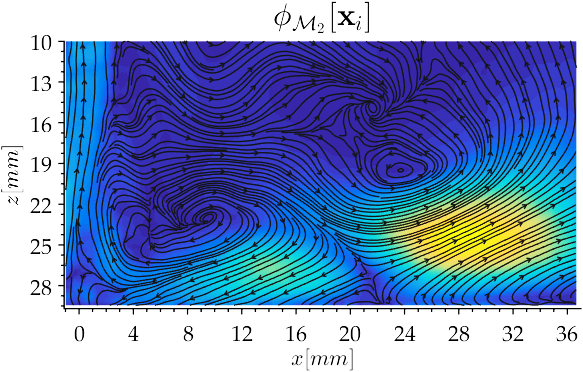}\\
\vspace{1mm}
\includegraphics[width=4.7cm]{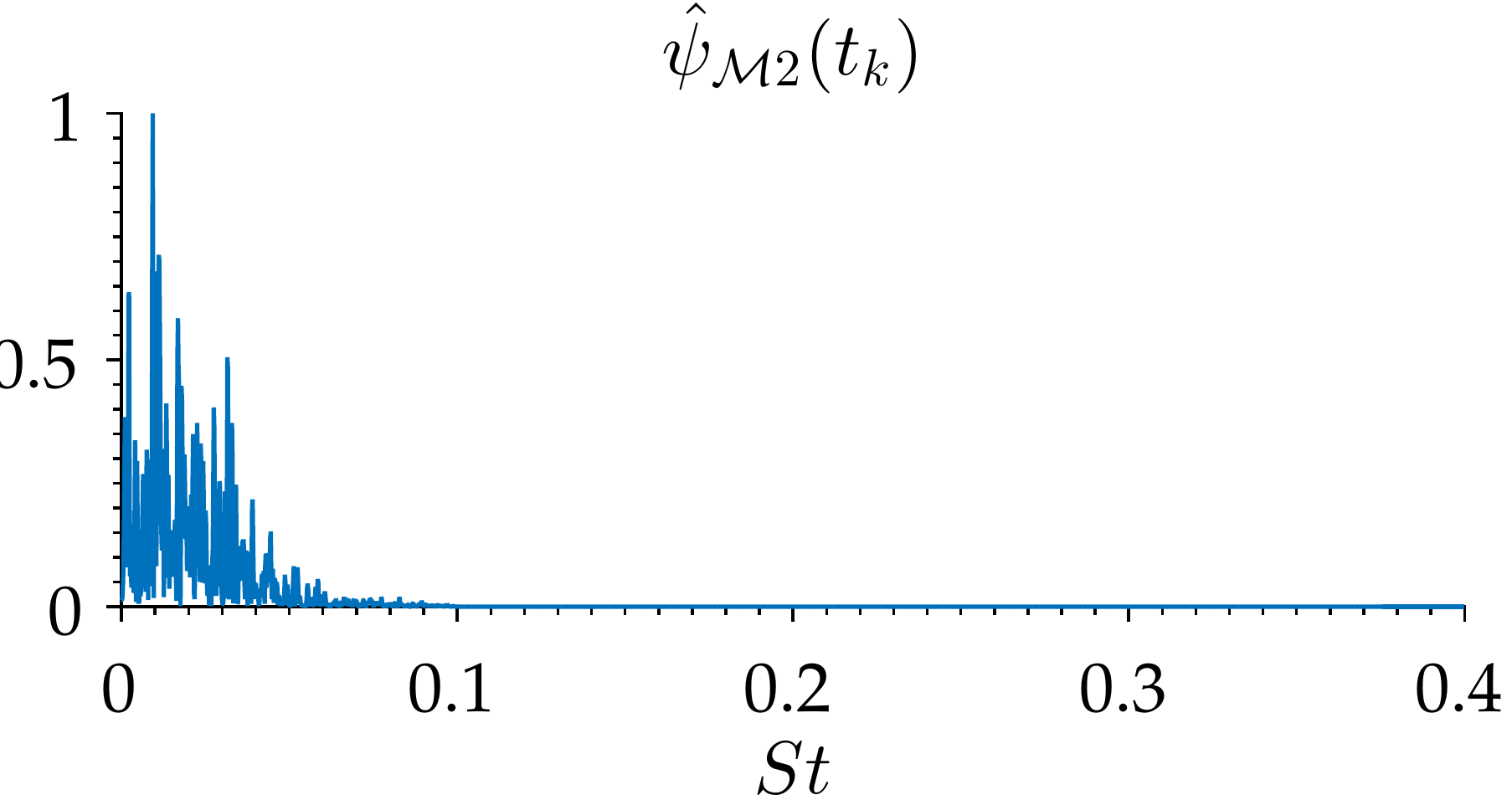}
\end{subfigure}
\begin{subfigure}{.49\textwidth}
  \centering
  \includegraphics[width=5.5cm]{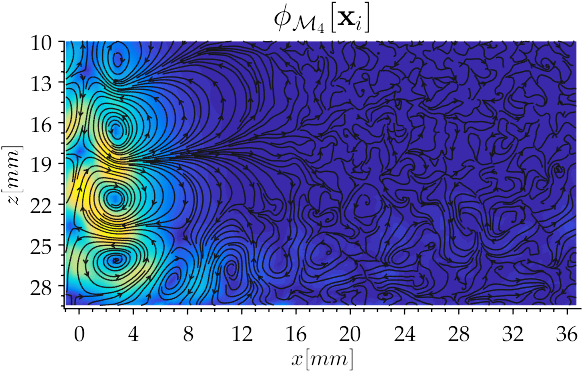}\\
\vspace{1mm}
\includegraphics[width=4.7cm]{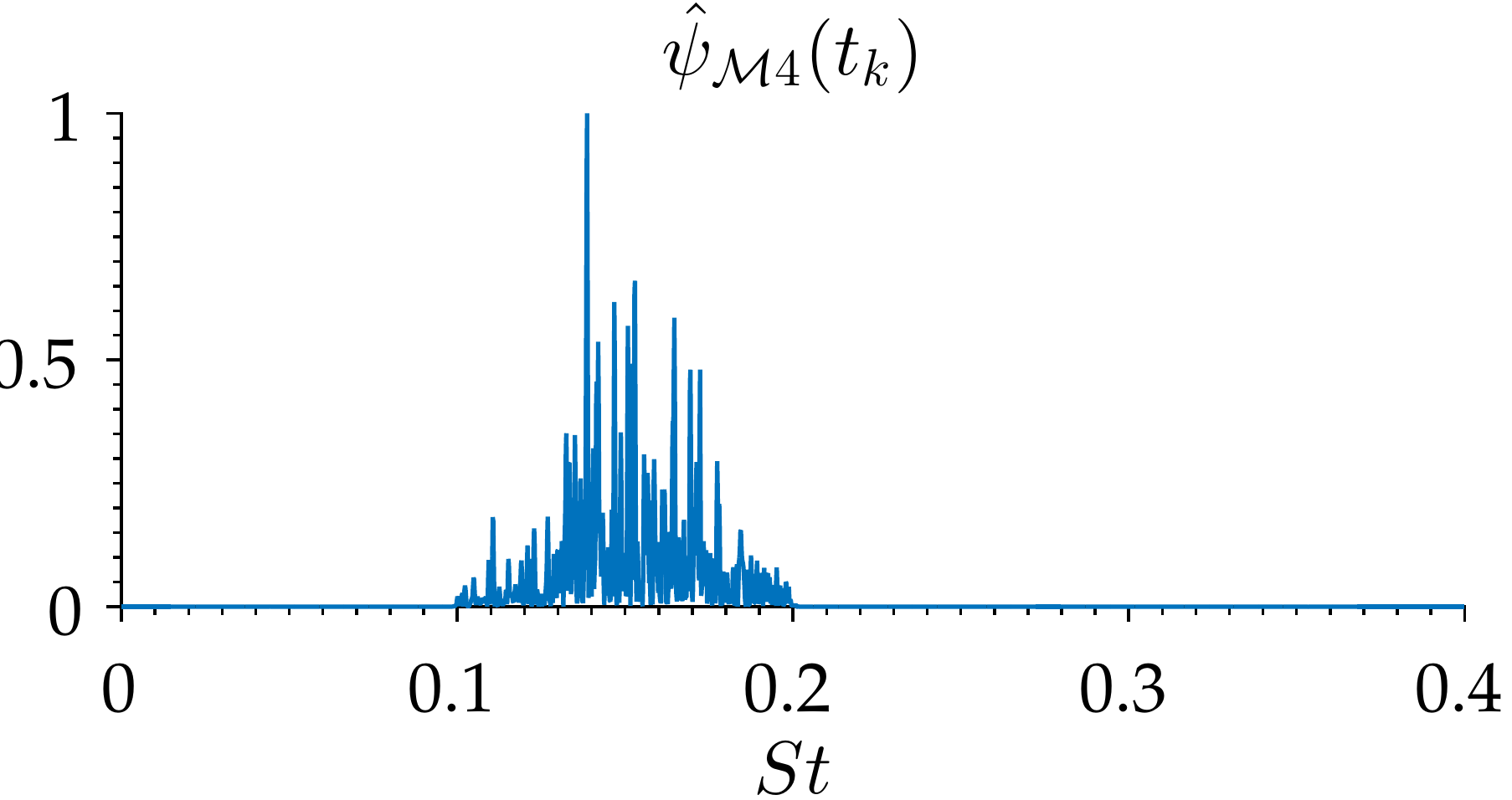}
\end{subfigure}
\vspace{1mm}\\
\begin{subfigure}{.49\textwidth}
  \centering
  \includegraphics[width=5.5cm]{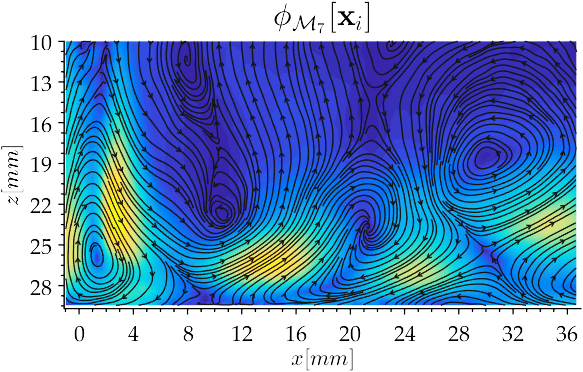}\\
\vspace{1mm}
\includegraphics[width=4.7cm]{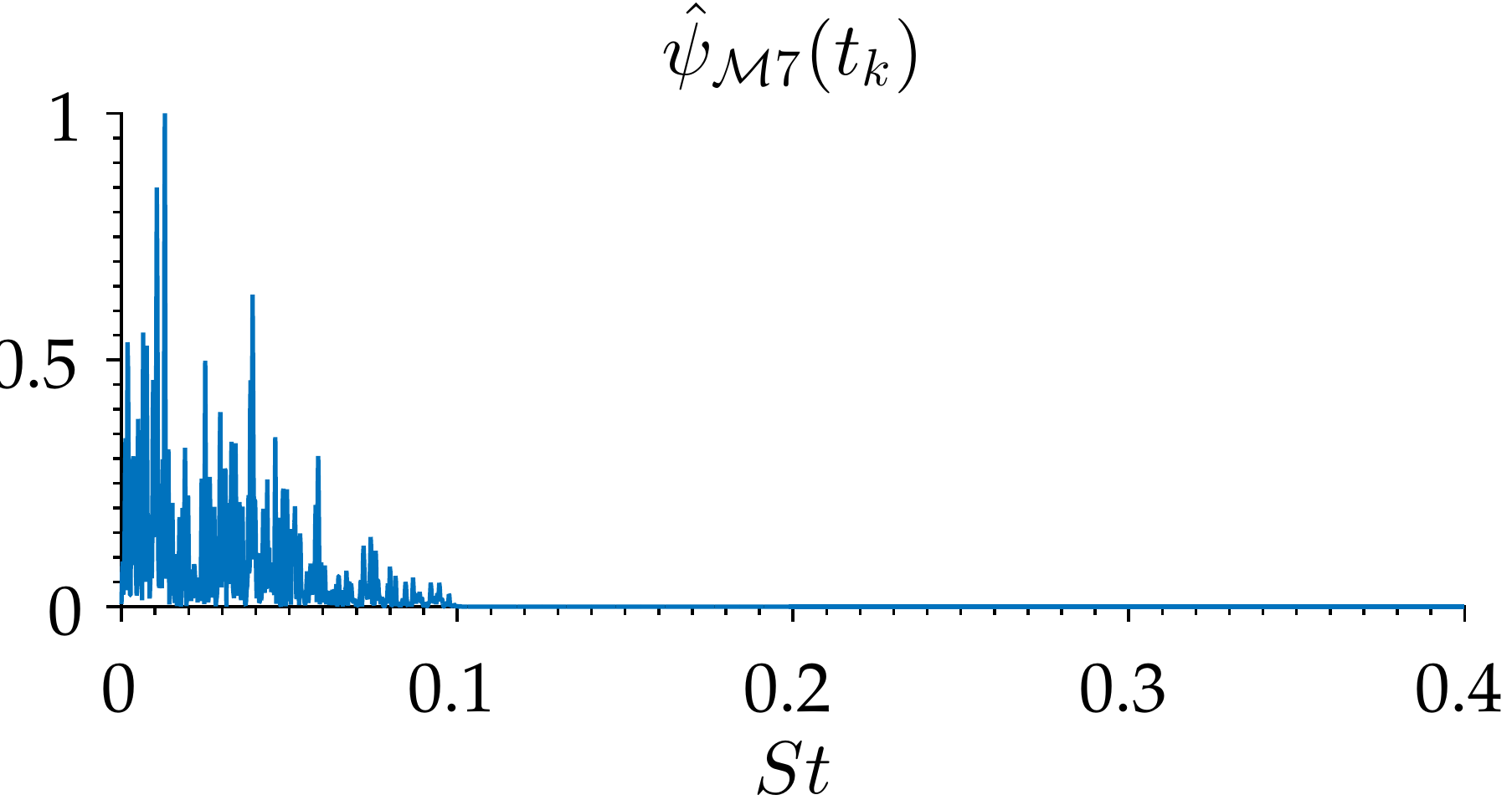}
\end{subfigure}
\begin{subfigure}{.49\textwidth}
  \centering
  \includegraphics[width=5.5cm]{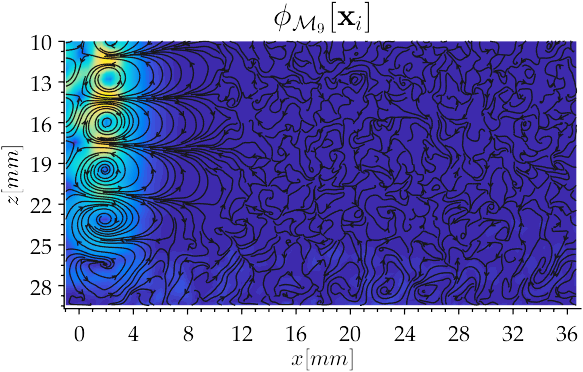}\\
\vspace{1mm}
\includegraphics[width=4.7cm]{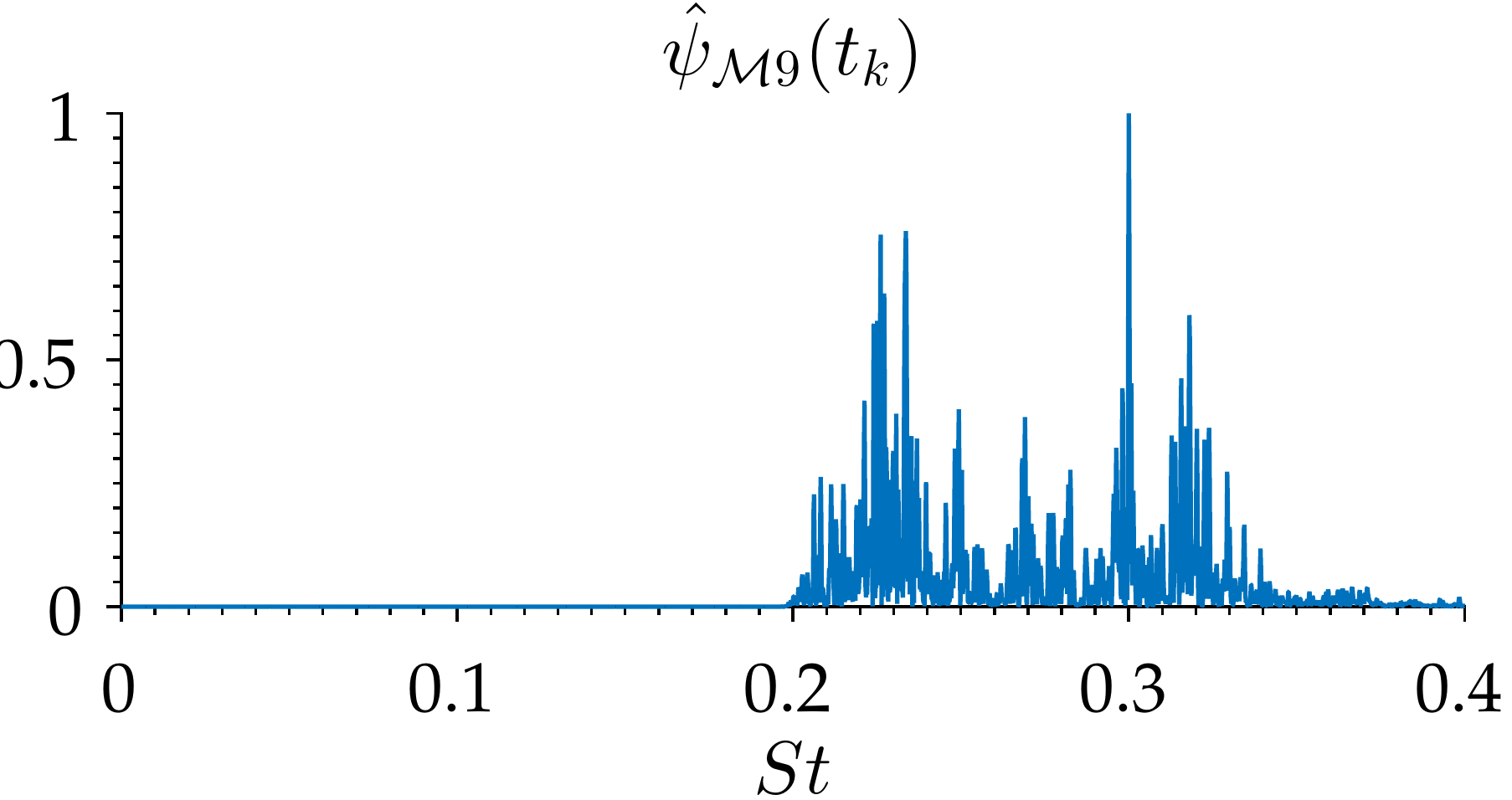}
\end{subfigure}
\caption{Selected mPOD modes corresponding to the POD modes in Figure \ref{POD_RES_EXP}.}
\label{mPOD_RES_EXP}
\end{figure*}

As for the previous two test cases, the spectral separation produced by the MRA is sufficient to isolate the contribution of different scales at the cost of a minor reduction in the decomposition convergence (cf. figure \ref{Conv_3}). Four representative mPOD modes, including spatial structures and related frequency content, are shown in Figure \ref{mPOD_RES_EXP}. These are constructed by setting $F_{V1}(m)=[0.1, 0.2, 0.5]$, producing the $L^2$ convergence labeled as mPOD1 in Figure \ref{Conv_3}. The resulting mPOD modes are very similar to the POD counterparts, showing that these spectral constraints have a minor impact on the decomposition. Yet, no spectral mixing is produced. The second mode captures \emph{only} the dynamics of the separating wall jet, the fourth is related \emph{only} to the evolution of the primary vortex advected towards the wall, while the ninth captures \emph{only} the formation of the primary vortices in the free jet region. The seventh mode describes the large scale oscillation of the jet which is associated with the stagnation region and the initial stage of the wall jet development. 

By increasing the number of frequency bandwidths, one could further narrow the frequency content of each mode, improving the frequency resolution but decreasing the $L^2$ convergence of the method as shown in Figure \ref{Conv_3} by the second mPOD with finer frequency splitting vector. This can alter the ranking of different modes, giving priority to mechanisms that have narrower frequency content, and eventually move the mPOD towards the DFT for unreasonably fine frequency splitting vectors.

\section{Conclusions}\label{X}

The state of the art on data-driven decomposition has been extensively discussed, with particular emphasis on the need for bridging the gap between the energy optimality of the Proper Orthogonal Decomposition (POD) and the spectral purity of the Dynamic Mode Decomposition (DMD). A generalized algebraic framework is introduced, and all the investigated decompositions are presented as matrix factorizations. This framework is then used to derive a novel decomposition, referred to as Multiscale Proper Orthogonal Decomposition (mPOD). 

The derivation of the mPOD starts with an investigation on the link between the frequency content (the DFT) of the correlation matrix and that of its eigenvectors (the POD modes). This link is used to study the impact of a time filter in the dataset on the frequency of the correlation matrix and thus the related POD modes. Specifically, it is shown how a filter bank can be used to perform multi-resolution analysis (MRA) of the correlation matrix, splitting it into the contribution of different scales retaining non-overlapping portions of the frequency spectra of the data. These scales are equipped with their optimal eigenbasis which can be kept mutually orthogonal and finally assembled into a single mPOD basis.
 
The novel decomposition is tested on three selected test cases and its performances compared to those of a Discrete Fourier Transform (DFT), a standard POD and the DMD in its two standard formulations: the Arnoldi-based approach (cDMD) and the SVD-based approach (sDMD).

The first of the proposed test case is a synthetic dataset with three predefined modes. These have equal energy, identical but shifted spatial structure, and temporal evolution with different frequency and time localization. None of the popular decompositions presented proved capable of distinguishing the introduced modes. Harmonic decompositions such as DFT and DMD allows for proper frequency identification, but yields poor convergence, high redundancy in the spatial structures, and a complete lack of temporal localization due to the infinite duration of the time basis. On the contrary, the POD allows for the best convergence, with only three modes required to describe correctly (i.e., with zero $L^2$ errors) the dataset. However, it produces a `spectral mixing' between different modes. This mixing results in modes that are linear combinations of the expected ones since their similar energy content results in almost repeated eigenvalues of the correlation matrix and thus almost undefined (non-unique) POD. 

The proposed mPOD proved capable of distinguishing the three introduced modes, with no spectral mixing, excellent time localization, and almost optimal convergence.

The second investigated test case is a 2D numerical simulation of the Navier-Stokes equation, with prescribed coherent, random and pseudo-random sources. The complexity of the data, further polluted with white noise, strongly accentuate all the limits highlighted by the previous test case, with the DMD $L^2$ relative error of the order of $200\%$ within the first modes and up to three different mechanisms sharing the same POD mode. Neither the DFT/DMD nor the POD are capable of identifying and localizing in space and time the different coherent source terms, while the mPOD succeeds at the scope with a negligible loss of convergence over the POD.

Finally, the third case is the time-resolved flow field of an impinging jet experimentally investigated via TR-PIV. Although the data is stationary and statistically well converged, the spectral mixing of the POD is observed while the harmonic constraints of the DFT and DMD yield much poorer convergence and noisy spatial structures. The mPOD naturally handles both problems.

 To conclude, the discussed test cases prove that a minor spectral constraint in the diagonalization of the correlation matrix allows overcoming the frequency limits of the POD, and show that the proposed mPOD provides an excellent compromise between energy optimality and spectral purity. While this work has focused on data-driven decompositions for fluid flows, the possibility of constraining the POD optimality within specific scales of interest gives to the mPOD more flexibility than POD or DMD/DFT for a large variety of applications. In data filtering/compression, for example, the mPOD is more robust against random noise (equally spread over all the POD/DFT spectra) and allows to better focus on particular features (e.g. allowing for different levels of data compression/filtering for different scales). This could find useful applications, for example, in the background/foreground separation problem in video analysis, where the slow motion of large objects is modeled by few dominant POD modes (or slow DMD ones) while the dynamics of smaller objects is captured by the remaining modes (see \citealt{Oliver2000,Sobral2015,app2,Mendez_J_1}). The mPOD provides an additional degree of freedom in this separation by also including considerations on the frequencies involved. In the data-driven identification of coherent patterns, the mPOD allows overcoming the POD non-uniqueness problem in case of phenomena having similar energy content (e.g., \citealt{Mendez_Journal_2}).

More generally, by providing bases that are spectrally cleaner than POD bases but energetically more relevant than DFT/DMD ones, the mPOD can be useful in any `projection-based' application (from pattern recognition to low order modeling of dynamical systems) in which the constraints of energy optimality or spectral purity are both unnecessarily extreme.

\section*{Acknowledements}

This research project has been partially funded by Arcelor-Mittal. The authors gratefully acknowledge the collaboration of the VUB-MECH laboratories at the Vrije Universiteit Brussel (VUB), for providing the time-resolved PIV system, and the contribution of Y. Aksoy, who participated in the TR-PIV experiments in the framework of his Research Master program at the von Karman Institute.

\appendix

\section{Notes on Filter Design and Wavelet Theory for MRA}\label{Annex1}

 The MRA presented in \S\ref{SUB2} can operate with any digital filter, provided that the spectral overlapping between the transfer function of adjacent scales is reasonably small and the sum of the modulus of all the transfer functions in the filter bank yields unity. In the formulation of the mPOD presented in this work, we have used the simplest possible filters, namely Finite Impulse Response (FIR) linear filters. 
 
 These are characterized by their impulse response $h[k]$, that is the inverse Fourier transform of the associated transfer function. The impulse response (or kernel) is the output of the filter when acting on a Dirac impulse $\delta[k]$, and its design can be carried out using standard methods presented in classical textbooks on signal analysis \citep{Signal_1,Signal_2}. Among these, the current mPOD algorithm uses the windowing method, which consists in smoothing an ideal transfer function
  
\begin{equation}
\label{ideal}
H^{I}_{\mathcal{L}m}(\omega)=\begin{cases}
        e^{-\alpha \omega \mathrm {j}} & \text{for } |\omega_n|  \leq 2\pi f^c_{m}\\
        0 & \text{otherwise } 
              \end{cases}\,,
\end{equation}

{\parindent0pt with a kernel of size $N$, introducing} a constant phase delay $\alpha=N/2$. The ideal transfer function for each scale is constructed from the frequency splitting vector $F_V$. This leads to an unstable filter with infinite impulse response:

\begin{equation}
\label{Id}
\begin{split}
h^{I}_{\mathcal{L}m}[k]=\frac{1}{2\pi}\int^{\pi}_{-\pi}H^{I}_{\mathcal{L}m}(\omega) e^{-\mathrm{j}\omega k }d\omega=\frac{sin[2 \pi\,f^c_{m}/f_s\, (k-\alpha)]}{\pi [k-\alpha]}\,,
\end{split}
\end{equation}
 
 {\parindent0pt but the smoothing in the frequency domain} corresponds to a regularizing multiplication in the time domain:

\begin{equation}
\label{hlowp}
h_{\mathcal{L}m}[k]=h^{I}_{\mathcal{L}m}[k]\, w[k]\,.
\end{equation}

The linear phase response is provided by taking the window kernel to be symmetric about its midpoint, that is $w[n]=w[N-n]$. The choice of the window, and most importantly its size $N$, controls how much the real transfer function differs from the ideal one, in terms of pass-band/stop-band deviation, and transition band \citep{Signal_1,Signal_2}. Once the window is chosen, the low-pass transfer function corresponding to the impulse response in \eqref{hlowp} is:
	
\begin{equation}
\label{acca}
H_{\mathcal{L}_m}=\frac{1}{2\pi}\int^{\pi}_{-\pi}H^{I}_{\mathcal{L}m}(\omega)\,\widehat{w}(\omega-\theta) d\theta
\end{equation}

{\parindent0ptwith $\omega=2\pi f_n/f_s$ the normalized frequency (radiant/samples) and $\widehat{w}=\Psi_\mathcal{F}\,w$ the DFT of the window}.  Observe that \eqref{acca} has an analytical representation and thus the transfer function of appropriate size $H_{\mathcal{L}_m}\in \mathbb{C}^{1\times n_t}$, here constructed as a row vector for later convenience, can be computed from \eqref{acca} on an arbitrary frequency discretization. In this work, the default window chosen is the Hanning window, which gives a stop-band attenuation of $-44 dB$ (decibels) and transition width of $3.1/N$.

 Concerning the filter order, one should observe that a sharp frequency cut-off becomes more and more difficult when lower frequencies are considered. This parameter is therefore left as a function of the frequency bandwidth considered, as it is common in the filter bank formulation of the Continuous Wavelet Transform \citep{Wavelet1,Strang_WAVELET}. For the test case in \S \ref{VIII}, for example, the filter orders $N_m$ is
\begin{equation}
\label{N_f}
N_m=
\begin{cases}
       0.5 n_t& \text{for } F_V[m]  \leq f_s/20\\
   n_t\, f_s/(40\, F_V[m])& \text{for } Fs/20\leq F_V[m]  \leq f_s/2
              \end{cases}\,.
\end{equation}

 A zero-phase implementation of these filters, using for example the Matlab function $\emph{filtfilt}$, is based on backward and forward filtering, in order to cancel the phase delay \citep{Filtfilt,Matlab_Toolbox}. This operation doubles the filter order, squares the modulus of its transfer function but renders it non-causal.

Finally, it is worth noticing that in previous formulations of the mPOD (see \citealt{Mendez_Journal_2,Mendez_ICNAM}), the filter bank architecture is presented in terms of Discrete (dyadic) Wavelet Transform (DWT). This formulation replaces the concept of impulse response with that of wavelet (for high-pass) and scale (for low-pass) functions, constructed from dilatation and shifting of reference functions. The link between wavelet theory and filter banks is extensively discussed by \cite{Wavelet1} and \cite{Strang_WAVELET}.

For the purposes of this work, it suffices to notice that the MRA of $K$ can by carried out via Wavelet toolboxes for image compression/denoising \citep{Gonzalez,Wavelet1}, based on the pyramid algorithm proposed by \cite{Mallat89}. In this case, however, both the scaling and the shift of the reference wavelet/scale functions are fixed and the frequency splitting vector $F_V$ becomes: $F_V=[2^{-M},\dots 2^{-3},2^{-2},2^{-1}]\,fs/2$ with $f_s$ the sampling frequency and $M=log_2(n_t)$ the number of scales.


\bibliographystyle{jfm}
\bibliography{REFERENCES}

%
%
%
%

\end{document}